\documentclass[superscriptaddress,aps,prb,reprint,twocolumn,amsmath,amssymb]{revtex4-2}
\usepackage{graphicx}
\usepackage{hyperref}
\usepackage{amssymb}
\usepackage{slashed}
\usepackage{dcolumn}
\usepackage{amsmath}
\usepackage{bm}
\usepackage{colordvi}
\usepackage{algorithm}
\usepackage{algpseudocode}
\usepackage{multirow}
\usepackage{newtxtext,newtxmath}
\usepackage{dsfont}
\usepackage{xcolor}
\usepackage{mathbbol}

\usepackage{hyperref}
\hypersetup{
    colorlinks=true,
    linkcolor=blue,
    citecolor=blue,     
    urlcolor=blue,
}

\allowdisplaybreaks

\usepackage{mathrsfs}
\makeatletter

\newcommand{\Rmnum}[1]{\expandafter\@slowromancap\romannumeral #1@}
\makeatother

\begin{document}

\title{Revealing THz optical signatures of Shiba-state-induced  gapped and gapless superconductivity}

\author{F. Yang}
\email{fzy5099@psu.edu}

\affiliation{Department of Materials Science and Engineering and Materials Research Institute, The Pennsylvania State University, University Park, PA 16802, USA}

\author{R. Y. Fang}
\affiliation{Department of Engineering Science and Mechanics, The Pennsylvania State University, University Park, PA 16802, USA}

\author{S. L. Zhang}
\affiliation{Department of Engineering Science and Mechanics, The Pennsylvania State University, University Park, PA 16802, USA}

\author{L. Q. Chen}
\email{lqc3@psu.edu}

\affiliation{Department of Materials Science and Engineering and Materials Research Institute, The Pennsylvania State University, University Park, PA 16802, USA}

\date{\today}

\begin{abstract}   
  We report a fully self-consistent calculation of the complex renormalization by exchange interactions and hence the complete phase diagram of conventional $s$-wave superconductors with magnetic impurities as well as the related physical properties including the optical response. We show that a small amount of magnetic disorder can drive the system into a gapless superconducting state, where the single-particle excitation gap vanishes whereas the superconducting order parameter $\Delta_0$ remains finite. In this phase, the linear optical conductivity exhibits a finite absorption over the low-frequency regime, particularly for photon energies below the conventional threshold $2|\Delta_0|$, even at low temperatures, in sharp contrast to the gapped state. The nonlinear response, however, remains coherent and is dominated by the Higgs-mode dynamics rather than gapless quasiparticle background. These findings reveal a fundamental distinction between dissipative single-particle excitations and  coherent collective dynamics of the condensate, a feature likely general to other gapless superconductors, and introduces a fundamentally different detection scheme, using THz spectroscopy to probe the signatures of Shiba states.

\end{abstract}

\maketitle  

\section{Introduction}

Conventional $s$-wave superconductivity is characterized by the formation of Cooper pairs and the opening of a full energy gap $|\Delta_0|$ near Fermi surface~\cite{bardeen1957theory,schrieffer1964theory}, such that the fermionic quasiparticle states are pushed by $|\Delta_0|$ above and below the Fermi surface and no electronic states remain inside the gap. As a result, as demonstrated in recent advances in THz optics~\cite{PhysRevLett.109.187002,PhysRevB.105.014506,yang2018terahertz}, the linear optical response  of superconductors at low temperatures shows complete suppression of absorption for photon energies below the threshold $2|\Delta_0|$, reflecting the absence of available single-particle states within the gap. This transparency allows the intense THz pulses to coherently excite the Higgs mode in the nonlinear regime~\cite{matsunaga2013higgs,matsunaga2014light,shimano2020higgs,Kim2024Tracing,PhysRevLett.122.257001,PhysRevB.105.L100508,PhysRevLett.124.207003,chu2020phase,PhysRevLett.120.117001,yang2019lightwave,luo2023quantum,yang2018terahertz,PhysRevB.102.054510},  which represents the amplitude fluctuations of the superconducting (SC) order parameter and appears as a gapped collective excitation at $\hbar\omega_H = 2|\Delta_0|$  in the long-wavelength limit~\cite{yang2019gauge,yang2023optical,pekker2015amplitude,cui2019impact,PhysRevB.102.054517}. This coherent excitation enables a transient modulation of the superfluid density~\cite{shimano2020higgs,Kim2024Tracing,PhysRevLett.122.257001,PhysRevLett.124.207003,PhysRevLett.120.117001,yang2019lightwave,luo2023quantum,yang2018terahertz}, making it appealing for controlled ultrafast manipulation of the condensate.

Experimental and theoretical studies over past few decades have revealed the existence of in-gap excitations induced by external perturbations~\cite{abrikosov1957magnetic,caroli1964bound,luh1965bound,shiba1968classical,rusinov1969superconductivity,RevModPhys.80.1083,PhysRevLett.100.096407,PhysRevLett.105.077001,PhysRevLett.105.177002,annurev,RevModPhys.78.373}. These excitations, though energetically confined within the SC gap, remain robust and coherent, protected by the gap itself, making them promising candidates for applications in SC quantum devices~\cite{RevModPhys.80.1083,PhysRevLett.100.096407,PhysRevLett.105.077001,annurev}. A paradigmatic example is a single magnetic impurity embedded in superconductors~\cite{menard2015coherent,hatter2015magnetic,PhysRevB.83.054512,hudson2001interplay}, which produces a pair of localized Yu–Shiba–Rusinov (YSR) bound states~\cite{luh1965bound,shiba1968classical,rusinov1969superconductivity,RevModPhys.78.373} via local exchange interactions that localize Cooper pairs.  Extensive scanning tunneling microscopy (STM) studies have mapped YSR states around individual impurities with atomic resolution~\cite{PhysRevLett.126.017001,doi:10.1126/science.1259327,PhysRevLett.100.226801,kuster2021long,choi2017mapping,beck2021spin,franke2011competition,PhysRevB.97.195429,PhysRevLett.118.117001,PhysRevB.103.235437,PhysRevB.102.174504,PhysRevLett.117.186801}. Notably, at finite   concentrations, hybridization of YSR states is predicted to generate impurity bands within the gap~\cite{shiba1968classical}, significantly altering SC properties. For example, by tailoring the spatial arrangement of magnetic atoms, the induced impurity bands can acquire non-trivial topology and host Majorana bound states at their ends~\cite{poyhonen2018amorphous,kim2018toward,schneider2020controlling,schneider2021topological,PhysRevB.100.075420,PhysRevB.88.020407,PhysRevB.84.195442,kezilebieke2020topological}, offering a promising route toward fault-tolerant quantum computation.
 
When the impurity-band bandwidth becomes comparable to the parent SC gap, early theoretical studies based on critical analysis~\cite{shiba1968classical,osti_4097498,gulian2002nonequilibrium}, predicted an intriguing  crossover from a gapped to a gapless SC state. In this state, the SC order parameter remains finite despite the closure of the single-particle excitation gap, leading to gapless superconductivity in $s$-wave systems.  {In general, the proliferation of low-energy quasiparticles in  SC state tends to undermine the stability and long-range phase coherence of Cooper pairs, thereby strongly influencing the associated properties of superconductivity. Nevertheless, existing theoretical studies have largely focused on isolated (one or two) impurities~\cite{PhysRevB.110.205404,PhysRevLett.78.3761,rg1x-bztv,PhysRevLett.127.186804,PhysRevB.105.174517,PhysRevB.56.11213,PhysRevB.99.174502} instead of a realistic continuous distribution, or on asymptotic and critical behaviors of the SC phase~\cite{PhysRevLett.26.428,Schuh1978,PhysRevB.41.4815,MULLERHARTMANN1972401,MULLERHARTMANN1976439,MATSUURA19761583,10.1143/PTP.57.1823}. Consequently, the regime far from the critical point, where transport, tunneling, non-equilibrium and optical measurements are most relevant, has remained essentially unexplored.}
 Resolving this issue requires a fully self-consistent numerical treatment involving complex-valued spectral functions and intricate branch-cut structures in the complex plane~\cite{shiba1968classical} across the full parameter space (temperature and impurity concentration). This challenge has rendered the issue a long-standing conjecture. In addition, techniques capable of identifying a coherent, topologically robust SC phase are essential to confirm that Majorana bound states~\cite{poyhonen2018amorphous,kim2018toward,schneider2020controlling,schneider2021topological,PhysRevB.100.075420,PhysRevB.88.020407,PhysRevB.84.195442,kezilebieke2020topological} are hosted by a stable condensate rather than trivial low-energy excitations. However, to date, experimental insight has relied almost exclusively on STM–based local probes~\cite{RevModPhys.78.373}, and a bulk-sensitive, non-contact spectroscopic technique to detect YSR bands and the gapless phase at the macroscopic level is still lacking.

For this purpose, we propose a numerically successful algorithm that enables a fully self-consistent calculation of the complete phase diagram of conventional $s$-wave superconductors with magnetic impurities, as well as related physical properties such as the optical response. We expect that linear and nonlinear THz spectroscopies can disentangle the dissipative (quasiparticle) and reactive (condensate) channels without interference from contacts or vortex motion, and that tracking these signatures as the single-particle excitation gap closes provides one of the most stringent tests of whether SC coherent dynamics~\cite{shimano2020higgs} can persist in the presence of a finite density of low-energy excitations and even gapless superconductivity.

The simulation results uncover a gapless SC phase induced by a small amount of magnetic disorder, where the single-particle excitation gap vanishes, yet the SC order parameter remains finite. This quantitatively confirms Shiba's conjecture regarding the existence of gapless superconductivity driven by magnetic impurities~\cite{shiba1968classical}. The linear optical conductivity of this phase exhibits clear signatures of gapless superconductivity, including finite absorption that fills the low-frequency regime, particularly for photon energies below the conventional threshold $2|\Delta_0|$, even at low temperatures. This behavior, reflecting the modification of the single-particle spectrum and the enhancement of dissipative channels, stands in stark contrast to the gapped phase~\cite{PhysRevLett.109.187002,PhysRevB.105.014506,yang2018terahertz,PhysRevB.109.064508}, which displays vanishing low-frequency absorption at zero temperature alongside sharp, temperature-dependent resonances.
However, the non-linear optical response remains largely intact and coherent, dominated by the collective dynamics of the SC condensate (i.e., Higgs-mode dynamics) rather than the gapless quasiparticle background. Thus, while magnetic impurities strongly modify linear dissipative processes, they leave the nonlinear, coherent response of the condensate nearly unaffected. These findings reveal a fundamental distinction between single-particle excitations, which govern dissipation, and collective condensate dynamics, which preserve SC order and coherence.

\section{Model of impurity Shiba bands}

{We begin with the full SC Hamiltonian that includes the $s$–$d$ interaction between electrons and magnetic impurities~\cite{PhysRevB.109.064508,shiba1968classical}:
\begin{eqnarray}
  H&=&\frac{1}{2}\sum_{\bf k}\Psi^{\dagger}_{\bf k}(\xi_{\bf k}\tau_3\!-\!\Delta_0\tau_2\sigma_2)\Psi_{\bf k}\!-\!\frac{1}{2}J\sum_{\bf kk'}\Psi^{\dagger}_{\bf k}\tilde{\bm \sigma}\Psi_{\bf k'}\!\cdot\!{\bf S},~~~~\label{Ham}
\end{eqnarray}
where $\Psi_{\bf k} = (\psi_{{\bf k}\uparrow}, \psi_{{\bf k}\downarrow}, \psi^{\dagger}_{-{\bf k}\uparrow}, \psi^{\dagger}_{-{\bf k}\downarrow})^{T}$ is the Nambu$\otimes$spin-space spinor, and $\xi_{\bf k} = k^2/(2m) - \mu$ is the normal-state dispersion with effective mass $m$ and chemical potential $\mu$, while $\sigma_i$ and $\tau_i$ are the Pauli matrices in Nambu and spin particle-hole space, respectively.  
The operator ${\tilde{\bm \sigma}}={\bm \sigma}{(1+\tau_3)/}{2}+\sigma_2{\bm \sigma}\sigma_2(1-\tau_3)/{2}$ represents the spin structure in the Nambu space, while ${\bf S}$ and $J$ denote the local spin and exchange interaction strength in the $s$–$d$ model, respectively.}

The generalized Green function, which in momentum-frequency space {takes the form} $G_{\bf k}(\omega)=-i\langle\Psi_{\bf k}(\omega)\Psi_{\bf k}^{\dagger}(\omega)\rangle$, satisfies the Dyson equation~\cite{abrikosov2012methods}:
\begin{equation}\label{Greenfunction}
G_{\bf k}(\omega)=G_{0{\bf k}}(\omega)+G_{0{\bf k}}(\omega)\Sigma(\omega)G_{\bf k}(\omega),  
\end{equation}
where the bare Green function is written as~\cite{abrikosov2012methods,PhysRevB.99.224511,yang2024optical,yang2023optical} 
\begin{equation}\label{G0}
G_{0{\bf k}}(\omega)=\frac{\omega\!+\!\xi_{\bf k}\tau_3\!-\!\Delta_0\tau_2\sigma_2}{\omega^2-\xi_{\bf k}^2-\Delta_0^2}.  
\end{equation} 
The self-energy $\Sigma(\omega)$ due to exchange interactions, evaluated within the random phase approximation (to incorporate both the spatial randomness and random orientations of the impurity spins), takes the form~\cite{abrikosov2012methods} 
\begin{align}
&\Sigma(\omega)=n_i{J}({\bf S}\cdot{\tilde{\bm \sigma}})\Big[\sum_{\bf k}G_{\bf k}(\omega)\Big]{J}({\bf S}\cdot{\tilde{\bm \sigma}})\nonumber\\
&\mbox{}+{J}({\bf S}\cdot{\tilde{\bm \sigma}})\Big[\sum_{\bf k}G_{\bf k}(\omega)\Big]\Sigma(\omega)\Big[\sum_{\bf k}G_{\bf k}(\omega)\Big]{J}({\bf S}\cdot{\tilde{\bm \sigma}}).~~~\label{SE}
\end{align} 
Here, $n_i$ is the magnetic-impurity density.

{To determine the superconducting order parameter $\Delta_0$ in the presence of magnetic disorder}, the self-consistent gap equation within the Green-function framework reads~\cite{PhysRevB.109.064508,PhysRevB.95.235403,RevModPhys.77.935,PhysRevLett.25.507,PhysRevLett.116.127002,yang2024optical}:
\begin{equation}\label{GGEE}
\frac{\Delta_0}{g}=i\int\frac{d\omega}{2\pi}\sum_{\bf k}\tanh\Big(\frac{\omega}{2k_BT}\Big)F_{\bf k}(\omega+i0^+,\Delta_0),  
\end{equation}
where $g$ denotes the pairing potential, $T$ is the temperature, and $F_{\bf k}(\omega+i0^+,\Delta_0)$ is the SC anomalous Green function obtained from Eq.~(\ref{Greenfunction}) in the presence of the self-energies.

To self-consistently calculate the Green function from Eqs.~(\ref{Greenfunction}) and (\ref{SE}), one can follow the self-energy renormalization theory~\cite{abrikosov2012methods} and Shiba's derivation~\cite{shiba1968classical}, assuming that the analytical structure of the Green function remains unchanged in the presence of the $s$–$d$ interaction. This leads to a renormalized Green function of the form  in frequency-momentum space~\cite{abrikosov2012methods,PhysRevB.109.064508,PhysRevB.95.235403,RevModPhys.77.935,PhysRevLett.25.507,PhysRevLett.116.127002,yang2023optical,yang2024optical} 
\begin{equation}\label{RG}
G_{\bf k}(\omega)=\frac{\tilde\omega\!+\!\xi_{\bf k}\tau_3\!-\!\tilde\Delta_0\tau_2\sigma_2}{\tilde\omega^2-\xi_{\bf k}^2-\tilde\Delta_0^2}.  
\end{equation}
{The parameters $\tilde{\omega}$ and $\tilde{\Delta}_0$ are the renormalized frequency and the gap in this renormalized Green function, respectively, and explicitly, based on the Dyson equation [Eq.~(\ref{Greenfunction})] and the self-energy arising from external perturbations [Eq.~(\ref{SE})], they are functionals of $\omega$ and $\Delta_0$, i.e. $\tilde{\omega}=\tilde{\omega}(\omega,\Delta_0)$ and $\tilde{\Delta}_0=\tilde{\Delta}_0(\omega,\Delta_0)$.} It should be emphasized here that  $\tilde{\Delta}_0$ is not a directly observable quantity but an auxiliary renormalization function in Shiba’s self-consistent formalism, i.e., an intermediate function that encodes the renormalization of the pairing self-energy and must be solved simultaneously with the renormalized frequency function $\tilde{\omega}(\omega,\Delta_0)$. Then, one needs to self-consistently determine the real physical SC gap $\Delta_0$ through the gap equation [Eq.~(\ref{GGEE})] using Eq.~(\ref{RG}).

Nonmagnetic impurities give rise to null renormalization~\cite{suhl1959impurity,skalski1964properties,andersen2020generalized}, known as the Anderson theorem~\cite{anderson1959theory}. Magnetic impurities induce complex renormalization due to exchange interactions~\cite{shiba1968classical,RevModPhys.78.373,skalski1964properties}. As derived by Shiba, this renormalization leads to two coupled equations. {Specifically, assuming $S_x^2=S_y^2=S_z^2=S^2/3$ and $S_xS_y=S_xS_z=S_yS_z=0$, using Eq.~(\ref{RG}), the self-energy in Eq.~(\ref{SE}) becomes
\begin{equation}\label{FSE}
\Sigma(\omega)=\frac{n_i(JS/2)^2Z(\omega)}{1-[JSZ(\omega)/2]^2},  
\end{equation}
with
\begin{equation}\label{Zfunction}
Z(\omega)=\sum_{\bf k}G_{\bf k}(\omega)=-\pi{D}\frac{\tilde\omega-\tilde\Delta_0\sigma_2\tau_2}{\sqrt{\tilde\Delta_0^2-\tilde\omega^2}}.  
\end{equation}
Here, $D$ denotes the normal state density of states at the Fermi level. Substituting Eqs.~(\ref{RG})–(\ref{FSE}) into the Dyson equation, one obtains Shiba's original renormalization equations~\cite{abrikosov2012methods}:
\begin{equation}\label{RE}
\frac{\omega}{\Delta_0}=\frac{\tilde\omega}{\tilde\Delta_0}\bigg[1\!-\!\frac{\gamma_s}{\Delta_0}\frac{\sqrt{1\!-\!(\frac{\tilde\omega}{\tilde\Delta_0})^2}}{\eta^2\!-\!(\frac{\tilde\omega}{\tilde\Delta_0})^2}\bigg],  
\end{equation}
and
\begin{equation}\label{RD}
{\tilde\Delta_0}=\Big[1-\frac{1-(JSD{\pi}/2)^2}{2}\Big(1-\frac{{\omega}/{\Delta_0}}{{\tilde\omega}/{\tilde\Delta_0}}\Big)\Big]\Delta_0.
\end{equation}
Here, $\gamma_s=2n_iD\pi(JS/2)^2/[1+(JSD{\pi}/2)^2]^2$ denotes the quasiparticle relaxation rate due to the $s$–$d$ interaction, and $D$ is the normal-state density of states at the Fermi level. The parameter $\eta$ is written as 
\begin{equation}
\eta=\frac{1-(JSD{\pi}/2)^2}{1+(JSD{\pi}/2)^2},
\end{equation}
which characterizes the energy splitting $\pm\eta \Delta_0$ of the pair of YSR bound states induced by a single magnetic impurity in an $s$-wave superconductor~\cite{luh1965bound,shiba1968classical,rusinov1969superconductivity}.}

Complex solutions to Eq.~(\ref{RE}) naturally emerge for $\omega>\Delta_0$, reflecting the expected damping of Bogoliubov quasiparticles by magnetic impurities. Remarkably, a qualitative analysis by Shiba proposed~\cite{shiba1968classical} that complex solutions can also appear within the SC gap ($\omega < \Delta_0$), and result in finite density of states, indicating the formation of in-gap impurity bands centered at $\pm\eta|\Delta_0|$, with a bandwidth that scales as the square root of the impurity concentration. This implies that increasing the impurity concentration can, in principle, drive the system into a gapless SC state. However, it also suppresses the SC gap $\Delta_0$ through the self-consistent feedback encoded in the full gap equation. Whether superconductivity survives in this regime therefore hinges on the fully self-consistent solution of the coupled equations, which requires advanced numerical techniques, as both the relevant functions and the coupled parameter space are intrinsically complex-valued and involve nontrivial branch cuts in the complex frequency plane.

\section{Numerical algorithm}

{Here we propose a numerical algorithm to solve this problem. First, we transform the complex-valued equation in Eq.~(\ref{RE}), which contains nontrivial branch cuts, into a multi-valued form, specifically, a sixth-order polynomial equation:}
\begin{eqnarray}
&&y^6-2xy^5+(x^2-2\eta^2+r^2)y^4+4\eta^2xy^3\nonumber\\
&&+(\eta^4-2\eta^2x^2-r^2){x^2}-2\eta^4xy+\eta^4x^2=0,\label{sixE}
\end{eqnarray}
where $y={\tilde\omega}/{\tilde\Delta_0}$, $x={\omega+i0^+}/{\Delta_0}$ and $r=\gamma_s/\Delta_0$. We then utilize a sophisticated, robust and general numerical scheme that is well-suited for solving the complex polynomial equations of arbitrary degree~\cite{skowron2012generalcomplexpolynomialroot}.

{However, the sixth-order equation in Eq.~(\ref{sixE}) generally yields six complex roots, among which only one corresponds to a physically meaningful solution. This ambiguity arises because, although our transformation successfully eliminates the nontrivial branch cuts, it inevitably introduces the issue of multiple mathematical solutions. To identify the physical root, we apply the following two-step strategy to ensure the analytical consistency of the Green function. First, since we are considering the retarded Green function, we retain only the solutions with ${\rm Im}(y) \geq 0$. Specifically, for Bogoliubov quasiparticles (when $|x| > 1$), the imaginary part of $y$ must be strictly positive (${ \rm Im}(y) > 0$) because these solutions correspond to real excitations with a finite decay rate. In contrast, for in-gap states (when $0 < |x| \leq 1$), the imaginary part of $y$ is allowed to be zero or non-negative, as these states correspond to localized states. Second, after the first-step procedure, we select the root that is closest to $y=x$ in the real part, allowing us to obtain the unique solution to the equation in Eq.~(\ref{sixE}). The last procedure is necessary because the solution $y=x$ corresponds to the case with no magnetic impurities, where the system is in its unperturbed state. In this case, $y=x$ is a simple, non-complex solution that represents the absence of impurity effects. When impurities are introduced, the solution typically deviates from this condition. To ensure continuity (i.e., a smooth and continuous transition as the system evolves from a state without impurities to a system with impurity effects), we select the root that is closest to $y=x$ in the real part. This choice ensures that the solution varies smoothly and avoids erratic jumps between different branches of the multi-valued function, thus acting as a {\it stability criterion} of the solution in the mathematical context.}

After solving for the unique solution of ${\tilde\omega}/{\tilde\Delta_0}$ form Eq.~(\ref{sixE}), we substitute this solution into Eq.~(\ref{RD}) to obtain the solutions for $\tilde{\Delta}=\tilde{\Delta}(\omega,\Delta_0)$, and consequently $\tilde{\omega}=\tilde{\omega}(\omega,\Delta_0)$. These solutions are then substituted into the SC anomalous Green function $F_{\bf k}(\omega,\Delta_0)$:
\begin{equation}
F_{\bf k}(\omega,\Delta_0)=-\frac{\tilde\Delta_0}{\tilde\omega^2-\xi_{\bf k}^2-\tilde\Delta_0^2}.
\end{equation}
By performing the $\omega$-integration and momentum summation, we can self-consistently solve for the real  physical SC gap $\Delta_0$ in the presence of magnetic impurities from the gap equation {[Eq.~(\ref{GGEE})]}:
\begin{eqnarray}
  \frac{\Delta_0}{g}&=&i\int\frac{d\omega}{2\pi}\sum_{\bf k}\tanh\Big(\frac{\omega}{2k_BT}\Big){F_{\bf k}(\omega+i0^+,\Delta_0)}\nonumber\\
 \Rightarrow \frac{1}{g}&=&D\int\frac{d\omega}{2}\tanh\Big(\frac{\omega}{2k_BT}\Big)\frac{\tilde\Delta_0/\Delta_0}{\sqrt{\tilde\omega^2-\tilde\Delta_0^2}}.  
\end{eqnarray}
Consequently, with this approach, we are able to rigorously and self-consistently solve the gap equation, thereby determining the full phase diagram as well as the related physical properties.

{By systematically eliminating the redundant  unphysical roots and selecting the physically unique solution, the algorithm we developed exhibits highly stable convergence across the entire parameter space, ensuring reliable numerical results.  Notably, the method is computationally efficient, requiring minimal resources, which enables rapid calculations typically completed in just a few minutes for the entire phase diagram under various conditions. As a result, we can easily obtain the full phase diagram for temperature and magnetic-impurity concentration, making this approach highly suitable for subsequent calculations of various properties in superconductors with magnetic impurities, as the full equilibrium Green function is exactly obtained. 
 In fact, once the full equilibrium  Green function is obtained, a wide range of observables, such as transport coefficients, tunneling properties, dynamic susceptibilities, various collective mode spectra, and the RKKY interaction, can, in principle, be computed within the same framework. In the following, we provide several examples, including the single-particle excitation density of states, linear and non-linear optical responses. }

\begin{center}
\begin{table}[htb]
  \caption{{Specific parameters used in the numerical simulations. For the gap equation, the pairing interaction strength $g$ (i.e., the dimensionless one $gD$) is determined by matching the zero-temperature gap value in the absence of magnetic impurities, $\Delta(T = 0, n_i = 0)$. The frequency integrals are carried out with a cutoff at the Debye frequency $\omega_D$, following the conventions established by Abrikosov and Gorkov ~\cite{abrikosov2012methods,PhysRevB.99.224511,RevModPhys.58.323,yang2023optical,yang2024optical}. Under the normalization convention here, our results are unaffected by the specific value of the normal-state density of states. }}  
\renewcommand\arraystretch{1.5}   \label{parameter}
\begin{tabular}{l l l l l l l}
    \hline
    \hline
      &\quad\quad~$\Delta(T=0,n_i=0)$&\quad\quad~$\omega_D$&\quad\quad~$0^+$\\  
    &\quad\quad\quad2~meV&\quad\quad$12.32~\mathrm{meV}$&\quad\quad$0.02~\mathrm{meV}$\\ 
    \hline
    \hline
\end{tabular}\\
\end{table}
\end{center}

\section{Phase diagram}

We choose $JS{\pi}D/2=0.5$, a moderate exchange coupling, which corresponds to $\eta = 0.6$, consistent with values observed in STM experiments~\cite{PhysRevLett.100.226801,kuster2021long,choi2017mapping,beck2021spin,franke2011competition,PhysRevB.97.195429,PhysRevB.102.174504,PhysRevLett.117.186801}.  {As the imaginary part of the $\sigma_0\tau_0$ component of the retarded Green function corresponds to the spectra function, we can calculate the single-particle-excitation density of states as~\cite{PhysRevB.109.064508,shiba1968classical,abrikosov2012methods}:
\begin{equation}
\rho(\omega)={\rm Im}{\rm Tr}\Big[\sum_{\bf k}G_{\bf k}(\omega+i0^+)\Big]/(4\pi)=-{\rm Im}\Big[\frac{{D}\tilde\omega}{\sqrt{\tilde\Delta_0^2-\tilde\omega^2}}\Big].\label{DOE}  
\end{equation}
In the absence of magnetic impurities, and thus without the renormalization, the density of states $\rho(\omega)$ from Eq.~(\ref{DOE}) recovers the conventional result from BCS theory~\cite{abrikosov2012methods,PhysRevB.99.224511}. In this case, it becomes finite for $\omega\ge\Delta_0$ but vanishes for $0<\omega<\Delta_0$, as expected. This behavior arises because the continuum of Bogoliubov quasiparticles lies above the SC gap.}

{The superfluid density $N_s(T)$, whose expression has been derived in our previous work using a diagrammatic formulation~\cite{PhysRevB.109.064508}, is written as 
\begin{equation}
\frac{N_s}{N}=\int\frac{d\omega}{2\pi}\tanh\Big(\frac{\omega}{2k_BT}\Big)\frac{\pi\tilde\Delta_0^2}{(\tilde\omega^2-\tilde\Delta_0^2)^{3/2}},  
\end{equation}
with $N$ being the electron density.}

\begin{figure}[htb]
  {\includegraphics[width=8.6cm]{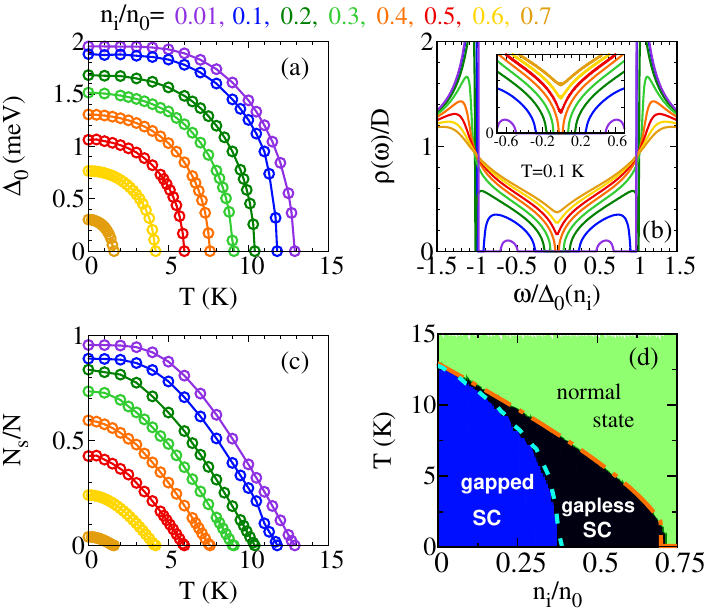}}
  \caption{(a) Temperature dependence of SC gap, (b) single-particle-excitation density of states, {and (c) temperature dependence of superfluid density,} at different impurity concentrations $n_i$. The inset in (b) provides an enlarged view of the data at $n_i/n_0=0.1,0.4,0.7$. {(d) The calculated $T$-$n_i$ phase diagram, including the gapped SC [$\Delta_0\ne0$ and $\rho(0^+)=0$] and gapless SC [determined by $\Delta_0\ne0$ and $\rho(0^+)\ne0$] states as well as the normal state ($\Delta_0=0$). The orange chain curve in (d) denotes the critical temperature $T_c(n_i)$ obtained from the critical theory by Abrikosov and Gorkov ~\cite{osti_4097498,gulian2002nonequilibrium}, 
  while the light-blue dashed curve in (d) denotes the temperature $T^*$ at which 
  $\gamma_s(n_i)=0.85\Delta_0(T^*,n_i)\exp(-\pi/4)$.} Here, $n_0/(2\pi{D})=2.14$~meV. Other used parameters including pairing potential and Debye cutoff frequency are addressed in Table~\ref{parameter}.}    
\label{figyc1}
\end{figure}

Then, the numerically calculated SC gap $\Delta_0(T)$,  single-particle-excitation density of states  $\rho(\omega,T=0.1~\text{K})$, and superfluid density $N_s(T)$ for several impurity concentrations $n_i$ are plotted in Fig.~\ref{figyc1}.  As shown in Fig.~\ref{figyc1}(a) {and (c)}, increasing $n_i$ leads to progressive localization of Cooper pairs by magnetic impurities, resulting in a continuous suppression of both the zero-temperature gap $\Delta_0(T=0)$ and superfluid density $N_s$, thereby suppressing the critical temperature $T_c$. Superconductivity ultimately vanishes at a quantum critical point near $n_i \simeq 0.72n_0$. On the other hand, as seen from $\rho(\omega)$ in Fig.~\ref{figyc1}(b), as the concentration rises from zero, a pair of impurity bands emerges inside the gap at low $n_i$, centered at $\pm\eta\Delta_0$, in agreement with previous theoretical studies~\cite{shiba1968classical} of the YSR state and STM measurements~\cite{PhysRevLett.126.017001,doi:10.1126/science.1259327,PhysRevLett.100.226801,kuster2021long,choi2017mapping,beck2021spin,franke2011competition,PhysRevB.97.195429,PhysRevLett.118.117001,PhysRevB.103.235437,PhysRevB.102.174504,PhysRevLett.117.186801}.  

With further increase in $n_i$, both the bandwidth and the spectral weight of the impurity bands grow. Notably, when $n_i$ reaches a relatively low concentration of $n_i \approx 0.4n_0$ (critical value), the impurity bands broaden sufficiently to fill the low-energy spectrum (i.e., the bands extend over the entire low-energy window) and hence, completely close the single-particle excitation gap, signaling the crossover to a gapless SC state. Remarkably, the zero-temperature order-parameter amplitude [$\Delta_0(T=0)\approx1.3~$meV at $n_i=0.4n_0$] {and superfluid density [$N_s(T=0)\approx0.6{N}$ at $n_i=0.4n_0$]} remain finite and large in this case.  The coexistence of a robust condensate with a vanishing  single-particle excitation gap provides direct, quantitative confirmation of Shiba's prediction~\cite{shiba1968classical} that strong magnetic impurities can render an $s$-wave superconductor {\sl gapless} without destroying superconductivity. Thus, even after the excitation gap collapses, the system retains a finite superfluid stiffness and continues to exhibit zero resistivity.

{The corresponding $T$--$n_i$ phase diagram, including the gapped  and gapless SC states as well as the normal state, is shown in Fig.~\ref{figyc1}(d). It has been shown by Shiba~\cite{shiba1968classical} that the self-consistent renormalization theory reduces to the Abrikosov--Gorkov critical theory~\cite{osti_4097498,gulian2002nonequilibrium} in the limiting case of vanishing gap ($\Delta \to 0$), yielding identical $T_c(n_i)$. Therefore, for determining $T_c(n_i)$~\cite{discussion}, our rigorous calculations of the Shiba’s self-consistent renormalization theory  naturally recover the Abrikosov--Gorkov results. As seen from the figure, the calculated $T_c(n_i)$ obtained from our self-consistent theory is in good agreement with the prediction of the classical Abrikosov--Gorkov critical theory~\cite{osti_4097498,gulian2002nonequilibrium} [orange chain curve in Fig.~\ref{figyc1}(d)],
\begin{equation}
\ln\Big(\frac{T_c}{T_{c0}}\Big)=\psi\Big(\frac{1}{2}\Big)-\psi\Big(\frac{1}{2}+\rho\Big),
\end{equation}
where $\psi$ is the digamma function and $
\rho = \frac{\gamma_s}{2\pi T_c}$. 
Here, $T_{c0}$ is the critical temperature in the absence of impurities. Moreover, we find that the crossover temperature $T^*$ from the gapped to the gapless SC state 
in our self-consistently calculation can be well captured by the condition 
\begin{equation}
\gamma_s(n_i)=0.85\,\Delta_0(T^*,n_i)\,e^{-\pi/4},
\end{equation}
shown by dashed curve in Fig.~\ref{figyc1}(d). This critical behavior is close to the theoretical estimate $\gamma_s \sim \Delta_0 e^{-\pi/4}$ by Abrikosov–Gorkov critical theory~\cite{osti_4097498,gulian2002nonequilibrium}, and in exact agreement with Shiba's critical analysis and analytical conclusion~\cite{shiba1968classical} that the fully self-consistent solution systematically gives a crossover line smaller than the Abrikosov–Gorkov prediction. We emphasize that the aim of the present study is not on reproducing this  established critical behavior, but on achieving a comprehensive self-consistent calculation deep in the noncritical regime, where nonequilibrium  properties of superconductors are most relevant.}

\section{Linear optical response}

As shown in Fig.~\ref{figyc1}(b), once the system enters the gapless regime, the impurity bands fill the low-energy spectrum, and the sharp gap edges characteristic of a gapped superconductor are completely washed out. In this regime, probing intrinsic order-parameter magnitude $\Delta_0$ requires techniques that are sensitive to the condensate rather than single-particle tunneling measurements.  We therefore examine the optical response. 

{We first examine the linear-response optical absorption in conventional superconductors with magnetic impurities. Building on the Mattis-Bardeen theory for superconductors in the anomalous skin-effect regime~\cite{PhysRev.111.412,PhysRev.156.470,PhysRevB.102.144508}, where the mean free path $l=v_F\tau$ exceeds the skin depth, with $\tau$ being the momentum-relaxation time that accounts for all relevant scattering processes, the induced optical current at a given spatial point depends not only on the optical field at that point but also on the surrounding fields. This non-local interaction gives rise to the following expression for the current~\cite{PhysRev.111.412,PhysRev.156.470,PhysRevB.102.144508}:
\begin{equation}
{\bf j}({\bf r})=\int\frac{{\bf R}[{\bf R}\cdot{\bf A}({\bf r}')]I(\Omega,{\bf R})e^{-R/l}}{R^4}d{\bf r'}.  
\end{equation}
Here,  ${\bf R}={\bf r}-{\bf r}'$; $I(\Omega,{\bf R})$ is the normalized linear-response coefficient and $\Omega$ representing the optical frequency. In the   strong-scattering limit, where the coherence length $\xi$ is much larger than the mean free path $l=v_F\tau$, one can apply the mean value theorem for integrals, leading to the conventional approximation:
\begin{equation}
  {\bf j}({\bf r}){\approx}I(\Omega,{\bf R}=0){\bf A}({\bf r})\!\!\int\!\frac{e^{-R/l}}{3R^2}d{\bf r'}.
\end{equation}
This directly results in the optical conductivity:
\begin{equation}\label{oc}
\sigma_{s}=\sigma_{s1}+i\sigma_{s2}=\frac{4\pi{l}}{3i\Omega}\sum_{\bf q}I(\Omega,{\bf q}).  
\end{equation} 
For the optical absorption, only the current-current ($\tau_0$-$\tau_0$) correlation contributes. As a result, the real part of the optical conductivity, in the Keldysh formalism~\cite{RevModPhys.58.323}, can be expressed as~\cite{PhysRevB.102.144508}: 
\begin{eqnarray}
  &&\sigma_{s1}={\rm Re}\Big[\frac{4\pi{l}}{i3\Omega}\sum_{\bf q}\frac{e^2}{m^2}\chi^K_{00}(\Omega,{\bf q})\Big]\nonumber\\
 &&=\frac{4\pi{l}e^2}{3m^2\Omega}\int\frac{dE}{2\pi}{\rm Re}\Big[\sum_{\bf kq}\frac{1}{4}{\rm Tr}[\tau_0{\hat G}_{\bf k+q}(E+\Omega)\tau_0{\hat G}_{\bf k}(E)]_{K}\Big].\nonumber\\
\end{eqnarray}
Here, the subscript ``K'' denotes the Keldysh component. The Green function matrices ${\hat G}_{\bf k}(E)$ is defined as~\cite{RevModPhys.58.323}
\begin{equation}
  {\hat G}_{\bf k}(E)=\left(\begin{array}{cc}
    G^R_{\bf k}(E) & G_{\bf k}^K(E) \\
    0 & G_{\bf k}^A(E)
  \end{array}\right),
\end{equation}
with the retarded (R), advanced (A), and Keldysh (K) components given by~\cite{RevModPhys.58.323}: {\small $G^R_{\bf k}(E)=G_{\bf k}(E+i0^+)$, $G_{\bf k}^A(E)=G_{\bf k}(E-i0^+)$} and {\small $G^K_{\bf k}(E)=h(E)[G^R_{\bf k}(E)-G^A_{\bf k}(E)]$}, where {\small $h(E)=\tanh(\beta{E}/2)$} is the distribution function and {\small $\beta=1/(k_BT)$} is the inverse temperature.  Using the relations {\small ${\rm Re}G^R_{\bf k}(E)={\rm Re}G^A_{\bf k}(E)$} and {\small ${\rm Im}G^R_{\bf k}(E)=-{\rm Im}G^A_{\bf k}(E)$}, the optical absorption becomes 
\begin{eqnarray}
  \sigma_{s1}\!\!&=&\frac{4e^2\pi{l}}{3\Omega{m^2}}\int\frac{dE}{2\pi}\sum_{{\bf kq}}{\rm Tr}[{\rm Im}G^R_{\bf k+q}(E+\Omega){\rm Im}G^R_{\bf k}(E)]\nonumber\\
  &&\times\frac{h(E+\Omega)\!-\!h(E)}{2},
\end{eqnarray}
and by replacing $\sum_{\bf kq}$ with $\sum_{\bf kk'}$ (where ${\bf k'}={\bf k+q}$), one has 
\begin{equation}\label{FOA}
\frac{\sigma_{s1}}{\sigma_n}=\int{dE}\frac{f(E)\!-\!f(E+\Omega)}{\Omega}\frac{{\rm Tr}[{\rm Im}Z^R(E+\Omega){\rm Im}Z^R(E)]}{(2{\pi}D)^2},~~~~
\end{equation}  
where $\sigma_{n}=\frac{ne^2\tau}{m}$ represents the electrical conductivity in normal metals, and $f(x)$ is the Fermi function. As a self-consistent check,  in the case without magnetic impurities, the vanishing renormalization leads to  ${\rm Im}Z^R(E)=\pi{D}\frac{(E-\Delta_0\sigma_2\tau_2){\rm sgn}(E)}{\sqrt{E^2-\Delta_0^2}}\theta(|E|-\Delta_0)$, with $\theta(x)$ being the step function. Then, as proved in Ref.~\cite{PhysRevB.109.064508}, $\sigma_{s1}$ in Eq.~(\ref{FOA}) exactly recovers the result from the Mattis-Bardeen theory~\cite{PhysRev.111.412,PhysRev.156.470}.  In this case, as well established in the literature~\cite{PhysRev.111.412,PhysRev.156.470,PhysRevB.102.144508,PhysRevLett.109.187002,PhysRevB.105.014506,yang2018terahertz}, at $T=0~$K with only the contribution from the interband transition, the optical absorption $\sigma_{s1}(\Omega)$ vanishes when {\small $\Omega<2\Delta_0$} but becomes finite above $2\Delta_0$, leading to a crossover point at $2\Delta_0$. At finite temperature, an additional quasiparticle contribution appears below $2\Delta_0$ due to the intraband transition.} 

\begin{figure}[htb]
  {\includegraphics[width=8.7cm]{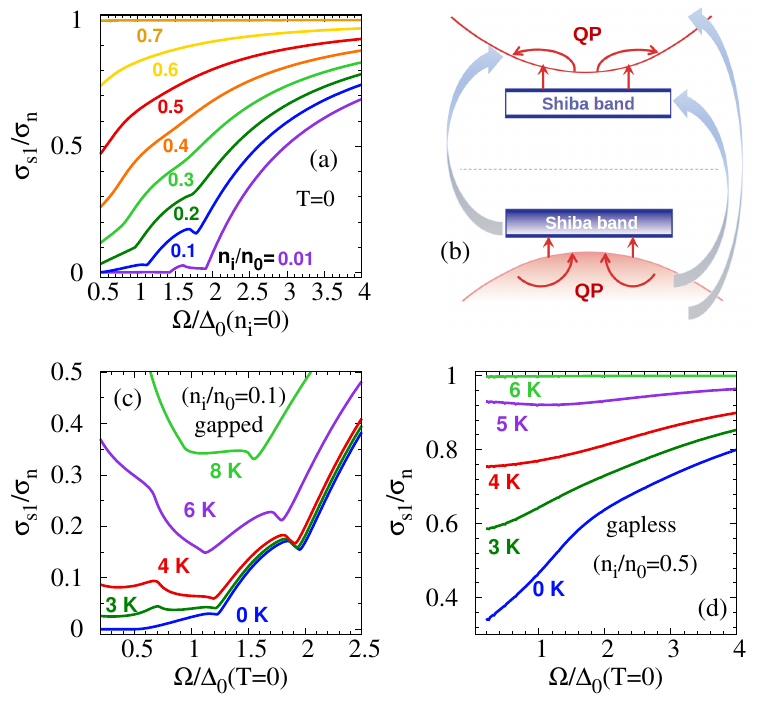}}
  \caption{(a) Zero-temperature optical absorption at different impurity concentrations $n_i$. (b) {Schematic illustration of the dominant interband transitions (gray arrows), which are present at all temperatures below $T_c$, including $T=0$, and the additional thermally activated transitions that appear only at finite temperatures (red arrows). The interband transition between impurity bands is usually weak and not shown here.} (c) and (d)  Optical absorption of gapped and gapless states at different $T$, respectively. The inset in (c) provides an enlarged view of the data at 4~K. Here, $n_0/(2\pi{D})=2.14$~meV. Other used parameters  are addressed in Table~\ref{parameter}.}    
\label{figyc2}
\end{figure}

{In the presence of magnetic impurities, one has to calculate $Z^{R}(E)=Z(E+i0^+)$ from Eq.~(\ref{Zfunction}) through the self-consistent calculation of Shiba's formalism, and substitute the obtained solution into Eq.~(\ref{FOA}) to obtain the optical absorption.} This formulation then naturally incorporates contributions from both impurity bands and quasiparticle continuum, yielding transitions  between states at $E$ and $E + \Omega$ [as illustrated in Fig.~\ref{figyc2}(b)].

Results from fully self-consistent numerical simulations are shown in Fig.~\ref{figyc2}. At $T = 0$, as seen in Fig.~\ref{figyc2}(a), the optical absorption in the gapped state (for dilute impurity concentrations $n_i/n_0 = 0.001$ and 0.1) displays two distinct features, as illustrated by gray arrows in Fig.~\ref{figyc2}(b). A sharp absorption onset appears at $\Omega > 2\Delta_0$, corresponding to interband transitions between Bogoliubov quasiparticles and quasiholes, consistent with the conventional absorption threshold~\cite{PhysRevLett.109.187002,PhysRevB.105.014506,yang2018terahertz,PhysRevB.84.224517}. {Inside the gap ($\Omega < 2\Delta_0$),  a small resonant absorption appears at $\Omega \sim 1.6\Delta_0$ for $n_i=0.1n_0$ and $n_i=0.01n_0$, associated with interband transitions from the hole-like impurity band to the Bogoliubov quasielectron continuum and from the  Bogoliubov quasihole continuum to the electron-like impurity band.} However, these distinct features are progressively suppressed with increasing impurity concentration. Once the system crosses into the gapless regime, a finite optical absorption $\sigma_{s1}(\Omega)$ persists at low frequencies even at $T = 0$, and increases smoothly with $\Omega$, in stark contrast to the gapped case where low-frequency absorption is absent.

At finite temperatures, as shown in Fig.~\ref{figyc2}(c), the gapped state exhibits additional spectral features,
 {due to the additional thermally activated
transitions that appear {\sl only} at finite temperature [as illustrated by red arrows in Fig.~\ref{figyc2}(b)]}.  These include a low-frequency upturn in $\sigma_{s1}(\Omega)$ as $\Omega$ drops below $\Delta_0$, due to thermally activated absorption within the quasiparticle band, and a distinct broad resonant peak around $\Omega \in[0.6\Delta_0,0.8\Delta_0]$, arising from interband transitions between the electron-like (hole-like) impurity band and the quasielectron (quasihole) continuum. These features are a direct consequence of the formation of impurity bands and the persistence of a finite single-particle excitation gap. By contrast, the gapless state remains nearly featureless at finite temperatures [Fig.~\ref{figyc2}(d)], lacking clear signatures of interband transitions or resonant absorption peaks. This reflects the complete filling of the low-energy spectrum by impurity states, and the associated disappearance of the excitation gap. The resulting optical response is smooth and monotonic across all frequencies, sharply differing and distinguishing it from the discrete, temperature-sensitive features of the gapped state.

\section{Nonlinear optical response}

We next turn to the nonlinear dynamics of the SC condensate under strong THz excitation. These dynamics can be understood by a self-consistent equation of motion for the SC order parameter $\Delta(t) = \Delta_0 + \delta|\Delta(t)|$: 
\begin{equation}
  u\partial_t^2\Delta+\gamma\partial_t\Delta\!=D\!\!\int\!\frac{\tilde\Delta\tanh\Big(\frac{\omega}{2k_BT}\Big)d\omega}{2\sqrt{\tilde\omega^2-\tilde\Delta^2}}-\frac{\Delta}{g}-\frac{{\lambda}e^2{\bf A}^2(t)\Delta}{m}. 
\end{equation}
 which can {phenomenologically} capture the amplitude fluctuations $\delta|\Delta(t)|$ (i.e., Higgs mode) during the optical excitations~\cite{pekker2015amplitude,yang2024optical,yang2023optical,yang2019gauge,shimano2020higgs,PhysRevB.111.144512}. Here, $\gamma$ is the damping coefficient that  can arise from the impurity scattering, as derived through both diagrammatic and kinetic approaches~\cite{cui2019impact,PhysRevB.102.144508,PhysRevB.109.054520,PhysRevB.106.144509,PhysRevB.109.L100503,Li_2025}.  The inertia-like coefficient $u$ reflects the retardation in the gap response. Both $\gamma$ and $u$, incorporating the renormalization effects arising from exchange interactions,  are microscopically derived from the Eilenberger formalism using fully renormalized Green functions, {with several simplifying approximations applied (such as neglecting the nonequilibrium time evolution of quasiparticles while fully retaining all equilibrium quasiparticle effects).} Their explicit expressions are provided in the Appendix~\ref{DEOP}.  The last term  describes the nonlinear coupling between the gap and the external THz field, represented by the vector potential ${\bf A}(t)$. This coupling term, phenomenologically present in time-dependent Ginzburg-Landau theory~\cite{pekker2015amplitude,yang2023optical}, has been microscopically derived within the BCS framework~\cite{yang2019gauge,abrikosov2012methods,yang2018gauge,yang2023optical,yang2024optical}. The response coefficient $\lambda$ quantifies the strength of this coupling, encapsulating the drive effect~\cite{shimano2020higgs,yang2018gauge}, which arises either from the direct acceleration of thermally excited quasiparticles~\cite{yang2019gauge,yang2023optical,yang2024optical} or from impurity-induced processes~\cite{PhysRevB.99.224511}. Notably, even weak non-magnetic disorder significantly enhances $\lambda$, making it the dominant mechanism for nonlinear optical coupling, far exceeding the contribution from  charge-density fluctuations~\cite{PhysRevB.99.224511}.
 {It should be emphasized that this driving term does not alter the intrinsic Higgs-mode dynamics, whose eigenfrequency and damping are determined by the self-consistently-renormalized equilibrium Green functions of the system. The source term only controls the excitation amplitude under external driving. Thus, we approximate response coefficient $\lambda$ as a constant here.}

\begin{figure}[htb]
  {\includegraphics[width=8.7cm]{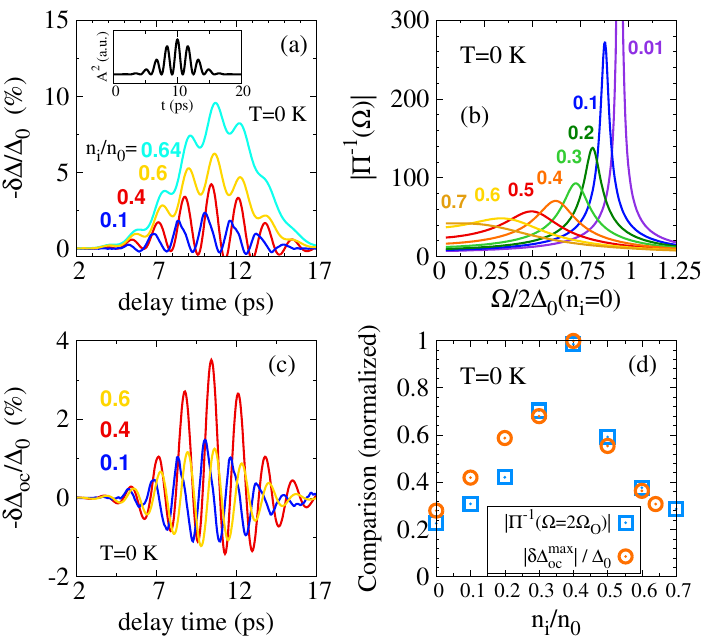}}
  \caption{(a) Dynamic simulation of gap dynamics  $\delta\Delta(t)$ and (b) diagrammatic formulation of the pole structure $|\Pi^{-1}(\Omega)|$ in Higgs-mode Green function as a function of $\Omega$,  at different $n_i$ for $T=0$. Inset of (a): the used THz field in the dynamic simulation. (c) Oscillatory component $\delta\Delta_{\rm oc}(t)$ in the gap dynamics $\delta\Delta(t)$ at several $n_i$. {(d) $|\Pi^{-1}(\Omega)|$ at fixed driving frequency $\Omega = 2\Omega_o$ and the maximum of $|\delta\Delta_{\rm oc}(t)|/\Delta_0$ (maximum magnitude of oscillation) during the temporal evolution, both plotted as functions of $n_i$ and normalized for comparison.} The  oscillatory part  $\delta\Delta_{\rm oc}(t)$ is isolated by directly removing the slowly varying background component of $\delta|\Delta(t)|$ in the simulation.  We approximate response coefficient $\lambda$ as a constant via setting the dimensionless driving strength to $\lambda e^2A_0^2/(mD)=0.01$. Other used parameters  are addressed in Table~\ref{parameter}.}    
\label{figyc3}
\end{figure}

This {phenomenological} self-consistent dynamic equation, incorporating the fully renormalized Green function, can describe the influence of magnetic impurities on condensate dynamics.  We consider a multi-cycle pulse field of the form  ${\bf A}(t)={\bf A}_0\cos(\Omega_o{t})\exp(-(t-t_0)^2/\tau^2)$ [inset of Fig.~\ref{figyc3}(a)], where $\Omega_o$ is the central frequency and $\tau=5~$ps is the pulse width. {Accordingly, the pump field enters the response through its intensity profile 
$A^2(t)=A_0^2\cos^2(\Omega_ot)\exp(-2(t-t_0)^2/\tau^2)
=A_0^2\exp(-2(t-t_0)^2/\tau^2)[1+\cos(2\Omega_ot)]/2$. Here, the first term generates a slowly varying background, while the second term proportional to $\cos(2\Omega_ot)$ is the genuinely oscillatory part that drives the Higgs-mode dynamics.}

Figure~\ref{figyc3}(a) shows the full dynamic simulation of the THz field response at an excitation frequency $\Omega_o \sim 0.3$~THz for different impurity concentrations. In both the gapped and gapless states, the THz field can coherently drive the SC gap dynamics into a strongly nonlinear state,   exhibiting clear oscillations on top of a non-oscillatory background after the THz stimulation.   The oscillatory component corresponds to the Higgs mode, i.e., coherent amplitude oscillations of the SC order parameter. This coherent excitation of the Higgs mode, characterized by the energy $\omega_H(n_i) = 2\Delta_0(n_i)$, is expected to reach its maximum when the resonant condition $2\Omega_o = \omega_H(n_i)$ is satisfied. In our case, the optical driving frequency is $\Omega_o \sim 0.3$~THz (approximately 1.24~meV), while the SC gap takes values $\Delta_0(n_i = 0.4n_0) = 1.3$~meV and $\Delta_0(n_i = 0.5n_0) = 1.1$~meV [as seen from Fig.~\ref{figyc1}(a)]. Accordingly, we expect the strongest coherent excitation of the SC gap dynamics in the second-order response to occur near $n_i = 0.4n_0$, {as demonstrated in Fig.~\ref{figyc3}(c) of the extracted oscillatory component from the time-resolved gap dynamics.} The non-oscillatory background arises from heating effects due to the energy input induced by the intense THz field~\cite{74d5-4hsw,cui2019impact,yang2018gauge}, which partially suppress the gap amplitude over longer timescales. This effect becomes particularly significant near $T_c$ with the  small gap (e.g., $n_i=0.64n_0$).

All of these features are in agreement with experimental observations in superconductors without magnetic impurities, where similar nonlinear responses, specifically the coherent oscillations corresponding to the Higgs mode and the non-oscillatory background due to heating effects, have been reported in conventional superconductors such as NbN under THz excitation~\cite{matsunaga2013higgs,matsunaga2014light,shimano2020higgs,PhysRevLett.122.257001}. This agreement suggests that while magnetic impurities can drive the system into a gapless SC state [Fig.~\ref{figyc1}(b)] and significantly modify the single-particle excitation spectrum and optical absorption in the linear response [Fig.~\ref{figyc2}(a)],  they leave the collective dynamics of the SC condensate largely intact and coherent. In other words, the amplitude (Higgs) mode, as a manifestation of the collective dynamics of the condensate, remains robust and coherent, despite the strong localization/correlation effects that significantly modify the single-particle excitation spectrum. This stands in sharp contrast to other symmetry-breaking channels, such as charge fluctuations, which are  more sensitive to disorder and tend to lose coherence under similar conditions.

\begin{figure}[htb]
  {\includegraphics[width=8.7cm]{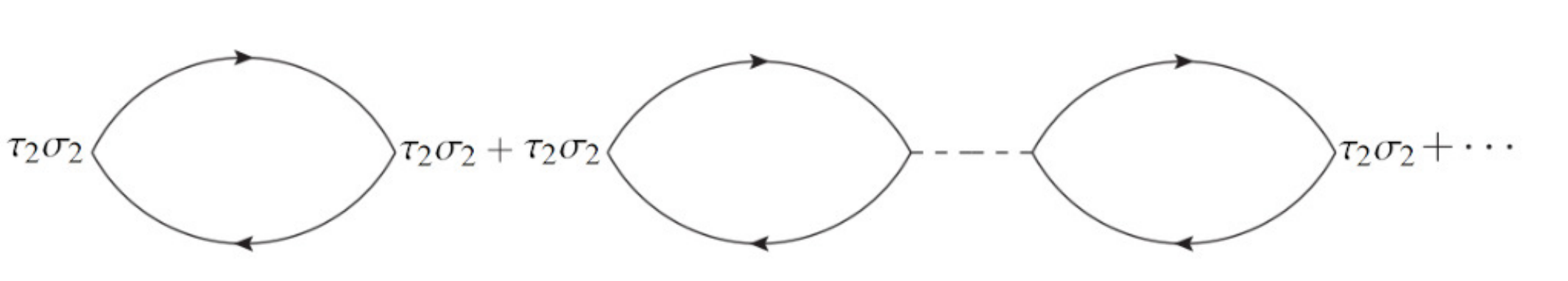}}
  \caption{Feynman diagram of the calculation of the Higgs-mode Green function. The dashed lines represent the SC pairing interaction $g$. The first diagram corresponds to the bare Higgs-mode Green function, i.e., the lowest-order contribution without interaction corrections. The second diagram depicts the first-order correction due to the pairing interaction, illustrating how the Higgs-mode propagator is renormalized by the interaction. The full Higgs-mode Green function requires the resummation of an infinite series of pairing interaction diagrams. This Dyson-type resummation leads to the full expression of the Higgs-mode propagator $D_{\rm H}(\Omega)$, from which the energy dispersion and damping of the mode can be extracted.}    
\label{SFIG1}
\end{figure}

{It should be emphasized that the dynamic-equation simulation, which necessarily involves several simplifying approximations (e.g., neglecting the nonequilibrium time evolution of quasiparticles), serves only as an illustrative extension for visualizing the time-domain response.} To rigorously confirm this conclusion, we  employ separately an independent approach based on a diagrammatic calculation of the  Higgs-mode  Green function~\cite{yanagisawa2018theory}, {defined by~\cite{yanagisawa2018theory,lee1974conductivity}
\begin{equation}
D_{\rm H}(x-y)=-i\langle{T_t{\hat \pi}_{\rm H}(x){\hat \pi}_{\rm H}(y)}\rangle,  
\end{equation}
where the Higgs-mode generator is given by ${\hat \pi}_{\rm H}=2\Psi^{\dagger}(x)\tau_2\sigma_2\Psi(x)$. This Green function can be evaluated diagrammatically using the Feynman diagram shown in Fig.~\ref{SFIG1}. In the frequency–momentum space and within the long-wavelength limit,} its expression is written as~\cite{yanagisawa2018theory} 
\begin{equation}
D_{\rm H}(\Omega)=\frac{\frac{1}{2}\sum_{{\bf k},\omega}{\rm Tr}[\tau_2\sigma_2G_{\bf k}(\omega+\Omega)\tau_2\sigma_2G_{\bf k}(\omega)]}{\Pi(\Omega)},  
\end{equation}
whose pole structure,  
\begin{equation}
\Pi(\Omega)=1-\frac{ig}{2}\sum_{{\bf k},\omega}{\rm Tr}[\tau_2\sigma_2G_{\bf k}(\omega+\Omega)\tau_2\sigma_2G_{\bf k}(\omega)],
\end{equation}
encodes the dynamics of the Higgs mode~\cite{yanagisawa2018theory,PhysRevResearch.2.023413,PhysRevB.106.144509,PhysRevB.99.224511,yang2019gauge,yang2023optical,yang2024optical},   
\begin{equation}
\Pi(\Omega)\propto\Omega^2-\omega_H^2-i\gamma_H\Omega,
\end{equation}
with $\omega_H$ being the Higgs-mode energy and $\gamma_H$ the Higgs-mode damping. The explicit formulation of $\Pi(\Omega)$, using the fully renormalized Green function incorporating exchange interactions, is presented in Fig.~\ref{figyc3}(b) where we show the evolution of $|\Pi^{-1}(\Omega)|$ as impurity concentration increases. The results clearly indicate that the Higgs-mode resonance remains well-defined and coherent even deep into the gapless regime, demonstrating that the collective amplitude dynamics of the SC condensate are robust against
the correlation effects by magnetic impurities, consistent with the dynamic simulation. {Notably, as shown in Fig.~\ref{figyc3}(d), the diagrammatic formulation of the Higgs-mode pole structure $|\Pi^{-1}(\Omega)|$ at the second harmonic of the driving frequency $\Omega=2\Omega_o$ (shown by squares), when tuning $n_i$, as anticipated, indicates a pronounced coherent excitation of the Higgs mode near $n_i = 0.4n_0$, consistent with the resonant condition $2\Omega_o = \omega_H(n_i)$ and also in agreement with the dynamic simulation (shown by circles).} The primary effect of magnetic impurities here is the broadening of the Higgs resonance, which reflects an enhanced damping of this collective mode~\cite{PhysRevB.111.174502}. Physically, this damping arises from the non-commutative relation between the exchange interaction and the Higgs-mode operator $\delta|\Delta|\tau_2\sigma_2$, thereby opening a decay channel and leading to finite lifetime broadening of the Higgs resonance. {It should also be emphasized that the same diagrammatic formalism can be applied to derive the energy spectra of other collective modes in the presence of magnetic impurities, such as the Nambu–Goldstone mode~\cite{schrieffer1964theory,yang2019gauge,PhysRevB.109.064508,PhysRevResearch.2.023413}, by appropriately choosing the corresponding symmetry generator.}

\section{Discussion}
{The interplay between magnetic disorder and superconductivity, together with the potential manipulation of the emerging quantum states, have continued to intrigue researchers for  decades. Hybridization of YSR states is known to create in-gap impurity bands that can acquire non-trivial topology and host Majorana bound states, as shown in recent experiments~\cite{poyhonen2018amorphous,kim2018toward,schneider2020controlling,schneider2021topological,PhysRevB.100.075420,kezilebieke2020topological}. When the impurity-band width rivals to the parent gap, an $s$-wave superconductor should cross from a gapped to a gapless phase in which the order parameter survives while the single-particle gap collapses.} 

{The present study obtained a fully self-consistent numerical solution of Shiba's complex renormalization for magnetic-impurity-doped $s$-wave superconductors. We have determined the full phase diagram, and show that robust superconductivity persists well beyond the closure of quasiparticle gap and a gapless superconductivity emerges at unexpectedly low impurity densities. We also computed the THz optical responses and identify distinctive, experimentally accessible signatures: (i) in the gapless SC state,  the low-frequency optical conductivity acquires finite absorption directly tied to the single-particle spectrum, providing a direct spectroscopic probe of the gapless regime; and (ii) in the nonlinear regime, the Higgs-mode dynamics dominates the response even in the presence of gapless quasiparticles, in contrast to the intuitive expectation that the gapless background would overwhelm collective effects.}

Our results reveal a fundamental distinction between dissipative channels driven by single-particle excitations and the coherent dynamics of the SC condensate in gapless superconductors. This distinction is expected to be general, applying to other types of gapless superconductivity such as in $d$-wave systems, where the density of states  remains finite inside the gap but THz-induced coherent nonlinear Higgs-mode dynamics has been both experimentally observed~\cite{chu2020phase,PhysRevLett.120.117001,PhysRevB.102.054510} and theoretically demonstrated~\cite{PhysRevB.102.014511,schwarz2020classification}. The study not only clarifies how magnetic impurities influence the single-particle-excitation  spectrum and the coherent collective dynamics of SC condensate, but also establishes a {\sl systematic}  theoretical framework which generalizes easily to transport coefficients, other collective modes, and engineered topological Shiba chains, for broadly studying the magnetic-impurity-induced phase transitions and the non-equilibrium phenomena under various conditions. The findings of the interplay between disorder and SC coherence provide a comprehensive set of quantitative criteria for the design of materials and quantum devices that support stable, magnetically disordered-based topological superconductivity in a robust condensate.

{It should be emphasized that Shiba's self-consistent renormalization scheme~\cite{shiba1968classical} adopts the random phase approximation (RPA)~\cite{abrikosov2012methods,RevModPhys.58.323}, which is primarily valid in spatially averaged, macroscopic systems.   This approximation neglects the possible effects arising from local magnetic domains or spin clusters  in realistic disordered environments. To address this limitation, i.e., to capture mesoscopic disorder effects beyond the assumptions of spatial uniformity in the RPA, we employ a domain-based simulation framework, by simulating the spatially inhomogeneous SC state using a macroscopic phase-field-like formulation of the gap equation~\cite{yang24thermodynamic}:
\begin{equation}
 \frac{\lambda_d\nabla_{\bf R}^2\Delta({\bf R})}{4m_e}=\frac{\Delta({\bf R})}{g}\!-\!i\sum_{{\bf k},\omega}\tanh\Big(\frac{\omega}{2k_BT}\Big)F_{\bf k}[\omega+i0^+,\Delta({\bf R})], 
\end{equation}
which can capture the spatial modulation of the order parameter $\Delta({\bf R})$ driven by the local impurity configurations. } 

{The microscopic derivation of this equation is provided in our recent work~\cite{yang24thermodynamic}, using a path-integral approach, where the $\nabla^2_{\bf R}\Delta({\bf R})$ term naturally emerges from the separation of center-of-mass and relative coordinates of Cooper pairs. In practice, we approximate $\lambda_d$ as a constant, following the standard  treatment in phase-field and  Ginzburg–Landau simulations of superconductors,  as $\lambda_d$ represents the gradient (or stiffness) coefficient associated with spatial variations of the order parameter, whose microscopic dependence on the local amplitude and gradients of $\Delta({\bf R})$ is generally weak. Retaining this weak dependence would only introduce higher-order corrections to the gradient energy without affecting the essential spatial evolution. In constant-$\lambda_d$ approximation, the characteristic spatial variation is governed primarily by the gauge-coupled Laplacian term and the free-energy landscape, rather than by a spatially varying stiffness parameter.}

\begin{widetext}
   \begin{center}
\begin{figure}[htb]
  {\includegraphics[width=17.6cm]{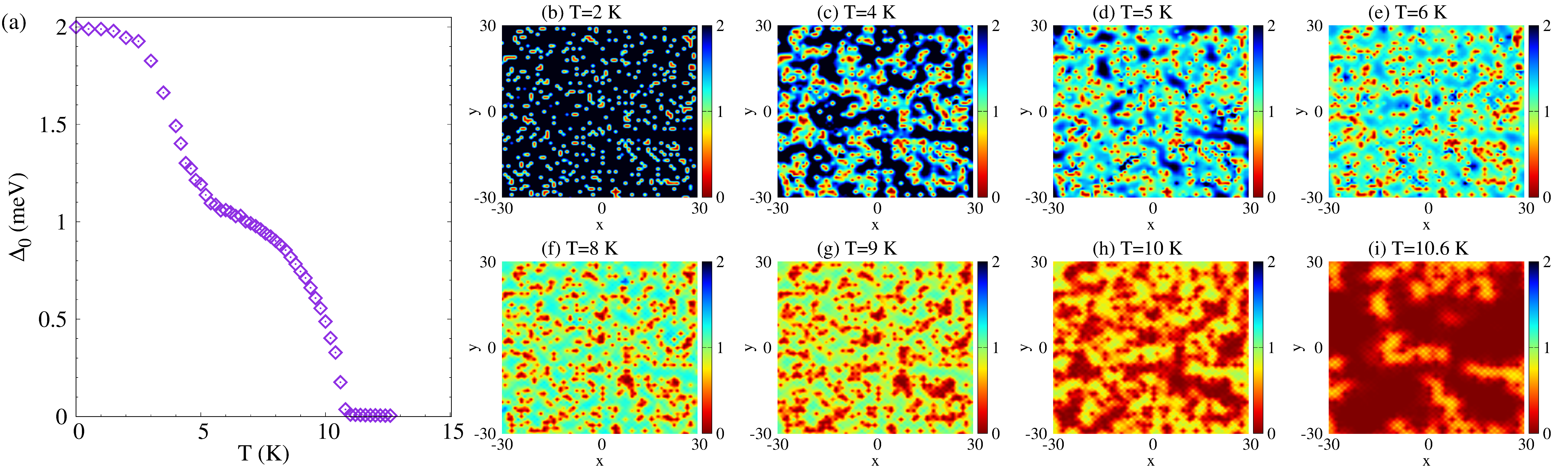}}
  \caption{(a):  Temperature dependence of the spatially averaged SC gap. (b)–(i): Spatial maps of the SC gap $\Delta({\bf R})$  (in meV) at different temperatures. The simulation incorporates 600 randomly distributed magnetic impurity domains on a $61\times61$ two-dimensional Cartesian grid with the periodic boundary conditions. In our simulation, we set the constant ${\lambda_d}/(4Dm_e)=(0.12)^2~\mu\text{{m}}^2$, which is a reasonable value for many disordered or thin-film $s$-wave superconductors. } 
\label{SFIG5}
\end{figure}
\end{center}
\end{widetext}

{We consider a $15~\mu$m $\times$ $15~\mu$m simulation region discretized utlizing a $61\times61$ two-dimensional Cartesian grid with periodic boundary conditions.  In each disorder configuration, magnetic-impurity domains are randomly distributed according to  $n_i({\bf R}) = n_{i,{\bf R}_i}\delta({\bf R} - {\bf R}_i)$, where the domain centers  ${\bf R}_i$ are randomly distributed across the simulation region, and the impurity concentrations  $n_{i,{\bf R}_i}\in (0, 20n_0)$ as well as the exchange interaction strengths $\eta \in (0.05, 0.95)$ in each domain are independently and randomly sampled within the physical ranges. This mimics the realistic disorder where local impurity environments fluctuate significantly across the system.  The anomalous Green function at each spatial point is obtained by solving the self-consistent renormalization equations using the local impurity parameters.  Meanwhile, the SC gap $\Delta({\bf R})$ is updated via a spatially resolved gap equation using phase-field methods~\cite{yang24thermodynamic}, enabling the capture of spatially varying SC properties across the entire system. The simulation then incorporates both the microscopic physics of impurity-induced renormalization and the mesoscopic structure of disorder. This hybrid approach allows us to explore how rare regions, impurity clustering, and inhomogeneous exchange coupling jointly impact the SC state. It reveals the emergence of spatial fluctuations, gap suppression, and the formation of a percolative SC network as the temperature increases. These inhomogeneous features go beyond what is captured by uniform RPA treatments and are relevant to understanding superconductivity in strongly disordered systems and thin films~\cite{yang24thermodynamic,kowal1994disorder,ghosal1998role,dubi2007nature,sacepe2020quantum,sacepe2011localization,sacepe2008disorder,pracht2016enhanced,dubouchet2019collective}.}

{The resulting temperature-dependent behavior is presented in Fig.~\ref{SFIG5}. As shown in Fig.~\ref{SFIG5}(a), the spatially averaged SC gap exhibits a generally monotonic but noticeably non-smooth evolution with temperature, in contrast to the smooth trend predicted by the RPA. This deviation originates from strong disorder in certain domains, which causes the local collapse of the SC gap even at zero temperature, leading to the formation of finite-sized normal-state regions.  As the temperature increases, these disorder-induced inhomogeneities become more pronounced, giving rise to a heterogeneous mixed state characterized by coexisting gapped SC islands and normal metallic regions, as visualized in the spatial maps of Fig.~\ref{SFIG5}(b)–(i). These spatial gap fluctuations are intrinsic features of the disordered SC landscape~\cite{yang24thermodynamic,kowal1994disorder,ghosal1998role,dubi2007nature,sacepe2020quantum,sacepe2011localization,sacepe2008disorder,pracht2016enhanced,dubouchet2019collective} and are beyond the scope of mean-field macroscopic RPA descriptions.}

{These findings through a phase-field-like formulation underscore the breakdown of spatial uniformity assumed in the RPA and highlight the crucial role of mesoscopic inhomogeneity in shaping SC properties in disordered systems. The local collapse of the gap due to impurity clustering leads to an emergent percolative SC network, wherein global phase coherence may still persist even when large regions are locally metallic or  the local SC gap can persist
above $T_c$ where the global SC phase coherence of the system is destroyed.  Such inhomogeneous SC states may manifest in experiments through broadened SC transitions, residual resistivity tails, and spatial variations in superfluid stiffness~\cite{crane2007survival,sacepe2008disorder,sacepe2011localization,noat2013unconventional}.}

{This inhomogeneous SC-phase structure closely resembles the granular superconductivity observed in high-$T_c$ cuprates and disordered thin films~\cite{yang24thermodynamic,kowal1994disorder,ghosal1998role,dubi2007nature,sacepe2020quantum,sacepe2011localization,sacepe2008disorder,pracht2016enhanced,dubouchet2019collective}, where SC  ``puddles'' are embedded within metallic or insulating backgrounds. To fully capture the THz optical responses in such systems, future investigations should consider the time-dependent gap fluctuations induced by external field on top of this emergent landscape background. More broadly, by combining the phase-field method with a time-dependent Ginzburg-Landau or Eilenberger-like formalism, one can quantitatively investigate the evolution of the SC coherence length, Josephson coupling between domains, and the vortex dynamics as a function of increasing magnetic disorder. These directions may offer dynamic insights into the nature of magnetic-disorder-driven SC-to-metal and SC-to-insulator transitions for the study of disordered quantum materials.} 

{\it Acknowledgments.---}F.Y. carried out the theoretical modeling and the numerical  simulations. R.Y.F. developed the numerical algorithm.  This work was supported by the U.S. Department of Energy, Office of Science, Basic Energy Sciences, under Award Number DE-SC0020145, as part of the Computational Materials Sciences Program. F.Y. and L.Q.C. also acknowledge the generous support of the Donald W. Hamer Foundation through a Hamer Professorship at Penn State.

\begin{appendix}

\begin{widetext}

\section{Dynamic equation of order parameter}
\label{DEOP}
To derive the equation of motion for the SC order parameter, we start from the microscopic dynamic description. For the complex renormalization by exchange interactions associated with Shiba states, the Eilenberger equation provides an efficient and tractable formalism. Specifically, the Eilenberger equation~\cite{eilenberger1968transformation} is derived from the Gorkov equation for the $\tau_3$-Green function via the quasiclassical approximation~\cite{RevModPhys.58.323}:
\begin{equation}\label{QG1}
g^{R/K/A}_{{\bf R,k_F}}(t,t')=\frac{i}{\pi}\!\!\int{d\xi_{\bf k}}\!\!\int{d{\bf r}}\tau_3G^{R/K/A}(x,x')e^{-i{\bf k}\cdot({\bf x}-{\bf x'})},  
\end{equation}
where ${\bf R}=({\bf x+x'})/{2}$ represents the center-of-mass spatial coordinate. The retarded (R), advanced (A), and Keldysh (K) Green functions are defined as~\cite{RevModPhys.58.323}:
\begin{eqnarray}
  G^R(x,x')&=&-i\langle\{\psi(x),\psi^{\dagger}(x')\}\rangle\theta(t-t'),\\
  G^A(x,x')&=&i\langle\{\psi(x),\psi^{\dagger}(x')\}\rangle\theta(t'-t),\\
  G^K(x,x')&=&-i\langle[\psi(x),\psi^{\dagger}(x')].
\end{eqnarray}
In the Nambu space and within the Keldysh formalism, the Eilenberger equation takes the form~\cite{PhysRevB.95.235403,PhysRevB.99.224511,yang2023optical,yang2024optical}:
\begin{equation}
  i\{\tau_3\partial_{t},{\hat g}\}_{t}-[(\Delta_0+\delta|\Delta|)\tau_1\tau_3,{\hat g}]_{t}\!=\!0, \label{ELE}
\end{equation}  
where the Green function matrix ${\hat g}$ is 
\begin{equation}
  {\hat g}=\left(\begin{array}{cc}
    g^R & g^K \\
    0 & g^A
  \end{array}\right).  
\end{equation}
The commutator and anticommutator are defined as {\small $[X,{\hat g}]_{t}=X(t_1){\hat g}(t_1,t_2)-{\hat g}(t_1,t_2)X(t_2)$} and {\small $\{X,{\hat g}\}_{t}=X(t_1){\hat g}(t_1,t_2)+{\hat g}(t_1,t_2)X(t_2)$}.

The self-consistent gap equation is given by~\cite{PhysRevB.95.235403,PhysRevB.99.224511,yang2023optical,yang2024optical}:
\begin{equation}\label{ELGE}
\frac{\Delta}{Dg}=\frac{\Delta_0+\delta|\Delta|}{Dg}=-i{\rm Tr}[\langle{g^K_{\bf R,k_F}(t,t)}\rangle_F\tau_2/2],  
\end{equation}
where $\langle...\rangle_F$ indicates angular averaging over the Fermi surface.

To formulate the dynamics of the order parameter, we expand the quasiclassical Green function as: 
\begin{equation}\label{RT}
{\hat g}={\hat g}^{(0)}+\delta{\hat g},
\end{equation}
where $\delta{\hat g}$ describes the nonequilibrium part on top of the equilibrium state ${\hat g}^{(0)}$. Then, taking the gap dynamics $\delta|\Delta(t)|=\delta|\Delta|e^{-i\Omega{t}}$, the Eilenberger equation in Eq.~(\ref{ELE}) becomes
\begin{eqnarray}
  &&\{\tau_3\partial_{t},\delta{\hat g}\}_{t}+[\Delta_0\tau_2,\delta{\hat g}]_{t}+[\delta|\Delta|e^{-i\Omega{t}}\tau_2,{\hat g}^{(0)}]_{t}\!=\!0, \label{ELE2}
\end{eqnarray}
which can be solved for  $\delta{\hat g}$  given the initial state ${\hat g}^{(0)}$. {Specifically, using Eq.~(\ref{RT}), one can find the solution of the retarded component from Eq.~(\ref{ELE2}) as follows~\cite{PhysRevB.99.224511,yang2023optical,yang2024optical}:}
 \begin{eqnarray}
  g^R(t,t')= g^{R(0)}(t,t')+ {\delta}g^R(t,t')=\!\!\!\int\!\frac{dE}{2\pi}[e^{-iEt}g^{R(0)}(E)e^{iEt}\!+\!e^{-i(E+\Omega)t}{\delta}g^{R}(E+\Omega,E)e^{iEt'}],
\end{eqnarray}
The equilibrium Green function is known from the Gorkov formalism~\cite{PhysRevB.95.235403,PhysRevB.99.224511,yang2023optical,yang2024optical}: 
\begin{equation}\label{gr0}
g^{R(0)}(E)=\int\frac{d\xi_{\bf k}}{\pi}\frac{i\tau_3(E\!+\!\xi_{\bf k}\tau_3\!+\!\Delta_0\tau_1)}{(E\!+\!i0^+)^2\!-\!\xi_{\bf k}^2\!-\!\Delta_0^2}=\frac{E\tau_3\!+\!i\Delta_0\tau_2}{S^R(E)},  
\end{equation}
with $S^R(E)=\sqrt{(E+i0^+)^2-\Delta_0^2}$. Plugging into Eq.~(\ref{ELE2}) gives: 
\begin{eqnarray}
&&\!\!\!\!\!\!\![(E+\Omega)\tau_3\!+\!i\Delta_0\tau_2]{\delta}g^{R}(E+\Omega,E)\!-\!{\delta}g^{R}(E+\Omega,E)(E\tau_3\!+\!i\Delta_0\tau_2)=i\delta|\Delta|[g^{R(0)}(E+\Omega)\tau_2\!-\!\tau_2g^{R(0)}(E)],~~~~~ \label{gr2}  
\end{eqnarray} 
which can be re-written as
\begin{equation}
S^R(E+\Omega)g^{R(0)}(E+\Omega){\delta}g^{R}(E+\Omega,E)\!-\!{\delta}g^{R}(E+\Omega,E)S^R(E)g^{R(0)}(E)=i\delta|\Delta|[g^{R(0)}(E+\Omega)\tau_2\!-\!\tau_2g^{R(0)}(E)],
\end{equation}
leading to the solution:
\begin{eqnarray}
  {\delta}g^{R}(E+\Omega,E)=i\delta|\Delta|\frac{\tau_2\!-\!g^{R(0)}(E+\Omega)\tau_2g^{R(0)}(E)}{S^R(E+\Omega)+S^R(E)}. \label{sgr2}
\end{eqnarray}

Substituting Eqs.~(\ref{gr0}) into Eq.~(\ref{sgr2}), the specific expression of the $\tau_2$ component of ${\delta}g^{R}(E)$ reads
\begin{eqnarray}
  {\delta}g^{R}_2(E+\Omega,E)&=&i\delta|\Delta|\frac{2\Delta_0^2\!+\!2E(E+\Omega)\!+\!2S^R(E)S^R(E+\Omega)}{4W(E)}.
\end{eqnarray}
with $W(E)=S^R(E+\Omega)S^R(E)[S^R(E+\Omega)+S^R(E)]/2\approx[S^R(E)]^3$. This can also be rewritten as: 
\begin{eqnarray}
  {\delta}g^{R}_2(E+\Omega,E)&=&i\delta|\Delta|\frac{4\Delta_0^2\!-\![E\!-\!(E+\Omega)]^2\!+\![S^R(E)\!+\!S^R(E+\Omega)]^2}{4[S^R(E)]^3}=i\delta|\Delta|\Big[\frac{4\Delta_0^2-\Omega^2}{4[S^R(E)]^3}\!+\!\frac{1}{S^R(E)}\Big].
\end{eqnarray}
Substituting into the gap equation Eq.~(\ref{ELGE}) yields: 
\begin{equation}
\frac{\Delta_0+\delta|\Delta|}{Dg}=\int\frac{dE}{2\pi}\tanh\Big(\frac{E}{2k_BT}\Big)\Big\{\frac{\Delta_0}{S^R(E)}+\delta|\Delta|\Big[\frac{4\Delta_0^2-\Omega^2}{4[S^R(E)]^3}\!+\!\frac{1}{S^R(E)}\Big]\Big\},
\end{equation}
{which is equivalent to:
\begin{eqnarray}
{\Omega^2\delta|\Delta|}\int\frac{dE}{2\pi}\tanh\Big(\frac{E}{2k_BT}\Big)\frac{1}{4[S^R(E)]^3}&=&\int\frac{dE}{2\pi}\tanh\Big(\frac{E}{2k_BT}\Big)\Big\{\delta|\Delta|\frac{\Delta_0^2}{[S^R(E)]^3}+\frac{\Delta}{S^R(E)}\Big\}-\frac{\Delta}{Dg}\nonumber\\
&=&\int\frac{dE}{2\pi}\tanh\Big(\frac{E}{2k_BT}\Big)\Big\{\delta|\Delta|\frac{\Delta_0^2}{((E+i0^+)^2-\Delta_0^2)^{3/2}}+\frac{\Delta}{\sqrt{(E+i0^+)^2-\Delta_0^2}}\Big\}-\frac{\Delta}{Dg}\nonumber\\
&\approx&\int\frac{dE}{2\pi}\tanh\Big(\frac{E}{2k_BT}\Big)\frac{\Delta}{\sqrt{(E+i0^+)^2-(\Delta_0+\delta|\Delta|)^2}}-\frac{\Delta}{Dg}\nonumber\\
&=&\int\frac{dE}{2\pi}\tanh\Big(\frac{E}{2k_BT}\Big)\frac{\Delta}{\sqrt{(E+i0^+)^2-\Delta^2}}-\frac{\Delta}{Dg},
\end{eqnarray}
Then, by including the coupling to the vector potential, we arrive at the equation of motion for the SC order parameter:}
\begin{eqnarray}
  u\partial_t^2\Delta+\gamma\partial_t\Delta=D\int\frac{dE}{2\pi}\tanh\Big(\frac{E}{2k_BT}\Big)\frac{\Delta}{\sqrt{(E+i0^+)^2-\Delta^2}}-\frac{\Delta}{g}-\lambda\frac{e^2{\bf A}^2(t)\Delta}{m}, 
\end{eqnarray}
with
\begin{eqnarray}
  u=-D\int\frac{dE}{2\pi}\tanh\Big(\frac{E}{2k_BT}\Big)\frac{1}{4[S^R(E)]^3},~~~~~~~~\text{and}~~~~~~\gamma=0^+.
\end{eqnarray}
{It is worth noting that the approximation $S^{-1}(E+\Omega)\approx{S^{-1}(E)}$ used here is only  to soley retain the time-dependent dynamics of the order parameter. In doing so, we neglect the nonequilibrium time evolution of quasiparticles, but all equilibrium quasiparticle effects remain fully encoded in the equilibrium quantities $S^{-1}(E)$. This is because 
    $S^{-1}(E+\Omega)\approx{S^{-1}(E)+\Omega\partial_ES^{-1}(E)}\rightarrow{S^{-1}(E)+i\partial_t[\partial_ES^{-1}(E)]}$. 
  The first term, $S^{-1}(E)$, contains the full equilibrium quasiparticle pole structure. The second term, proportional to $i\partial_t$, corresponds to the nonequilibrium time evolution of quasiparticles. Our approximation amounts to keeping the former and discarding the lattet, consistent with our goal which is to isolate the collective Higgs-mode dynamics rather than to formulate a full kinetic theory of quasiparticles.}

  Moreover, for clarity and analytical transparency, our initial derivation of the SC order parameter dynamics was based on the bare (i.e., unrenormalized) Green function. This pedagogical approach helps elucidate the underlying physics of SC dynamics and provides a tractable analytical framework. To go beyond and incorporate the full renormalization effects induced by Shiba states, we now perform a parallel derivation starting directly from the fully renormalized Green functions. This is achieved by utilizing the fully renormalized equilibrium Green function~\cite{PhysRevB.109.064508}:
\begin{equation}
g^{R(0)}(E)=\frac{\tilde{E}\tau_3\!+\!i\tilde{\Delta}_0\tau_2}{S^R(E)},    
\end{equation}
with the self-consistently determined quantities  $\tilde{E}=\tilde{\omega}(E,\Delta_0)$ and $\tilde\Delta_0=\tilde\Delta_0(E,\Delta_0)$ from Shiba's renormalization equations~\cite{shiba1968classical}, and substituting it to $\delta{g^R}(E)$ in Eq.~(\ref{sgr2}). Now, $S^R(E)=\sqrt{\tilde{E}^2-\tilde\Delta_0^2}$. 

Then, following the similar derivation procedure, we obtain the updated expressions for the inertia coefficient $u$ and the damping rate $\gamma$ in the presence of complex renormalization by exchange interactions: 
\begin{eqnarray}
  u&=&D\int\frac{dE}{2\pi}\tanh\Big(\frac{E}{2k_BT}\Big)\frac{\rm Re}{\Omega^2}\Big\{\frac{1\!-\!{\rm Tr}[g^{R(0)}(E+\Omega)\tau_2g^{R(0)}(E)\tau_2]/2}{S^R(E+\Omega)+S^R(E)}-\frac{1\!-\!{\rm Tr}[g^{R(0)}(E)\tau_2g^{R(0)}(E)\tau_2]/2}{S^R(E)+S^R(E)}\Bigg\}\nonumber\\
  &\approx&D\int\frac{dE}{2\pi}\tanh\Big(\frac{E}{2k_BT}\Big)\frac{\rm Re}{\Omega^2}\Big[\frac{(\tilde\Delta^+_0+\tilde\Delta_0)^2-(\tilde{E}-\tilde{E^+})^2}{4(\tilde{E}-\tilde\Delta)^{3/2}}-\frac{4\tilde\Delta_0^2}{4(\tilde{E}-\tilde\Delta)^{3/2}}\Big], \label{fullms}\\
  \gamma&=&D\int\frac{dE}{2\pi}\tanh\Big(\frac{E}{2k_BT}\Big)\frac{\rm Im}{\Omega}\Big\{\frac{1\!-\!{\rm Tr}[g^{R(0)}(E+\Omega)\tau_2g^{R(0)}(E)\tau_2]/2}{S^R(E+\Omega)+S^R(E)}-\frac{1\!-\!{\rm Tr}[g^{R(0)}(E)\tau_2g^{R(0)}(E)\tau_2]/2}{S^R(E)+S^R(E)}\Bigg\}\nonumber\\
  &\approx&D\int\frac{dE}{2\pi}\tanh\Big(\frac{E}{2k_BT}\Big)\frac{\rm Im}{\Omega}\Big[\frac{(\tilde\Delta^+_0+\tilde\Delta_0)^2-(\tilde{E}-\tilde{E^+})^2}{4(\tilde{E}-\tilde\Delta)^{3/2}}-\frac{4\tilde\Delta_0^2}{4(\tilde{E}-\tilde\Delta)^{3/2}}\Big], \label{damping}
\end{eqnarray}
where $\tilde{E}^+=\tilde{\omega}(E+\Omega,\Delta_0)$ and  $\tilde\Delta^+_0=\tilde\Delta_0(E+\Omega,\Delta_0)$. These provide a fully renormalized formulation of the SC dynamic coefficients in the presence of magnetic impurities. In practical calculations, we adopt the conventional limit $\Omega \rightarrow 0$ {(equivalent to the approximation $S^{-1}(E+\Omega)\approx{S^{-1}(E)}$ above)} for evaluating $u$ [Eq.~(\ref{fullms})] and $\gamma$ [Eq.~(\ref{damping})], a standard approximation in simulations of collective modes~\cite{pekker2015amplitude}. In addition, to account for additional relevant scattering channels (e.g., quasiparticles) beyond those captured by the microscopic damping derived in Eq.~(\ref{damping}), we include a phenomenological contribution $\gamma_d$ to the total damping rate~\cite{cui2019impact,PhysRevB.102.144508,PhysRevB.109.054520,PhysRevB.106.144509,PhysRevB.109.L100503,Li_2025}, and take $\gamma_d/D=0.05~\mathrm{ps}^{-1}$ in the dynamic simulation.

\section{Comparison with analytical solution}

In our previous work~\cite{PhysRevB.109.064508}, under the assumption of a {\sl fixed} SC gap (i.e., without enforcing self-consistency in the gap equation),  we derived an analytical solution for the Shiba-state impurity bands by performing a {\sl weak}-concentration expansion (i.e., treating the imaginary part of the complex renormalization as a small quantity). While this analytical solution cannot describe the SC phase transition due to the absence of gap renormalization, and becomes inaccurate at high magnetic-impurity concentrations, it remains valuable as a benchmark for validating our numerical calculations at low magnetic-impurity concentrations.

\begin{figure}[htb]
  {\includegraphics[width=12.0cm]{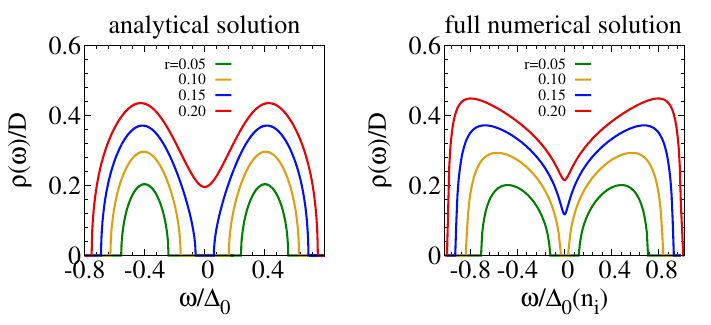}}
  \caption{Comparison between analytical (left panel) and numerical (right panel) solutions for the single-particle density of states at $T=0$ and $\eta=0.4$ for different magnetic-impurity concentrations. $r=\gamma_s/\Delta_0(T=0,n_i=0)$. The analytical results, obtained under the assumption of a fixed $\Delta_0$ at low impurity concentrations, are taken from our previous work in Ref.~\cite{PhysRevB.109.064508}.}    
\label{SFIG3}
\end{figure}

Figure~\ref{SFIG3} presents a comparison between the analytical and numerical results for the single-particle density of states. As seen in the figure, the two approaches show good agreement, including relatively matched peak values at each impurity concentration~$r$. The primary difference arises from the self-consistent treatment in the numerical approach, which yields slightly broader impurity bands and asymmetric peak shapes, exhibiting mild distortions toward higher energies.

\begin{figure}[htb]
  {\includegraphics[width=13.6cm]{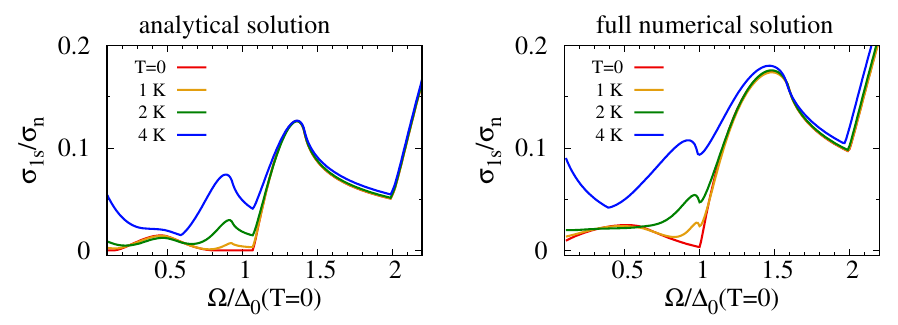}}
  \caption{Comparison between analytical (left panel) and numerical (right panel) solutions for the linear-response optical absorption at different temperatures for $\eta=0.25$ and $r=\gamma_s/\Delta_0(T=0)=0.05$. The analytical results, obtained under the assumption of a fixed $\Delta_0=2~$meV at low impurity concentrations, are taken from our previous work in Ref.~\cite{PhysRevB.109.064508}.  The figure reveals three resonance peaks and one threshold (as originally discussed in our previous work~\cite{PhysRevB.109.064508}). First, a sharp absorption onset appears at $\Omega>2\Delta_0$, corresponding to interband transitions between Bogoliubov quasielectrons and quasiholes. Second, the resonance peak around $2\eta\Delta_0=0.5\Delta_0$ at $T=0$ arises from interband transitions between impurity bands and becomes suppressed at higher temperatures. Third, the resonance peak near $(1+\eta)\Delta_0=1.25\Delta_0$ originates from interband transitions between impurity bands and the quasiparticle continuum (from hole-type to electron-type). Finally, an additional peak emerges around $(1-\eta)\Delta_0=0.75\Delta_0$ at finite temperatures, associated with interband transitions between impurity bands and the quasiparticle continuum (hole–hole and electron–electron types).} 
\label{SFIG4}
\end{figure}

Figure~\ref{SFIG4} compares the analytical and numerical results for the linear-response optical absorption. The two approaches exhibit consistent behavior across temperatures, capturing all temperature-dependent resonance features. While the analytical results exhibit sharper resonance peaks, the self-consistent treatment in the numerical calculation produces broader and smoother features, reflecting the effects of spectral renormalization and lifetime broadening during the self-consistent formulation. We would like to point out that our previous analytic results in Ref.~\cite{PhysRevB.109.064508} are derived under a set of simplifying approximations (e.g., treating the gap as fixed rather than fully self-consistent, and assuming the low-impurity limit). They are valuable in providing  qualitative analysis,  transparent physical understanding, but they cannot be expected to reproduce the numerical results in a strictly quantitative manner across the full parameter space. This justifies our emphasis on a rigorous self-consistent treatment.

\end{widetext}

\end{appendix}


\begin{thebibliography}{117}%
\makeatletter
\providecommand \@ifxundefined [1]{%
 \@ifx{#1\undefined}
}%
\providecommand \@ifnum [1]{%
 \ifnum #1\expandafter \@firstoftwo
 \else \expandafter \@secondoftwo
 \fi
}%
\providecommand \@ifx [1]{%
 \ifx #1\expandafter \@firstoftwo
 \else \expandafter \@secondoftwo
 \fi
}%
\providecommand \natexlab [1]{#1}%
\providecommand \enquote  [1]{``#1''}%
\providecommand \bibnamefont  [1]{#1}%
\providecommand \bibfnamefont [1]{#1}%
\providecommand \citenamefont [1]{#1}%
\providecommand \href@noop [0]{\@secondoftwo}%
\providecommand \href [0]{\begingroup \@sanitize@url \@href}%
\providecommand \@href[1]{\@@startlink{#1}\@@href}%
\providecommand \@@href[1]{\endgroup#1\@@endlink}%
\providecommand \@sanitize@url [0]{\catcode `\\12\catcode `\$12\catcode
  `\&12\catcode `\#12\catcode `\^12\catcode `\_12\catcode `\%12\relax}%
\providecommand \@@startlink[1]{}%
\providecommand \@@endlink[0]{}%
\providecommand \url  [0]{\begingroup\@sanitize@url \@url }%
\providecommand \@url [1]{\endgroup\@href {#1}{\urlprefix }}%
\providecommand \urlprefix  [0]{URL }%
\providecommand \Eprint [0]{\href }%
\providecommand \doibase [0]{https://doi.org/}%
\providecommand \selectlanguage [0]{\@gobble}%
\providecommand \bibinfo  [0]{\@secondoftwo}%
\providecommand \bibfield  [0]{\@secondoftwo}%
\providecommand \translation [1]{[#1]}%
\providecommand \BibitemOpen [0]{}%
\providecommand \bibitemStop [0]{}%
\providecommand \bibitemNoStop [0]{.\EOS\space}%
\providecommand \EOS [0]{\spacefactor3000\relax}%
\providecommand \BibitemShut  [1]{\csname bibitem#1\endcsname}%
\let\auto@bib@innerbib\@empty
\bibitem [{\citenamefont {Bardeen}\ \emph {et~al.}(1957)\citenamefont
  {Bardeen}, \citenamefont {Cooper},\ and\ \citenamefont
  {Schrieffer}}]{bardeen1957theory}%
  \BibitemOpen
  \bibfield  {author} {\bibinfo {author} {\bibfnamefont {J.}~\bibnamefont
  {Bardeen}}, \bibinfo {author} {\bibfnamefont {L.~N.}\ \bibnamefont
  {Cooper}},\ and\ \bibinfo {author} {\bibfnamefont {J.~R.}\ \bibnamefont
  {Schrieffer}},\ }\bibfield  {title} {\bibinfo {title} {Theory of
  superconductivity},\ }\href@noop {} {\bibfield  {journal} {\bibinfo
  {journal} {Phys. Rev.}\ }\textbf {\bibinfo {volume} {108}},\ \bibinfo {pages}
  {1175} (\bibinfo {year} {1957})}\BibitemShut {NoStop}%
\bibitem [{\citenamefont {Schrieffer}(1964)}]{schrieffer1964theory}%
  \BibitemOpen
  \bibfield  {author} {\bibinfo {author} {\bibfnamefont {J.}~\bibnamefont
  {Schrieffer}},\ }\href@noop {} {\emph {\bibinfo {title} {Theory of
  Superconductivity}}}\ (\bibinfo  {publisher} {W.A. Benjamin},\ \bibinfo
  {year} {1964})\BibitemShut {NoStop}%
\bibitem [{\citenamefont {Matsunaga}\ and\ \citenamefont
  {Shimano}(2012)}]{PhysRevLett.109.187002}%
  \BibitemOpen
  \bibfield  {author} {\bibinfo {author} {\bibfnamefont {R.}~\bibnamefont
  {Matsunaga}}\ and\ \bibinfo {author} {\bibfnamefont {R.}~\bibnamefont
  {Shimano}},\ }\bibfield  {title} {\bibinfo {title} {Nonequilibrium bcs state
  dynamics induced by intense terahertz pulses in a superconducting nbn film},\
  }\href@noop {} {\bibfield  {journal} {\bibinfo  {journal} {Phys. Rev. Lett.}\
  }\textbf {\bibinfo {volume} {109}},\ \bibinfo {pages} {187002} (\bibinfo
  {year} {2012})}\BibitemShut {NoStop}%
\bibitem [{\citenamefont {\ifmmode~\check{S}\else \v{S}\fi{}indler}\ \emph
  {et~al.}(2022)\citenamefont {\ifmmode~\check{S}\else \v{S}\fi{}indler},
  \citenamefont {Kadlec},\ and\ \citenamefont {Kadlec}}]{PhysRevB.105.014506}%
  \BibitemOpen
  \bibfield  {author} {\bibinfo {author} {\bibfnamefont {M.}~\bibnamefont
  {\ifmmode~\check{S}\else \v{S}\fi{}indler}}, \bibinfo {author} {\bibfnamefont
  {F.}~\bibnamefont {Kadlec}},\ and\ \bibinfo {author} {\bibfnamefont
  {C.}~\bibnamefont {Kadlec}},\ }\bibfield  {title} {\bibinfo {title} {Onset of
  a superconductor-insulator transition in an ultrathin nbn film under in-plane
  magnetic field studied by terahertz spectroscopy},\ }\href@noop {} {\bibfield
   {journal} {\bibinfo  {journal} {Phys. Rev. B}\ }\textbf {\bibinfo {volume}
  {105}},\ \bibinfo {pages} {014506} (\bibinfo {year} {2022})}\BibitemShut
  {NoStop}%
\bibitem [{\citenamefont {Yang}\ \emph {et~al.}(2018)\citenamefont {Yang},
  \citenamefont {Vaswani}, \citenamefont {Sundahl}, \citenamefont {Mootz},
  \citenamefont {Gagel}, \citenamefont {Luo}, \citenamefont {Kang},
  \citenamefont {Orth}, \citenamefont {Perakis}, \citenamefont {Eom} \emph
  {et~al.}}]{yang2018terahertz}%
  \BibitemOpen
  \bibfield  {author} {\bibinfo {author} {\bibfnamefont {X.}~\bibnamefont
  {Yang}}, \bibinfo {author} {\bibfnamefont {C.}~\bibnamefont {Vaswani}},
  \bibinfo {author} {\bibfnamefont {C.}~\bibnamefont {Sundahl}}, \bibinfo
  {author} {\bibfnamefont {M.}~\bibnamefont {Mootz}}, \bibinfo {author}
  {\bibfnamefont {P.}~\bibnamefont {Gagel}}, \bibinfo {author} {\bibfnamefont
  {L.}~\bibnamefont {Luo}}, \bibinfo {author} {\bibfnamefont {J.}~\bibnamefont
  {Kang}}, \bibinfo {author} {\bibfnamefont {P.}~\bibnamefont {Orth}}, \bibinfo
  {author} {\bibfnamefont {I.}~\bibnamefont {Perakis}}, \bibinfo {author}
  {\bibfnamefont {C.}~\bibnamefont {Eom}}, \emph {et~al.},\ }\bibfield  {title}
  {\bibinfo {title} {Terahertz-light quantum tuning of a metastable emergent
  phase hidden by superconductivity},\ }\href@noop {} {\bibfield  {journal}
  {\bibinfo  {journal} {Nat. Mat.}\ }\textbf {\bibinfo {volume} {17}},\
  \bibinfo {pages} {586} (\bibinfo {year} {2018})}\BibitemShut {NoStop}%
\bibitem [{\citenamefont {Matsunaga}\ \emph {et~al.}(2013)\citenamefont
  {Matsunaga}, \citenamefont {Hamada}, \citenamefont {Makise}, \citenamefont
  {Uzawa}, \citenamefont {Terai}, \citenamefont {Wang},\ and\ \citenamefont
  {Shimano}}]{matsunaga2013higgs}%
  \BibitemOpen
  \bibfield  {author} {\bibinfo {author} {\bibfnamefont {R.}~\bibnamefont
  {Matsunaga}}, \bibinfo {author} {\bibfnamefont {Y.~I.}\ \bibnamefont
  {Hamada}}, \bibinfo {author} {\bibfnamefont {K.}~\bibnamefont {Makise}},
  \bibinfo {author} {\bibfnamefont {Y.}~\bibnamefont {Uzawa}}, \bibinfo
  {author} {\bibfnamefont {H.}~\bibnamefont {Terai}}, \bibinfo {author}
  {\bibfnamefont {Z.}~\bibnamefont {Wang}},\ and\ \bibinfo {author}
  {\bibfnamefont {R.}~\bibnamefont {Shimano}},\ }\bibfield  {title} {\bibinfo
  {title} {Higgs amplitude mode in the {BCS} superconductors
  {Nb$_{1-x}$Ti$_{x}$N} induced by terahertz pulse excitation},\ }\href@noop {}
  {\bibfield  {journal} {\bibinfo  {journal} {Phys. Rev. Lett.}\ }\textbf
  {\bibinfo {volume} {111}},\ \bibinfo {pages} {057002} (\bibinfo {year}
  {2013})}\BibitemShut {NoStop}%
\bibitem [{\citenamefont {Matsunaga}\ \emph {et~al.}(2014)\citenamefont
  {Matsunaga}, \citenamefont {Tsuji}, \citenamefont {Fujita}, \citenamefont
  {Sugioka}, \citenamefont {Makise}, \citenamefont {Uzawa}, \citenamefont
  {Terai}, \citenamefont {Wang}, \citenamefont {Aoki},\ and\ \citenamefont
  {Shimano}}]{matsunaga2014light}%
  \BibitemOpen
  \bibfield  {author} {\bibinfo {author} {\bibfnamefont {R.}~\bibnamefont
  {Matsunaga}}, \bibinfo {author} {\bibfnamefont {N.}~\bibnamefont {Tsuji}},
  \bibinfo {author} {\bibfnamefont {H.}~\bibnamefont {Fujita}}, \bibinfo
  {author} {\bibfnamefont {A.}~\bibnamefont {Sugioka}}, \bibinfo {author}
  {\bibfnamefont {K.}~\bibnamefont {Makise}}, \bibinfo {author} {\bibfnamefont
  {Y.}~\bibnamefont {Uzawa}}, \bibinfo {author} {\bibfnamefont
  {H.}~\bibnamefont {Terai}}, \bibinfo {author} {\bibfnamefont
  {Z.}~\bibnamefont {Wang}}, \bibinfo {author} {\bibfnamefont {H.}~\bibnamefont
  {Aoki}},\ and\ \bibinfo {author} {\bibfnamefont {R.}~\bibnamefont
  {Shimano}},\ }\bibfield  {title} {\bibinfo {title} {Light-induced collective
  pseudospin precession resonating with {Higgs} mode in a superconductor},\
  }\href@noop {} {\bibfield  {journal} {\bibinfo  {journal} {Science}\ }\textbf
  {\bibinfo {volume} {345}},\ \bibinfo {pages} {1145} (\bibinfo {year}
  {2014})}\BibitemShut {NoStop}%
\bibitem [{\citenamefont {Shimano}\ and\ \citenamefont
  {Tsuji}(2020)}]{shimano2020higgs}%
  \BibitemOpen
  \bibfield  {author} {\bibinfo {author} {\bibfnamefont {R.}~\bibnamefont
  {Shimano}}\ and\ \bibinfo {author} {\bibfnamefont {N.}~\bibnamefont
  {Tsuji}},\ }\bibfield  {title} {\bibinfo {title} {Higgs mode in
  superconductors},\ }\href@noop {} {\bibfield  {journal} {\bibinfo  {journal}
  {Annu. Rev. Condens. Matter Phys.}\ }\textbf {\bibinfo {volume} {11}},\
  \bibinfo {pages} {103} (\bibinfo {year} {2020})}\BibitemShut {NoStop}%
\bibitem [{\citenamefont {Kim}\ \emph {et~al.}(2024)\citenamefont {Kim},
  \citenamefont {Kovalev}, \citenamefont {Udina}, \citenamefont {Haenel},
  \citenamefont {Kim}, \citenamefont {Puviani}, \citenamefont {Cristiani},
  \citenamefont {Ilyakov}, \citenamefont {de~Oliveira}, \citenamefont
  {Ponomaryov}, \citenamefont {Deinert}, \citenamefont {Logvenov},
  \citenamefont {Keimer}, \citenamefont {Manske}, \citenamefont {Benfatto},\
  and\ \citenamefont {Kaiser}}]{Kim2024Tracing}%
  \BibitemOpen
  \bibfield  {author} {\bibinfo {author} {\bibfnamefont {M.-J.}\ \bibnamefont
  {Kim}}, \bibinfo {author} {\bibfnamefont {S.}~\bibnamefont {Kovalev}},
  \bibinfo {author} {\bibfnamefont {M.}~\bibnamefont {Udina}}, \bibinfo
  {author} {\bibfnamefont {R.}~\bibnamefont {Haenel}}, \bibinfo {author}
  {\bibfnamefont {G.}~\bibnamefont {Kim}}, \bibinfo {author} {\bibfnamefont
  {M.}~\bibnamefont {Puviani}}, \bibinfo {author} {\bibfnamefont
  {G.}~\bibnamefont {Cristiani}}, \bibinfo {author} {\bibfnamefont
  {I.}~\bibnamefont {Ilyakov}}, \bibinfo {author} {\bibfnamefont {T.~V. A.~G.}\
  \bibnamefont {de~Oliveira}}, \bibinfo {author} {\bibfnamefont
  {A.}~\bibnamefont {Ponomaryov}}, \bibinfo {author} {\bibfnamefont {J.-C.}\
  \bibnamefont {Deinert}}, \bibinfo {author} {\bibfnamefont {G.}~\bibnamefont
  {Logvenov}}, \bibinfo {author} {\bibfnamefont {B.}~\bibnamefont {Keimer}},
  \bibinfo {author} {\bibfnamefont {D.}~\bibnamefont {Manske}}, \bibinfo
  {author} {\bibfnamefont {L.}~\bibnamefont {Benfatto}},\ and\ \bibinfo
  {author} {\bibfnamefont {S.}~\bibnamefont {Kaiser}},\ }\bibfield  {title}
  {\bibinfo {title} {Tracing the dynamics of superconducting order via
  transient terahertz third-harmonic generation},\ }\href@noop {} {\bibfield
  {journal} {\bibinfo  {journal} {Sci. Adv.}\ }\textbf {\bibinfo {volume}
  {10}},\ \bibinfo {pages} {eadi7598} (\bibinfo {year} {2024})}\BibitemShut
  {NoStop}%
\bibitem [{\citenamefont {Nakamura}\ \emph {et~al.}(2019)\citenamefont
  {Nakamura}, \citenamefont {Iida}, \citenamefont {Murotani}, \citenamefont
  {Matsunaga}, \citenamefont {Terai},\ and\ \citenamefont
  {Shimano}}]{PhysRevLett.122.257001}%
  \BibitemOpen
  \bibfield  {author} {\bibinfo {author} {\bibfnamefont {S.}~\bibnamefont
  {Nakamura}}, \bibinfo {author} {\bibfnamefont {Y.}~\bibnamefont {Iida}},
  \bibinfo {author} {\bibfnamefont {Y.}~\bibnamefont {Murotani}}, \bibinfo
  {author} {\bibfnamefont {R.}~\bibnamefont {Matsunaga}}, \bibinfo {author}
  {\bibfnamefont {H.}~\bibnamefont {Terai}},\ and\ \bibinfo {author}
  {\bibfnamefont {R.}~\bibnamefont {Shimano}},\ }\bibfield  {title} {\bibinfo
  {title} {Infrared activation of the higgs mode by supercurrent injection in
  superconducting nbn},\ }\href@noop {} {\bibfield  {journal} {\bibinfo
  {journal} {Phys. Rev. Lett.}\ }\textbf {\bibinfo {volume} {122}},\ \bibinfo
  {pages} {257001} (\bibinfo {year} {2019})}\BibitemShut {NoStop}%
\bibitem [{\citenamefont {Wang}\ \emph {et~al.}(2022)\citenamefont {Wang},
  \citenamefont {Xue}, \citenamefont {Shi}, \citenamefont {Jia}, \citenamefont
  {Lin}, \citenamefont {Shi}, \citenamefont {Dong}, \citenamefont {Wang},\ and\
  \citenamefont {Wang}}]{PhysRevB.105.L100508}%
  \BibitemOpen
  \bibfield  {author} {\bibinfo {author} {\bibfnamefont {Z.-X.}\ \bibnamefont
  {Wang}}, \bibinfo {author} {\bibfnamefont {J.-R.}\ \bibnamefont {Xue}},
  \bibinfo {author} {\bibfnamefont {H.-K.}\ \bibnamefont {Shi}}, \bibinfo
  {author} {\bibfnamefont {X.-Q.}\ \bibnamefont {Jia}}, \bibinfo {author}
  {\bibfnamefont {T.}~\bibnamefont {Lin}}, \bibinfo {author} {\bibfnamefont
  {L.-Y.}\ \bibnamefont {Shi}}, \bibinfo {author} {\bibfnamefont
  {T.}~\bibnamefont {Dong}}, \bibinfo {author} {\bibfnamefont {F.}~\bibnamefont
  {Wang}},\ and\ \bibinfo {author} {\bibfnamefont {N.-L.}\ \bibnamefont
  {Wang}},\ }\bibfield  {title} {\bibinfo {title} {Transient higgs oscillations
  and high-order nonlinear light-higgs coupling in a terahertz wave driven nbn
  superconductor},\ }\href@noop {} {\bibfield  {journal} {\bibinfo  {journal}
  {Phys. Rev. B}\ }\textbf {\bibinfo {volume} {105}},\ \bibinfo {pages}
  {L100508} (\bibinfo {year} {2022})}\BibitemShut {NoStop}%
\bibitem [{\citenamefont {Vaswani}\ \emph {et~al.}(2020)\citenamefont
  {Vaswani}, \citenamefont {Mootz}, \citenamefont {Sundahl}, \citenamefont
  {Mudiyanselage}, \citenamefont {Kang}, \citenamefont {Yang}, \citenamefont
  {Cheng}, \citenamefont {Huang}, \citenamefont {Kim}, \citenamefont {Liu},
  \citenamefont {Luo}, \citenamefont {Perakis}, \citenamefont {Eom},\ and\
  \citenamefont {Wang}}]{PhysRevLett.124.207003}%
  \BibitemOpen
  \bibfield  {author} {\bibinfo {author} {\bibfnamefont {C.}~\bibnamefont
  {Vaswani}}, \bibinfo {author} {\bibfnamefont {M.}~\bibnamefont {Mootz}},
  \bibinfo {author} {\bibfnamefont {C.}~\bibnamefont {Sundahl}}, \bibinfo
  {author} {\bibfnamefont {D.~H.}\ \bibnamefont {Mudiyanselage}}, \bibinfo
  {author} {\bibfnamefont {J.~H.}\ \bibnamefont {Kang}}, \bibinfo {author}
  {\bibfnamefont {X.}~\bibnamefont {Yang}}, \bibinfo {author} {\bibfnamefont
  {D.}~\bibnamefont {Cheng}}, \bibinfo {author} {\bibfnamefont
  {C.}~\bibnamefont {Huang}}, \bibinfo {author} {\bibfnamefont {R.~H.~J.}\
  \bibnamefont {Kim}}, \bibinfo {author} {\bibfnamefont {Z.}~\bibnamefont
  {Liu}}, \bibinfo {author} {\bibfnamefont {L.}~\bibnamefont {Luo}}, \bibinfo
  {author} {\bibfnamefont {I.~E.}\ \bibnamefont {Perakis}}, \bibinfo {author}
  {\bibfnamefont {C.~B.}\ \bibnamefont {Eom}},\ and\ \bibinfo {author}
  {\bibfnamefont {J.}~\bibnamefont {Wang}},\ }\bibfield  {title} {\bibinfo
  {title} {Terahertz second-harmonic generation from lightwave acceleration of
  symmetry-breaking nonlinear supercurrents},\ }\href@noop {} {\bibfield
  {journal} {\bibinfo  {journal} {Phys. Rev. Lett.}\ }\textbf {\bibinfo
  {volume} {124}},\ \bibinfo {pages} {207003} (\bibinfo {year}
  {2020})}\BibitemShut {NoStop}%
\bibitem [{\citenamefont {Chu}\ \emph {et~al.}(2020)\citenamefont {Chu},
  \citenamefont {Kim}, \citenamefont {Katsumi}, \citenamefont {Kovalev},
  \citenamefont {Dawson}, \citenamefont {Schwarz}, \citenamefont {Yoshikawa},
  \citenamefont {Kim}, \citenamefont {Putzky}, \citenamefont {Li} \emph
  {et~al.}}]{chu2020phase}%
  \BibitemOpen
  \bibfield  {author} {\bibinfo {author} {\bibfnamefont {H.}~\bibnamefont
  {Chu}}, \bibinfo {author} {\bibfnamefont {M.-J.}\ \bibnamefont {Kim}},
  \bibinfo {author} {\bibfnamefont {K.}~\bibnamefont {Katsumi}}, \bibinfo
  {author} {\bibfnamefont {S.}~\bibnamefont {Kovalev}}, \bibinfo {author}
  {\bibfnamefont {R.~D.}\ \bibnamefont {Dawson}}, \bibinfo {author}
  {\bibfnamefont {L.}~\bibnamefont {Schwarz}}, \bibinfo {author} {\bibfnamefont
  {N.}~\bibnamefont {Yoshikawa}}, \bibinfo {author} {\bibfnamefont
  {G.}~\bibnamefont {Kim}}, \bibinfo {author} {\bibfnamefont {D.}~\bibnamefont
  {Putzky}}, \bibinfo {author} {\bibfnamefont {Z.~Z.}\ \bibnamefont {Li}},
  \emph {et~al.},\ }\bibfield  {title} {\bibinfo {title} {Phase-resolved higgs
  response in superconducting cuprates},\ }\href@noop {} {\bibfield  {journal}
  {\bibinfo  {journal} {Nat. Commun.}\ }\textbf {\bibinfo {volume} {11}},\
  \bibinfo {pages} {1793} (\bibinfo {year} {2020})}\BibitemShut {NoStop}%
\bibitem [{\citenamefont {Katsumi}\ \emph {et~al.}(2018)\citenamefont
  {Katsumi}, \citenamefont {Tsuji}, \citenamefont {Hamada}, \citenamefont
  {Matsunaga}, \citenamefont {Schneeloch}, \citenamefont {Zhong}, \citenamefont
  {Gu}, \citenamefont {Aoki}, \citenamefont {Gallais},\ and\ \citenamefont
  {Shimano}}]{PhysRevLett.120.117001}%
  \BibitemOpen
  \bibfield  {author} {\bibinfo {author} {\bibfnamefont {K.}~\bibnamefont
  {Katsumi}}, \bibinfo {author} {\bibfnamefont {N.}~\bibnamefont {Tsuji}},
  \bibinfo {author} {\bibfnamefont {Y.~I.}\ \bibnamefont {Hamada}}, \bibinfo
  {author} {\bibfnamefont {R.}~\bibnamefont {Matsunaga}}, \bibinfo {author}
  {\bibfnamefont {J.}~\bibnamefont {Schneeloch}}, \bibinfo {author}
  {\bibfnamefont {R.~D.}\ \bibnamefont {Zhong}}, \bibinfo {author}
  {\bibfnamefont {G.~D.}\ \bibnamefont {Gu}}, \bibinfo {author} {\bibfnamefont
  {H.}~\bibnamefont {Aoki}}, \bibinfo {author} {\bibfnamefont {Y.}~\bibnamefont
  {Gallais}},\ and\ \bibinfo {author} {\bibfnamefont {R.}~\bibnamefont
  {Shimano}},\ }\bibfield  {title} {\bibinfo {title} {Higgs mode in the
  $d$-wave superconductor
  ${\mathrm{bi}}_{2}{\mathrm{sr}}_{2}{\mathrm{cacu}}_{2}{\mathrm{o}}_{8+x}$
  driven by an intense terahertz pulse},\ }\href@noop {} {\bibfield  {journal}
  {\bibinfo  {journal} {Phys. Rev. Lett.}\ }\textbf {\bibinfo {volume} {120}},\
  \bibinfo {pages} {117001} (\bibinfo {year} {2018})}\BibitemShut {NoStop}%
\bibitem [{\citenamefont {Yang}\ \emph {et~al.}(2019)\citenamefont {Yang},
  \citenamefont {Vaswani}, \citenamefont {Sundahl}, \citenamefont {Mootz},
  \citenamefont {Luo}, \citenamefont {Kang}, \citenamefont {Perakis},
  \citenamefont {Eom},\ and\ \citenamefont {Wang}}]{yang2019lightwave}%
  \BibitemOpen
  \bibfield  {author} {\bibinfo {author} {\bibfnamefont {X.}~\bibnamefont
  {Yang}}, \bibinfo {author} {\bibfnamefont {C.}~\bibnamefont {Vaswani}},
  \bibinfo {author} {\bibfnamefont {C.}~\bibnamefont {Sundahl}}, \bibinfo
  {author} {\bibfnamefont {M.}~\bibnamefont {Mootz}}, \bibinfo {author}
  {\bibfnamefont {L.}~\bibnamefont {Luo}}, \bibinfo {author} {\bibfnamefont
  {J.}~\bibnamefont {Kang}}, \bibinfo {author} {\bibfnamefont {I.}~\bibnamefont
  {Perakis}}, \bibinfo {author} {\bibfnamefont {C.}~\bibnamefont {Eom}},\ and\
  \bibinfo {author} {\bibfnamefont {J.}~\bibnamefont {Wang}},\ }\bibfield
  {title} {\bibinfo {title} {Lightwave-driven gapless superconductivity and
  forbidden quantum beats by terahertz symmetry breaking},\ }\href@noop {}
  {\bibfield  {journal} {\bibinfo  {journal} {Nat. Photonics}\ }\textbf
  {\bibinfo {volume} {13}},\ \bibinfo {pages} {707} (\bibinfo {year}
  {2019})}\BibitemShut {NoStop}%
\bibitem [{\citenamefont {Luo}\ \emph {et~al.}(2023)\citenamefont {Luo},
  \citenamefont {Mootz}, \citenamefont {Kang}, \citenamefont {Huang},
  \citenamefont {Eom}, \citenamefont {Lee}, \citenamefont {Vaswani},
  \citenamefont {Collantes}, \citenamefont {Hellstrom}, \citenamefont {Perakis}
  \emph {et~al.}}]{luo2023quantum}%
  \BibitemOpen
  \bibfield  {author} {\bibinfo {author} {\bibfnamefont {L.}~\bibnamefont
  {Luo}}, \bibinfo {author} {\bibfnamefont {M.}~\bibnamefont {Mootz}}, \bibinfo
  {author} {\bibfnamefont {J.-H.}\ \bibnamefont {Kang}}, \bibinfo {author}
  {\bibfnamefont {C.}~\bibnamefont {Huang}}, \bibinfo {author} {\bibfnamefont
  {K.}~\bibnamefont {Eom}}, \bibinfo {author} {\bibfnamefont {J.}~\bibnamefont
  {Lee}}, \bibinfo {author} {\bibfnamefont {C.}~\bibnamefont {Vaswani}},
  \bibinfo {author} {\bibfnamefont {Y.}~\bibnamefont {Collantes}}, \bibinfo
  {author} {\bibfnamefont {E.}~\bibnamefont {Hellstrom}}, \bibinfo {author}
  {\bibfnamefont {I.~E.}\ \bibnamefont {Perakis}}, \emph {et~al.},\ }\bibfield
  {title} {\bibinfo {title} {Quantum coherence tomography of light-controlled
  superconductivity},\ }\href@noop {} {\bibfield  {journal} {\bibinfo
  {journal} {Nat. Phys.}\ }\textbf {\bibinfo {volume} {19}},\ \bibinfo {pages}
  {201} (\bibinfo {year} {2023})}\BibitemShut {NoStop}%
\bibitem [{\citenamefont {Katsumi}\ \emph {et~al.}(2020)\citenamefont
  {Katsumi}, \citenamefont {Li}, \citenamefont {Raffy}, \citenamefont
  {Gallais},\ and\ \citenamefont {Shimano}}]{PhysRevB.102.054510}%
  \BibitemOpen
  \bibfield  {author} {\bibinfo {author} {\bibfnamefont {K.}~\bibnamefont
  {Katsumi}}, \bibinfo {author} {\bibfnamefont {Z.~Z.}\ \bibnamefont {Li}},
  \bibinfo {author} {\bibfnamefont {H.}~\bibnamefont {Raffy}}, \bibinfo
  {author} {\bibfnamefont {Y.}~\bibnamefont {Gallais}},\ and\ \bibinfo {author}
  {\bibfnamefont {R.}~\bibnamefont {Shimano}},\ }\bibfield  {title} {\bibinfo
  {title} {Superconducting fluctuations probed by the higgs mode in
  ${\mathrm{bi}}_{2}{\mathrm{sr}}_{2}\mathrm{Ca}{\mathrm{cu}}_{2}{\mathrm{o}}_{8+x}$
  thin films},\ }\href@noop {} {\bibfield  {journal} {\bibinfo  {journal}
  {Phys. Rev. B}\ }\textbf {\bibinfo {volume} {102}},\ \bibinfo {pages}
  {054510} (\bibinfo {year} {2020})}\BibitemShut {NoStop}%
\bibitem [{\citenamefont {Yang}\ and\ \citenamefont
  {Wu}(2019)}]{yang2019gauge}%
  \BibitemOpen
  \bibfield  {author} {\bibinfo {author} {\bibfnamefont {F.}~\bibnamefont
  {Yang}}\ and\ \bibinfo {author} {\bibfnamefont {M.}~\bibnamefont {Wu}},\
  }\bibfield  {title} {\bibinfo {title} {Gauge-invariant microscopic kinetic
  theory of superconductivity: {Application} to the optical response of
  {Nambu-Goldstone and Higgs modes}},\ }\href@noop {} {\bibfield  {journal}
  {\bibinfo  {journal} {Phys. Rev. B}\ }\textbf {\bibinfo {volume} {100}},\
  \bibinfo {pages} {104513} (\bibinfo {year} {2019})}\BibitemShut {NoStop}%
\bibitem [{\citenamefont {Yang}\ and\ \citenamefont
  {Wu}(2023)}]{yang2023optical}%
  \BibitemOpen
  \bibfield  {author} {\bibinfo {author} {\bibfnamefont {F.}~\bibnamefont
  {Yang}}\ and\ \bibinfo {author} {\bibfnamefont {M.}~\bibnamefont {Wu}},\
  }\bibfield  {title} {\bibinfo {title} {Optical response of higgs mode in
  superconductors at clean limit},\ }\href@noop {} {\bibfield  {journal}
  {\bibinfo  {journal} {Ann. Phys.}\ }\textbf {\bibinfo {volume} {453}},\
  \bibinfo {pages} {169312} (\bibinfo {year} {2023})}\BibitemShut {NoStop}%
\bibitem [{\citenamefont {Pekker}\ and\ \citenamefont
  {Varma}(2015)}]{pekker2015amplitude}%
  \BibitemOpen
  \bibfield  {author} {\bibinfo {author} {\bibfnamefont {D.}~\bibnamefont
  {Pekker}}\ and\ \bibinfo {author} {\bibfnamefont {C.}~\bibnamefont {Varma}},\
  }\bibfield  {title} {\bibinfo {title} {Amplitude/{Higgs} modes in condensed
  matter physics},\ }\href@noop {} {\bibfield  {journal} {\bibinfo  {journal}
  {Annu. Rev. Condens. Matter Phys.}\ }\textbf {\bibinfo {volume} {6}},\
  \bibinfo {pages} {269} (\bibinfo {year} {2015})}\BibitemShut {NoStop}%
\bibitem [{\citenamefont {Cui}\ \emph {et~al.}(2019)\citenamefont {Cui},
  \citenamefont {Yang}, \citenamefont {Vaswani}, \citenamefont {Wang},
  \citenamefont {Fernandes},\ and\ \citenamefont {Orth}}]{cui2019impact}%
  \BibitemOpen
  \bibfield  {author} {\bibinfo {author} {\bibfnamefont {T.}~\bibnamefont
  {Cui}}, \bibinfo {author} {\bibfnamefont {X.}~\bibnamefont {Yang}}, \bibinfo
  {author} {\bibfnamefont {C.}~\bibnamefont {Vaswani}}, \bibinfo {author}
  {\bibfnamefont {J.}~\bibnamefont {Wang}}, \bibinfo {author} {\bibfnamefont
  {R.~M.}\ \bibnamefont {Fernandes}},\ and\ \bibinfo {author} {\bibfnamefont
  {P.~P.}\ \bibnamefont {Orth}},\ }\bibfield  {title} {\bibinfo {title} {Impact
  of damping on the superconducting gap dynamics induced by intense terahertz
  pulses},\ }\href@noop {} {\bibfield  {journal} {\bibinfo  {journal} {Phys.
  Rev. B}\ }\textbf {\bibinfo {volume} {100}},\ \bibinfo {pages} {054504}
  (\bibinfo {year} {2019})}\BibitemShut {NoStop}%
\bibitem [{\citenamefont {Mootz}\ \emph {et~al.}(2020)\citenamefont {Mootz},
  \citenamefont {Wang},\ and\ \citenamefont {Perakis}}]{PhysRevB.102.054517}%
  \BibitemOpen
  \bibfield  {author} {\bibinfo {author} {\bibfnamefont {M.}~\bibnamefont
  {Mootz}}, \bibinfo {author} {\bibfnamefont {J.}~\bibnamefont {Wang}},\ and\
  \bibinfo {author} {\bibfnamefont {I.~E.}\ \bibnamefont {Perakis}},\
  }\bibfield  {title} {\bibinfo {title} {Lightwave terahertz quantum
  manipulation of nonequilibrium superconductor phases and their collective
  modes},\ }\href@noop {} {\bibfield  {journal} {\bibinfo  {journal} {Phys.
  Rev. B}\ }\textbf {\bibinfo {volume} {102}},\ \bibinfo {pages} {054517}
  (\bibinfo {year} {2020})}\BibitemShut {NoStop}%
\bibitem [{\citenamefont {Abrikosov}(1957)}]{abrikosov1957magnetic}%
  \BibitemOpen
  \bibfield  {author} {\bibinfo {author} {\bibfnamefont {A.~A.}\ \bibnamefont
  {Abrikosov}},\ }\bibfield  {title} {\bibinfo {title} {On the magnetic
  properties of superconductors of the second group},\ }\href@noop {}
  {\bibfield  {journal} {\bibinfo  {journal} {Sov. Phys. JETP}\ }\textbf
  {\bibinfo {volume} {5}},\ \bibinfo {pages} {1174} (\bibinfo {year}
  {1957})}\BibitemShut {NoStop}%
\bibitem [{\citenamefont {Caroli}\ \emph {et~al.}(1964)\citenamefont {Caroli},
  \citenamefont {De~Gennes},\ and\ \citenamefont {Matricon}}]{caroli1964bound}%
  \BibitemOpen
  \bibfield  {author} {\bibinfo {author} {\bibfnamefont {C.}~\bibnamefont
  {Caroli}}, \bibinfo {author} {\bibfnamefont {P.}~\bibnamefont {De~Gennes}},\
  and\ \bibinfo {author} {\bibfnamefont {J.}~\bibnamefont {Matricon}},\
  }\bibfield  {title} {\bibinfo {title} {Bound fermion states on a vortex line
  in a type ii superconductor},\ }\href@noop {} {\bibfield  {journal} {\bibinfo
   {journal} {Phys. Lett.}\ }\textbf {\bibinfo {volume} {9}},\ \bibinfo {pages}
  {307} (\bibinfo {year} {1964})}\BibitemShut {NoStop}%
\bibitem [{\citenamefont {Luh}(1965)}]{luh1965bound}%
  \BibitemOpen
  \bibfield  {author} {\bibinfo {author} {\bibfnamefont {Y.}~\bibnamefont
  {Luh}},\ }\bibfield  {title} {\bibinfo {title} {Bound state in
  superconductors with paramagnetic impurities},\ }\href@noop {} {\bibfield
  {journal} {\bibinfo  {journal} {Acta. Phys. Sin.}\ }\textbf {\bibinfo
  {volume} {21}},\ \bibinfo {pages} {75} (\bibinfo {year} {1965})}\BibitemShut
  {NoStop}%
\bibitem [{\citenamefont {Shiba}(1968)}]{shiba1968classical}%
  \BibitemOpen
  \bibfield  {author} {\bibinfo {author} {\bibfnamefont {H.}~\bibnamefont
  {Shiba}},\ }\bibfield  {title} {\bibinfo {title} {Classical spins in
  superconductors},\ }\href@noop {} {\bibfield  {journal} {\bibinfo  {journal}
  {Prog. Theor. Phys.}\ }\textbf {\bibinfo {volume} {40}},\ \bibinfo {pages}
  {435} (\bibinfo {year} {1968})}\BibitemShut {NoStop}%
\bibitem [{\citenamefont {Rusinov}(1969)}]{rusinov1969superconductivity}%
  \BibitemOpen
  \bibfield  {author} {\bibinfo {author} {\bibfnamefont {A.}~\bibnamefont
  {Rusinov}},\ }\bibfield  {title} {\bibinfo {title} {Superconductivity near a
  paramagnetic impurity},\ }\href@noop {} {\bibfield  {journal} {\bibinfo
  {journal} {JETP Lett.}\ }\textbf {\bibinfo {volume} {9}} (\bibinfo {year}
  {1969})}\BibitemShut {NoStop}%
\bibitem [{\citenamefont {Nayak}\ \emph {et~al.}(2008)\citenamefont {Nayak},
  \citenamefont {Simon}, \citenamefont {Stern}, \citenamefont {Freedman},\ and\
  \citenamefont {Das~Sarma}}]{RevModPhys.80.1083}%
  \BibitemOpen
  \bibfield  {author} {\bibinfo {author} {\bibfnamefont {C.}~\bibnamefont
  {Nayak}}, \bibinfo {author} {\bibfnamefont {S.~H.}\ \bibnamefont {Simon}},
  \bibinfo {author} {\bibfnamefont {A.}~\bibnamefont {Stern}}, \bibinfo
  {author} {\bibfnamefont {M.}~\bibnamefont {Freedman}},\ and\ \bibinfo
  {author} {\bibfnamefont {S.}~\bibnamefont {Das~Sarma}},\ }\bibfield  {title}
  {\bibinfo {title} {Non-abelian anyons and topological quantum computation},\
  }\href@noop {} {\bibfield  {journal} {\bibinfo  {journal} {Rev. Mod. Phys.}\
  }\textbf {\bibinfo {volume} {80}},\ \bibinfo {pages} {1083} (\bibinfo {year}
  {2008})}\BibitemShut {NoStop}%
\bibitem [{\citenamefont {Fu}\ and\ \citenamefont
  {Kane}(2008)}]{PhysRevLett.100.096407}%
  \BibitemOpen
  \bibfield  {author} {\bibinfo {author} {\bibfnamefont {L.}~\bibnamefont
  {Fu}}\ and\ \bibinfo {author} {\bibfnamefont {C.~L.}\ \bibnamefont {Kane}},\
  }\bibfield  {title} {\bibinfo {title} {Superconducting proximity effect and
  majorana fermions at the surface of a topological insulator},\ }\href@noop {}
  {\bibfield  {journal} {\bibinfo  {journal} {Phys. Rev. Lett.}\ }\textbf
  {\bibinfo {volume} {100}},\ \bibinfo {pages} {096407} (\bibinfo {year}
  {2008})}\BibitemShut {NoStop}%
\bibitem [{\citenamefont {Lutchyn}\ \emph {et~al.}(2010)\citenamefont
  {Lutchyn}, \citenamefont {Sau},\ and\ \citenamefont
  {Das~Sarma}}]{PhysRevLett.105.077001}%
  \BibitemOpen
  \bibfield  {author} {\bibinfo {author} {\bibfnamefont {R.~M.}\ \bibnamefont
  {Lutchyn}}, \bibinfo {author} {\bibfnamefont {J.~D.}\ \bibnamefont {Sau}},\
  and\ \bibinfo {author} {\bibfnamefont {S.}~\bibnamefont {Das~Sarma}},\
  }\bibfield  {title} {\bibinfo {title} {Majorana fermions and a topological
  phase transition in semiconductor-superconductor heterostructures},\
  }\href@noop {} {\bibfield  {journal} {\bibinfo  {journal} {Phys. Rev. Lett.}\
  }\textbf {\bibinfo {volume} {105}},\ \bibinfo {pages} {077001} (\bibinfo
  {year} {2010})}\BibitemShut {NoStop}%
\bibitem [{\citenamefont {Oreg}\ \emph {et~al.}(2010)\citenamefont {Oreg},
  \citenamefont {Refael},\ and\ \citenamefont {von
  Oppen}}]{PhysRevLett.105.177002}%
  \BibitemOpen
  \bibfield  {author} {\bibinfo {author} {\bibfnamefont {Y.}~\bibnamefont
  {Oreg}}, \bibinfo {author} {\bibfnamefont {G.}~\bibnamefont {Refael}},\ and\
  \bibinfo {author} {\bibfnamefont {F.}~\bibnamefont {von Oppen}},\ }\bibfield
  {title} {\bibinfo {title} {Helical liquids and majorana bound states in
  quantum wires},\ }\href@noop {} {\bibfield  {journal} {\bibinfo  {journal}
  {Phys. Rev. Lett.}\ }\textbf {\bibinfo {volume} {105}},\ \bibinfo {pages}
  {177002} (\bibinfo {year} {2010})}\BibitemShut {NoStop}%
\bibitem [{\citenamefont {Beenakker}(2013)}]{annurev}%
  \BibitemOpen
  \bibfield  {author} {\bibinfo {author} {\bibfnamefont {C.}~\bibnamefont
  {Beenakker}},\ }\bibfield  {title} {\bibinfo {title} {Search for majorana
  fermions in superconductors},\ }\href@noop {} {\bibfield  {journal} {\bibinfo
   {journal} {Annu. Rev. Condens. Matter Phys.}\ }\textbf {\bibinfo {volume}
  {4}},\ \bibinfo {pages} {113} (\bibinfo {year} {2013})}\BibitemShut {NoStop}%
\bibitem [{\citenamefont {Balatsky}\ \emph {et~al.}(2006)\citenamefont
  {Balatsky}, \citenamefont {Vekhter},\ and\ \citenamefont
  {Zhu}}]{RevModPhys.78.373}%
  \BibitemOpen
  \bibfield  {author} {\bibinfo {author} {\bibfnamefont {A.~V.}\ \bibnamefont
  {Balatsky}}, \bibinfo {author} {\bibfnamefont {I.}~\bibnamefont {Vekhter}},\
  and\ \bibinfo {author} {\bibfnamefont {J.-X.}\ \bibnamefont {Zhu}},\
  }\bibfield  {title} {\bibinfo {title} {Impurity-induced states in
  conventional and unconventional superconductors},\ }\href@noop {} {\bibfield
  {journal} {\bibinfo  {journal} {Rev. Mod. Phys.}\ }\textbf {\bibinfo {volume}
  {78}},\ \bibinfo {pages} {373} (\bibinfo {year} {2006})}\BibitemShut
  {NoStop}%
\bibitem [{\citenamefont {M{\'e}nard}\ \emph {et~al.}(2015)\citenamefont
  {M{\'e}nard}, \citenamefont {Guissart}, \citenamefont {Brun}, \citenamefont
  {Pons}, \citenamefont {Stolyarov}, \citenamefont {Debontridder},
  \citenamefont {Leclerc}, \citenamefont {Janod}, \citenamefont {Cario},
  \citenamefont {Roditchev} \emph {et~al.}}]{menard2015coherent}%
  \BibitemOpen
  \bibfield  {author} {\bibinfo {author} {\bibfnamefont {G.~C.}\ \bibnamefont
  {M{\'e}nard}}, \bibinfo {author} {\bibfnamefont {S.}~\bibnamefont
  {Guissart}}, \bibinfo {author} {\bibfnamefont {C.}~\bibnamefont {Brun}},
  \bibinfo {author} {\bibfnamefont {S.}~\bibnamefont {Pons}}, \bibinfo {author}
  {\bibfnamefont {V.~S.}\ \bibnamefont {Stolyarov}}, \bibinfo {author}
  {\bibfnamefont {F.}~\bibnamefont {Debontridder}}, \bibinfo {author}
  {\bibfnamefont {M.~V.}\ \bibnamefont {Leclerc}}, \bibinfo {author}
  {\bibfnamefont {E.}~\bibnamefont {Janod}}, \bibinfo {author} {\bibfnamefont
  {L.}~\bibnamefont {Cario}}, \bibinfo {author} {\bibfnamefont
  {D.}~\bibnamefont {Roditchev}}, \emph {et~al.},\ }\bibfield  {title}
  {\bibinfo {title} {Coherent long-range magnetic bound states in a
  superconductor},\ }\href@noop {} {\bibfield  {journal} {\bibinfo  {journal}
  {Nat. Phys.}\ }\textbf {\bibinfo {volume} {11}},\ \bibinfo {pages} {1013}
  (\bibinfo {year} {2015})}\BibitemShut {NoStop}%
\bibitem [{\citenamefont {Hatter}\ \emph {et~al.}(2015)\citenamefont {Hatter},
  \citenamefont {Heinrich}, \citenamefont {Ruby}, \citenamefont {Pascual},\
  and\ \citenamefont {Franke}}]{hatter2015magnetic}%
  \BibitemOpen
  \bibfield  {author} {\bibinfo {author} {\bibfnamefont {N.}~\bibnamefont
  {Hatter}}, \bibinfo {author} {\bibfnamefont {B.~W.}\ \bibnamefont
  {Heinrich}}, \bibinfo {author} {\bibfnamefont {M.}~\bibnamefont {Ruby}},
  \bibinfo {author} {\bibfnamefont {J.~I.}\ \bibnamefont {Pascual}},\ and\
  \bibinfo {author} {\bibfnamefont {K.~J.}\ \bibnamefont {Franke}},\ }\bibfield
   {title} {\bibinfo {title} {Magnetic anisotropy in shiba bound states across
  a quantum phase transition},\ }\href@noop {} {\bibfield  {journal} {\bibinfo
  {journal} {Nat. Commun.}\ }\textbf {\bibinfo {volume} {6}},\ \bibinfo {pages}
  {8988} (\bibinfo {year} {2015})}\BibitemShut {NoStop}%
\bibitem [{\citenamefont {\ifmmode~\check{z}\else \v{z}\fi{}itko}\ \emph
  {et~al.}(2011)\citenamefont {\ifmmode~\check{z}\else \v{z}\fi{}itko},
  \citenamefont {Bodensiek},\ and\ \citenamefont
  {Pruschke}}]{PhysRevB.83.054512}%
  \BibitemOpen
  \bibfield  {author} {\bibinfo {author} {\bibfnamefont {R.}~\bibnamefont
  {\ifmmode~\check{z}\else \v{z}\fi{}itko}}, \bibinfo {author} {\bibfnamefont
  {O.}~\bibnamefont {Bodensiek}},\ and\ \bibinfo {author} {\bibfnamefont
  {T.}~\bibnamefont {Pruschke}},\ }\bibfield  {title} {\bibinfo {title}
  {Effects of magnetic anisotropy on the subgap excitations induced by quantum
  impurities in a superconducting host},\ }\href@noop {} {\bibfield  {journal}
  {\bibinfo  {journal} {Phys. Rev. B}\ }\textbf {\bibinfo {volume} {83}},\
  \bibinfo {pages} {054512} (\bibinfo {year} {2011})}\BibitemShut {NoStop}%
\bibitem [{\citenamefont {Hudson}\ \emph {et~al.}(2001)\citenamefont {Hudson},
  \citenamefont {Lang}, \citenamefont {Madhavan}, \citenamefont {Pan},
  \citenamefont {Eisaki}, \citenamefont {Uchida},\ and\ \citenamefont
  {Davis}}]{hudson2001interplay}%
  \BibitemOpen
  \bibfield  {author} {\bibinfo {author} {\bibfnamefont {E.}~\bibnamefont
  {Hudson}}, \bibinfo {author} {\bibfnamefont {K.}~\bibnamefont {Lang}},
  \bibinfo {author} {\bibfnamefont {V.}~\bibnamefont {Madhavan}}, \bibinfo
  {author} {\bibfnamefont {S.}~\bibnamefont {Pan}}, \bibinfo {author}
  {\bibfnamefont {H.}~\bibnamefont {Eisaki}}, \bibinfo {author} {\bibfnamefont
  {S.}~\bibnamefont {Uchida}},\ and\ \bibinfo {author} {\bibfnamefont
  {J.}~\bibnamefont {Davis}},\ }\bibfield  {title} {\bibinfo {title} {Interplay
  of magnetism and high-t c superconductivity at individual ni impurity atoms
  in bi2sr2cacu2o8+ $\delta$},\ }\href@noop {} {\bibfield  {journal} {\bibinfo
  {journal} {Nature}\ }\textbf {\bibinfo {volume} {411}},\ \bibinfo {pages}
  {920} (\bibinfo {year} {2001})}\BibitemShut {NoStop}%
\bibitem [{\citenamefont {Rubio-Verd\'u}\ \emph {et~al.}(2021)\citenamefont
  {Rubio-Verd\'u}, \citenamefont {Zald\'{\i}var}, \citenamefont
  {\ifmmode~\check{Z}\else \v{Z}\fi{}itko},\ and\ \citenamefont
  {Pascual}}]{PhysRevLett.126.017001}%
  \BibitemOpen
  \bibfield  {author} {\bibinfo {author} {\bibfnamefont {C.}~\bibnamefont
  {Rubio-Verd\'u}}, \bibinfo {author} {\bibfnamefont {J.}~\bibnamefont
  {Zald\'{\i}var}}, \bibinfo {author} {\bibfnamefont {R.}~\bibnamefont
  {\ifmmode~\check{Z}\else \v{Z}\fi{}itko}},\ and\ \bibinfo {author}
  {\bibfnamefont {J.~I.}\ \bibnamefont {Pascual}},\ }\bibfield  {title}
  {\bibinfo {title} {Coupled yu-shiba-rusinov states induced by a many-body
  molecular spin on a superconductor},\ }\href@noop {} {\bibfield  {journal}
  {\bibinfo  {journal} {Phys. Rev. Lett.}\ }\textbf {\bibinfo {volume} {126}},\
  \bibinfo {pages} {017001} (\bibinfo {year} {2021})}\BibitemShut {NoStop}%
\bibitem [{\citenamefont {Nadj-Perge}\ \emph {et~al.}(2014)\citenamefont
  {Nadj-Perge}, \citenamefont {Drozdov}, \citenamefont {Li}, \citenamefont
  {Chen}, \citenamefont {Jeon}, \citenamefont {Seo}, \citenamefont {MacDonald},
  \citenamefont {Bernevig},\ and\ \citenamefont
  {Yazdani}}]{doi:10.1126/science.1259327}%
  \BibitemOpen
  \bibfield  {author} {\bibinfo {author} {\bibfnamefont {S.}~\bibnamefont
  {Nadj-Perge}}, \bibinfo {author} {\bibfnamefont {I.~K.}\ \bibnamefont
  {Drozdov}}, \bibinfo {author} {\bibfnamefont {J.}~\bibnamefont {Li}},
  \bibinfo {author} {\bibfnamefont {H.}~\bibnamefont {Chen}}, \bibinfo {author}
  {\bibfnamefont {S.}~\bibnamefont {Jeon}}, \bibinfo {author} {\bibfnamefont
  {J.}~\bibnamefont {Seo}}, \bibinfo {author} {\bibfnamefont {A.~H.}\
  \bibnamefont {MacDonald}}, \bibinfo {author} {\bibfnamefont {B.~A.}\
  \bibnamefont {Bernevig}},\ and\ \bibinfo {author} {\bibfnamefont
  {A.}~\bibnamefont {Yazdani}},\ }\bibfield  {title} {\bibinfo {title}
  {Observation of majorana fermions in ferromagnetic atomic chains on a
  superconductor},\ }\href@noop {} {\bibfield  {journal} {\bibinfo  {journal}
  {Science}\ }\textbf {\bibinfo {volume} {346}},\ \bibinfo {pages} {602}
  (\bibinfo {year} {2014})}\BibitemShut {NoStop}%
\bibitem [{\citenamefont {Ji}\ \emph {et~al.}(2008)\citenamefont {Ji},
  \citenamefont {Zhang}, \citenamefont {Fu}, \citenamefont {Chen},
  \citenamefont {Ma}, \citenamefont {Li}, \citenamefont {Duan}, \citenamefont
  {Jia},\ and\ \citenamefont {Xue}}]{PhysRevLett.100.226801}%
  \BibitemOpen
  \bibfield  {author} {\bibinfo {author} {\bibfnamefont {S.-H.}\ \bibnamefont
  {Ji}}, \bibinfo {author} {\bibfnamefont {T.}~\bibnamefont {Zhang}}, \bibinfo
  {author} {\bibfnamefont {Y.-S.}\ \bibnamefont {Fu}}, \bibinfo {author}
  {\bibfnamefont {X.}~\bibnamefont {Chen}}, \bibinfo {author} {\bibfnamefont
  {X.-C.}\ \bibnamefont {Ma}}, \bibinfo {author} {\bibfnamefont
  {J.}~\bibnamefont {Li}}, \bibinfo {author} {\bibfnamefont {W.-H.}\
  \bibnamefont {Duan}}, \bibinfo {author} {\bibfnamefont {J.-F.}\ \bibnamefont
  {Jia}},\ and\ \bibinfo {author} {\bibfnamefont {Q.-K.}\ \bibnamefont {Xue}},\
  }\bibfield  {title} {\bibinfo {title} {High-resolution scanning tunneling
  spectroscopy of magnetic impurity induced bound states in the superconducting
  gap of pb thin films},\ }\href@noop {} {\bibfield  {journal} {\bibinfo
  {journal} {Phys. Rev. Lett.}\ }\textbf {\bibinfo {volume} {100}},\ \bibinfo
  {pages} {226801} (\bibinfo {year} {2008})}\BibitemShut {NoStop}%
\bibitem [{\citenamefont {K{\"u}ster}\ \emph {et~al.}(2021)\citenamefont
  {K{\"u}ster}, \citenamefont {Brinker}, \citenamefont {Lounis}, \citenamefont
  {Parkin},\ and\ \citenamefont {Sessi}}]{kuster2021long}%
  \BibitemOpen
  \bibfield  {author} {\bibinfo {author} {\bibfnamefont {F.}~\bibnamefont
  {K{\"u}ster}}, \bibinfo {author} {\bibfnamefont {S.}~\bibnamefont {Brinker}},
  \bibinfo {author} {\bibfnamefont {S.}~\bibnamefont {Lounis}}, \bibinfo
  {author} {\bibfnamefont {S.~S.}\ \bibnamefont {Parkin}},\ and\ \bibinfo
  {author} {\bibfnamefont {P.}~\bibnamefont {Sessi}},\ }\bibfield  {title}
  {\bibinfo {title} {Long range and highly tunable interaction between local
  spins coupled to a superconducting condensate},\ }\href@noop {} {\bibfield
  {journal} {\bibinfo  {journal} {Nat. Commun.}\ }\textbf {\bibinfo {volume}
  {12}},\ \bibinfo {pages} {6722} (\bibinfo {year} {2021})}\BibitemShut
  {NoStop}%
\bibitem [{\citenamefont {Choi}\ \emph {et~al.}(2017)\citenamefont {Choi},
  \citenamefont {Rubio-Verd{\'u}}, \citenamefont {De~Bruijckere}, \citenamefont
  {Ugeda}, \citenamefont {Lorente},\ and\ \citenamefont
  {Pascual}}]{choi2017mapping}%
  \BibitemOpen
  \bibfield  {author} {\bibinfo {author} {\bibfnamefont {D.-J.}\ \bibnamefont
  {Choi}}, \bibinfo {author} {\bibfnamefont {C.}~\bibnamefont
  {Rubio-Verd{\'u}}}, \bibinfo {author} {\bibfnamefont {J.}~\bibnamefont
  {De~Bruijckere}}, \bibinfo {author} {\bibfnamefont {M.~M.}\ \bibnamefont
  {Ugeda}}, \bibinfo {author} {\bibfnamefont {N.}~\bibnamefont {Lorente}},\
  and\ \bibinfo {author} {\bibfnamefont {J.~I.}\ \bibnamefont {Pascual}},\
  }\bibfield  {title} {\bibinfo {title} {Mapping the orbital structure of
  impurity bound states in a superconductor},\ }\href@noop {} {\bibfield
  {journal} {\bibinfo  {journal} {Nat. Commun.}\ }\textbf {\bibinfo {volume}
  {8}},\ \bibinfo {pages} {15175} (\bibinfo {year} {2017})}\BibitemShut
  {NoStop}%
\bibitem [{\citenamefont {Beck}\ \emph {et~al.}(2021)\citenamefont {Beck},
  \citenamefont {Schneider}, \citenamefont {R{\'o}zsa}, \citenamefont
  {Palot{\'a}s}, \citenamefont {L{\'a}szl{\'o}ffy}, \citenamefont {Szunyogh},
  \citenamefont {Wiebe},\ and\ \citenamefont {Wiesendanger}}]{beck2021spin}%
  \BibitemOpen
  \bibfield  {author} {\bibinfo {author} {\bibfnamefont {P.}~\bibnamefont
  {Beck}}, \bibinfo {author} {\bibfnamefont {L.}~\bibnamefont {Schneider}},
  \bibinfo {author} {\bibfnamefont {L.}~\bibnamefont {R{\'o}zsa}}, \bibinfo
  {author} {\bibfnamefont {K.}~\bibnamefont {Palot{\'a}s}}, \bibinfo {author}
  {\bibfnamefont {A.}~\bibnamefont {L{\'a}szl{\'o}ffy}}, \bibinfo {author}
  {\bibfnamefont {L.}~\bibnamefont {Szunyogh}}, \bibinfo {author}
  {\bibfnamefont {J.}~\bibnamefont {Wiebe}},\ and\ \bibinfo {author}
  {\bibfnamefont {R.}~\bibnamefont {Wiesendanger}},\ }\bibfield  {title}
  {\bibinfo {title} {Spin-orbit coupling induced splitting of yu-shiba-rusinov
  states in antiferromagnetic dimers},\ }\href@noop {} {\bibfield  {journal}
  {\bibinfo  {journal} {Nat. Commun.}\ }\textbf {\bibinfo {volume} {12}},\
  \bibinfo {pages} {2040} (\bibinfo {year} {2021})}\BibitemShut {NoStop}%
\bibitem [{\citenamefont {Franke}\ \emph {et~al.}(2011)\citenamefont {Franke},
  \citenamefont {Schulze},\ and\ \citenamefont
  {Pascual}}]{franke2011competition}%
  \BibitemOpen
  \bibfield  {author} {\bibinfo {author} {\bibfnamefont {K.}~\bibnamefont
  {Franke}}, \bibinfo {author} {\bibfnamefont {G.}~\bibnamefont {Schulze}},\
  and\ \bibinfo {author} {\bibfnamefont {J.}~\bibnamefont {Pascual}},\
  }\bibfield  {title} {\bibinfo {title} {Competition of superconducting
  phenomena and kondo screening at the nanoscale},\ }\href@noop {} {\bibfield
  {journal} {\bibinfo  {journal} {Science}\ }\textbf {\bibinfo {volume}
  {332}},\ \bibinfo {pages} {940} (\bibinfo {year} {2011})}\BibitemShut
  {NoStop}%
\bibitem [{\citenamefont {Brand}\ \emph {et~al.}(2018)\citenamefont {Brand},
  \citenamefont {Gozdzik}, \citenamefont {N\'eel}, \citenamefont {Lado},
  \citenamefont {Fern\'andez-Rossier},\ and\ \citenamefont
  {Kr\"oger}}]{PhysRevB.97.195429}%
  \BibitemOpen
  \bibfield  {author} {\bibinfo {author} {\bibfnamefont {J.}~\bibnamefont
  {Brand}}, \bibinfo {author} {\bibfnamefont {S.}~\bibnamefont {Gozdzik}},
  \bibinfo {author} {\bibfnamefont {N.}~\bibnamefont {N\'eel}}, \bibinfo
  {author} {\bibfnamefont {J.~L.}\ \bibnamefont {Lado}}, \bibinfo {author}
  {\bibfnamefont {J.}~\bibnamefont {Fern\'andez-Rossier}},\ and\ \bibinfo
  {author} {\bibfnamefont {J.}~\bibnamefont {Kr\"oger}},\ }\bibfield  {title}
  {\bibinfo {title} {Electron and cooper-pair transport across a single
  magnetic molecule explored with a scanning tunneling microscope},\
  }\href@noop {} {\bibfield  {journal} {\bibinfo  {journal} {Phys. Rev. B}\
  }\textbf {\bibinfo {volume} {97}},\ \bibinfo {pages} {195429} (\bibinfo
  {year} {2018})}\BibitemShut {NoStop}%
\bibitem [{\citenamefont {Island}\ \emph {et~al.}(2017)\citenamefont {Island},
  \citenamefont {Gaudenzi}, \citenamefont {de~Bruijckere}, \citenamefont
  {Burzur\'{\i}}, \citenamefont {Franco}, \citenamefont {Mas-Torrent},
  \citenamefont {Rovira}, \citenamefont {Veciana}, \citenamefont {Klapwijk},
  \citenamefont {Aguado},\ and\ \citenamefont {van~der
  Zant}}]{PhysRevLett.118.117001}%
  \BibitemOpen
  \bibfield  {author} {\bibinfo {author} {\bibfnamefont {J.~O.}\ \bibnamefont
  {Island}}, \bibinfo {author} {\bibfnamefont {R.}~\bibnamefont {Gaudenzi}},
  \bibinfo {author} {\bibfnamefont {J.}~\bibnamefont {de~Bruijckere}}, \bibinfo
  {author} {\bibfnamefont {E.}~\bibnamefont {Burzur\'{\i}}}, \bibinfo {author}
  {\bibfnamefont {C.}~\bibnamefont {Franco}}, \bibinfo {author} {\bibfnamefont
  {M.}~\bibnamefont {Mas-Torrent}}, \bibinfo {author} {\bibfnamefont
  {C.}~\bibnamefont {Rovira}}, \bibinfo {author} {\bibfnamefont
  {J.}~\bibnamefont {Veciana}}, \bibinfo {author} {\bibfnamefont {T.~M.}\
  \bibnamefont {Klapwijk}}, \bibinfo {author} {\bibfnamefont {R.}~\bibnamefont
  {Aguado}},\ and\ \bibinfo {author} {\bibfnamefont {H.~S.~J.}\ \bibnamefont
  {van~der Zant}},\ }\bibfield  {title} {\bibinfo {title} {Proximity-induced
  shiba states in a molecular junction},\ }\href@noop {} {\bibfield  {journal}
  {\bibinfo  {journal} {Phys. Rev. Lett.}\ }\textbf {\bibinfo {volume} {118}},\
  \bibinfo {pages} {117001} (\bibinfo {year} {2017})}\BibitemShut {NoStop}%
\bibitem [{\citenamefont {Friedrich}\ \emph {et~al.}(2021)\citenamefont
  {Friedrich}, \citenamefont {Boshuis}, \citenamefont {Bode},\ and\
  \citenamefont {Odobesko}}]{PhysRevB.103.235437}%
  \BibitemOpen
  \bibfield  {author} {\bibinfo {author} {\bibfnamefont {F.}~\bibnamefont
  {Friedrich}}, \bibinfo {author} {\bibfnamefont {R.}~\bibnamefont {Boshuis}},
  \bibinfo {author} {\bibfnamefont {M.}~\bibnamefont {Bode}},\ and\ \bibinfo
  {author} {\bibfnamefont {A.}~\bibnamefont {Odobesko}},\ }\bibfield  {title}
  {\bibinfo {title} {Coupling of yu-shiba-rusinov states in one-dimensional
  chains of fe atoms on nb(110)},\ }\href@noop {} {\bibfield  {journal}
  {\bibinfo  {journal} {Phys. Rev. B}\ }\textbf {\bibinfo {volume} {103}},\
  \bibinfo {pages} {235437} (\bibinfo {year} {2021})}\BibitemShut {NoStop}%
\bibitem [{\citenamefont {Odobesko}\ \emph {et~al.}(2020)\citenamefont
  {Odobesko}, \citenamefont {Di~Sante}, \citenamefont {Kowalski}, \citenamefont
  {Wilfert}, \citenamefont {Friedrich}, \citenamefont {Thomale}, \citenamefont
  {Sangiovanni},\ and\ \citenamefont {Bode}}]{PhysRevB.102.174504}%
  \BibitemOpen
  \bibfield  {author} {\bibinfo {author} {\bibfnamefont {A.}~\bibnamefont
  {Odobesko}}, \bibinfo {author} {\bibfnamefont {D.}~\bibnamefont {Di~Sante}},
  \bibinfo {author} {\bibfnamefont {A.}~\bibnamefont {Kowalski}}, \bibinfo
  {author} {\bibfnamefont {S.}~\bibnamefont {Wilfert}}, \bibinfo {author}
  {\bibfnamefont {F.}~\bibnamefont {Friedrich}}, \bibinfo {author}
  {\bibfnamefont {R.}~\bibnamefont {Thomale}}, \bibinfo {author} {\bibfnamefont
  {G.}~\bibnamefont {Sangiovanni}},\ and\ \bibinfo {author} {\bibfnamefont
  {M.}~\bibnamefont {Bode}},\ }\bibfield  {title} {\bibinfo {title}
  {Observation of tunable single-atom yu-shiba-rusinov states},\ }\href@noop {}
  {\bibfield  {journal} {\bibinfo  {journal} {Phys. Rev. B}\ }\textbf {\bibinfo
  {volume} {102}},\ \bibinfo {pages} {174504} (\bibinfo {year}
  {2020})}\BibitemShut {NoStop}%
\bibitem [{\citenamefont {Ruby}\ \emph {et~al.}(2016)\citenamefont {Ruby},
  \citenamefont {Peng}, \citenamefont {von Oppen}, \citenamefont {Heinrich},\
  and\ \citenamefont {Franke}}]{PhysRevLett.117.186801}%
  \BibitemOpen
  \bibfield  {author} {\bibinfo {author} {\bibfnamefont {M.}~\bibnamefont
  {Ruby}}, \bibinfo {author} {\bibfnamefont {Y.}~\bibnamefont {Peng}}, \bibinfo
  {author} {\bibfnamefont {F.}~\bibnamefont {von Oppen}}, \bibinfo {author}
  {\bibfnamefont {B.~W.}\ \bibnamefont {Heinrich}},\ and\ \bibinfo {author}
  {\bibfnamefont {K.~J.}\ \bibnamefont {Franke}},\ }\bibfield  {title}
  {\bibinfo {title} {Orbital picture of yu-shiba-rusinov multiplets},\
  }\href@noop {} {\bibfield  {journal} {\bibinfo  {journal} {Phys. Rev. Lett.}\
  }\textbf {\bibinfo {volume} {117}},\ \bibinfo {pages} {186801} (\bibinfo
  {year} {2016})}\BibitemShut {NoStop}%
\bibitem [{\citenamefont {P{\"o}yh{\"o}nen}\ \emph {et~al.}(2018)\citenamefont
  {P{\"o}yh{\"o}nen}, \citenamefont {Sahlberg}, \citenamefont {Weststr{\"o}m},\
  and\ \citenamefont {Ojanen}}]{poyhonen2018amorphous}%
  \BibitemOpen
  \bibfield  {author} {\bibinfo {author} {\bibfnamefont {K.}~\bibnamefont
  {P{\"o}yh{\"o}nen}}, \bibinfo {author} {\bibfnamefont {I.}~\bibnamefont
  {Sahlberg}}, \bibinfo {author} {\bibfnamefont {A.}~\bibnamefont
  {Weststr{\"o}m}},\ and\ \bibinfo {author} {\bibfnamefont {T.}~\bibnamefont
  {Ojanen}},\ }\bibfield  {title} {\bibinfo {title} {Amorphous topological
  superconductivity in a shiba glass},\ }\href@noop {} {\bibfield  {journal}
  {\bibinfo  {journal} {Nat. Commun.}\ }\textbf {\bibinfo {volume} {9}},\
  \bibinfo {pages} {2103} (\bibinfo {year} {2018})}\BibitemShut {NoStop}%
\bibitem [{\citenamefont {Kim}\ \emph {et~al.}(2018)\citenamefont {Kim},
  \citenamefont {Palacio-Morales}, \citenamefont {Posske}, \citenamefont
  {R{\'o}zsa}, \citenamefont {Palot{\'a}s}, \citenamefont {Szunyogh},
  \citenamefont {Thorwart},\ and\ \citenamefont
  {Wiesendanger}}]{kim2018toward}%
  \BibitemOpen
  \bibfield  {author} {\bibinfo {author} {\bibfnamefont {H.}~\bibnamefont
  {Kim}}, \bibinfo {author} {\bibfnamefont {A.}~\bibnamefont
  {Palacio-Morales}}, \bibinfo {author} {\bibfnamefont {T.}~\bibnamefont
  {Posske}}, \bibinfo {author} {\bibfnamefont {L.}~\bibnamefont {R{\'o}zsa}},
  \bibinfo {author} {\bibfnamefont {K.}~\bibnamefont {Palot{\'a}s}}, \bibinfo
  {author} {\bibfnamefont {L.}~\bibnamefont {Szunyogh}}, \bibinfo {author}
  {\bibfnamefont {M.}~\bibnamefont {Thorwart}},\ and\ \bibinfo {author}
  {\bibfnamefont {R.}~\bibnamefont {Wiesendanger}},\ }\bibfield  {title}
  {\bibinfo {title} {Toward tailoring majorana bound states in artificially
  constructed magnetic atom chains on elemental superconductors},\ }\href@noop
  {} {\bibfield  {journal} {\bibinfo  {journal} {Sci. Adv.}\ }\textbf {\bibinfo
  {volume} {4}},\ \bibinfo {pages} {eaar5251} (\bibinfo {year}
  {2018})}\BibitemShut {NoStop}%
\bibitem [{\citenamefont {Schneider}\ \emph {et~al.}(2020)\citenamefont
  {Schneider}, \citenamefont {Brinker}, \citenamefont {Steinbrecher},
  \citenamefont {Hermenau}, \citenamefont {Posske}, \citenamefont {dos
  Santos~Dias}, \citenamefont {Lounis}, \citenamefont {Wiesendanger},\ and\
  \citenamefont {Wiebe}}]{schneider2020controlling}%
  \BibitemOpen
  \bibfield  {author} {\bibinfo {author} {\bibfnamefont {L.}~\bibnamefont
  {Schneider}}, \bibinfo {author} {\bibfnamefont {S.}~\bibnamefont {Brinker}},
  \bibinfo {author} {\bibfnamefont {M.}~\bibnamefont {Steinbrecher}}, \bibinfo
  {author} {\bibfnamefont {J.}~\bibnamefont {Hermenau}}, \bibinfo {author}
  {\bibfnamefont {T.}~\bibnamefont {Posske}}, \bibinfo {author} {\bibfnamefont
  {M.}~\bibnamefont {dos Santos~Dias}}, \bibinfo {author} {\bibfnamefont
  {S.}~\bibnamefont {Lounis}}, \bibinfo {author} {\bibfnamefont
  {R.}~\bibnamefont {Wiesendanger}},\ and\ \bibinfo {author} {\bibfnamefont
  {J.}~\bibnamefont {Wiebe}},\ }\bibfield  {title} {\bibinfo {title}
  {Controlling in-gap end states by linking nonmagnetic atoms and
  artificially-constructed spin chains on superconductors},\ }\href@noop {}
  {\bibfield  {journal} {\bibinfo  {journal} {Nat. Commun.}\ }\textbf {\bibinfo
  {volume} {11}},\ \bibinfo {pages} {4707} (\bibinfo {year}
  {2020})}\BibitemShut {NoStop}%
\bibitem [{\citenamefont {Schneider}\ \emph {et~al.}(2021)\citenamefont
  {Schneider}, \citenamefont {Beck}, \citenamefont {Posske}, \citenamefont
  {Crawford}, \citenamefont {Mascot}, \citenamefont {Rachel}, \citenamefont
  {Wiesendanger},\ and\ \citenamefont {Wiebe}}]{schneider2021topological}%
  \BibitemOpen
  \bibfield  {author} {\bibinfo {author} {\bibfnamefont {L.}~\bibnamefont
  {Schneider}}, \bibinfo {author} {\bibfnamefont {P.}~\bibnamefont {Beck}},
  \bibinfo {author} {\bibfnamefont {T.}~\bibnamefont {Posske}}, \bibinfo
  {author} {\bibfnamefont {D.}~\bibnamefont {Crawford}}, \bibinfo {author}
  {\bibfnamefont {E.}~\bibnamefont {Mascot}}, \bibinfo {author} {\bibfnamefont
  {S.}~\bibnamefont {Rachel}}, \bibinfo {author} {\bibfnamefont
  {R.}~\bibnamefont {Wiesendanger}},\ and\ \bibinfo {author} {\bibfnamefont
  {J.}~\bibnamefont {Wiebe}},\ }\bibfield  {title} {\bibinfo {title}
  {Topological shiba bands in artificial spin chains on superconductors},\
  }\href@noop {} {\bibfield  {journal} {\bibinfo  {journal} {Nat. Phys.}\
  }\textbf {\bibinfo {volume} {17}},\ \bibinfo {pages} {943} (\bibinfo {year}
  {2021})}\BibitemShut {NoStop}%
\bibitem [{\citenamefont {Sticlet}\ and\ \citenamefont
  {Morari}(2019)}]{PhysRevB.100.075420}%
  \BibitemOpen
  \bibfield  {author} {\bibinfo {author} {\bibfnamefont {D.}~\bibnamefont
  {Sticlet}}\ and\ \bibinfo {author} {\bibfnamefont {C.}~\bibnamefont
  {Morari}},\ }\bibfield  {title} {\bibinfo {title} {Topological
  superconductivity from magnetic impurities on monolayer
  ${\mathrm{nbse}}_{2}$},\ }\href@noop {} {\bibfield  {journal} {\bibinfo
  {journal} {Phys. Rev. B}\ }\textbf {\bibinfo {volume} {100}},\ \bibinfo
  {pages} {075420} (\bibinfo {year} {2019})}\BibitemShut {NoStop}%
\bibitem [{\citenamefont {Nadj-Perge}\ \emph {et~al.}(2013)\citenamefont
  {Nadj-Perge}, \citenamefont {Drozdov}, \citenamefont {Bernevig},\ and\
  \citenamefont {Yazdani}}]{PhysRevB.88.020407}%
  \BibitemOpen
  \bibfield  {author} {\bibinfo {author} {\bibfnamefont {S.}~\bibnamefont
  {Nadj-Perge}}, \bibinfo {author} {\bibfnamefont {I.~K.}\ \bibnamefont
  {Drozdov}}, \bibinfo {author} {\bibfnamefont {B.~A.}\ \bibnamefont
  {Bernevig}},\ and\ \bibinfo {author} {\bibfnamefont {A.}~\bibnamefont
  {Yazdani}},\ }\bibfield  {title} {\bibinfo {title} {Proposal for realizing
  majorana fermions in chains of magnetic atoms on a superconductor},\
  }\href@noop {} {\bibfield  {journal} {\bibinfo  {journal} {Phys. Rev. B}\
  }\textbf {\bibinfo {volume} {88}},\ \bibinfo {pages} {020407} (\bibinfo
  {year} {2013})}\BibitemShut {NoStop}%
\bibitem [{\citenamefont {Choy}\ \emph {et~al.}(2011)\citenamefont {Choy},
  \citenamefont {Edge}, \citenamefont {Akhmerov},\ and\ \citenamefont
  {Beenakker}}]{PhysRevB.84.195442}%
  \BibitemOpen
  \bibfield  {author} {\bibinfo {author} {\bibfnamefont {T.-P.}\ \bibnamefont
  {Choy}}, \bibinfo {author} {\bibfnamefont {J.~M.}\ \bibnamefont {Edge}},
  \bibinfo {author} {\bibfnamefont {A.~R.}\ \bibnamefont {Akhmerov}},\ and\
  \bibinfo {author} {\bibfnamefont {C.~W.~J.}\ \bibnamefont {Beenakker}},\
  }\bibfield  {title} {\bibinfo {title} {Majorana fermions emerging from
  magnetic nanoparticles on a superconductor without spin-orbit coupling},\
  }\href@noop {} {\bibfield  {journal} {\bibinfo  {journal} {Phys. Rev. B}\
  }\textbf {\bibinfo {volume} {84}},\ \bibinfo {pages} {195442} (\bibinfo
  {year} {2011})}\BibitemShut {NoStop}%
\bibitem [{\citenamefont {Kezilebieke}\ \emph {et~al.}(2020)\citenamefont
  {Kezilebieke}, \citenamefont {Huda}, \citenamefont {Va{\v{n}}o},
  \citenamefont {Aapro}, \citenamefont {Ganguli}, \citenamefont {Silveira},
  \citenamefont {G{\l}odzik}, \citenamefont {Foster}, \citenamefont {Ojanen},\
  and\ \citenamefont {Liljeroth}}]{kezilebieke2020topological}%
  \BibitemOpen
  \bibfield  {author} {\bibinfo {author} {\bibfnamefont {S.}~\bibnamefont
  {Kezilebieke}}, \bibinfo {author} {\bibfnamefont {M.~N.}\ \bibnamefont
  {Huda}}, \bibinfo {author} {\bibfnamefont {V.}~\bibnamefont {Va{\v{n}}o}},
  \bibinfo {author} {\bibfnamefont {M.}~\bibnamefont {Aapro}}, \bibinfo
  {author} {\bibfnamefont {S.~C.}\ \bibnamefont {Ganguli}}, \bibinfo {author}
  {\bibfnamefont {O.~J.}\ \bibnamefont {Silveira}}, \bibinfo {author}
  {\bibfnamefont {S.}~\bibnamefont {G{\l}odzik}}, \bibinfo {author}
  {\bibfnamefont {A.~S.}\ \bibnamefont {Foster}}, \bibinfo {author}
  {\bibfnamefont {T.}~\bibnamefont {Ojanen}},\ and\ \bibinfo {author}
  {\bibfnamefont {P.}~\bibnamefont {Liljeroth}},\ }\bibfield  {title} {\bibinfo
  {title} {Topological superconductivity in a van der waals heterostructure},\
  }\href@noop {} {\bibfield  {journal} {\bibinfo  {journal} {Nature}\ }\textbf
  {\bibinfo {volume} {588}},\ \bibinfo {pages} {424} (\bibinfo {year}
  {2020})}\BibitemShut {NoStop}%
\bibitem [{\citenamefont {Abrikosov}\ and\ \citenamefont
  {Gor'kov}(1961)}]{osti_4097498}%
  \BibitemOpen
  \bibfield  {author} {\bibinfo {author} {\bibfnamefont {A.~A.}\ \bibnamefont
  {Abrikosov}}\ and\ \bibinfo {author} {\bibfnamefont {L.~P.}\ \bibnamefont
  {Gor'kov}},\ }\bibfield  {title} {\bibinfo {title} {Contribution to the
  theory of superconducting alloys with paramagnetic impurities},\ }\href@noop
  {} {\bibfield  {journal} {\bibinfo  {journal} {Soviet Phys. JETP}\ }\textbf
  {\bibinfo {volume} {12}},\ \bibinfo {pages} {1243} (\bibinfo {year}
  {1961})}\BibitemShut {NoStop}%
\bibitem [{\citenamefont {Gulian}\ and\ \citenamefont
  {Zharkov}(2002)}]{gulian2002nonequilibrium}%
  \BibitemOpen
  \bibfield  {author} {\bibinfo {author} {\bibfnamefont {A.~M.}\ \bibnamefont
  {Gulian}}\ and\ \bibinfo {author} {\bibfnamefont {G.~F.}\ \bibnamefont
  {Zharkov}},\ }\href@noop {} {\emph {\bibinfo {title} {Nonequilibrium
  Electrons and Phonons in Superconductors: Selected Topics in
  Superconductivity}}}\ (\bibinfo  {publisher} {Springer},\ \bibinfo {year}
  {2002})\BibitemShut {NoStop}%
\bibitem [{\citenamefont {Jyoti}\ \emph {et~al.}(2024)\citenamefont {Jyoti},
  \citenamefont {Choi},\ and\ \citenamefont {Lorente}}]{PhysRevB.110.205404}%
  \BibitemOpen
  \bibfield  {author} {\bibinfo {author} {\bibfnamefont {D.}~\bibnamefont
  {Jyoti}}, \bibinfo {author} {\bibfnamefont {D.-J.}\ \bibnamefont {Choi}},\
  and\ \bibinfo {author} {\bibfnamefont {N.}~\bibnamefont {Lorente}},\
  }\bibfield  {title} {\bibinfo {title} {In-gap states induced by magnetic
  impurities on wide-band $s$-wave superconductors: Self-consistent
  calculations},\ }\href@noop {} {\bibfield  {journal} {\bibinfo  {journal}
  {Phys. Rev. B}\ }\textbf {\bibinfo {volume} {110}},\ \bibinfo {pages}
  {205404} (\bibinfo {year} {2024})}\BibitemShut {NoStop}%
\bibitem [{\citenamefont {Flatt\'e}\ and\ \citenamefont
  {Byers}(1997{\natexlab{a}})}]{PhysRevLett.78.3761}%
  \BibitemOpen
  \bibfield  {author} {\bibinfo {author} {\bibfnamefont {M.~E.}\ \bibnamefont
  {Flatt\'e}}\ and\ \bibinfo {author} {\bibfnamefont {J.~M.}\ \bibnamefont
  {Byers}},\ }\bibfield  {title} {\bibinfo {title} {Local electronic structure
  of a single magnetic impurity in a superconductor},\ }\href@noop {}
  {\bibfield  {journal} {\bibinfo  {journal} {Phys. Rev. Lett.}\ }\textbf
  {\bibinfo {volume} {78}},\ \bibinfo {pages} {3761} (\bibinfo {year}
  {1997}{\natexlab{a}})}\BibitemShut {NoStop}%
\bibitem [{\citenamefont {Kattel}\ \emph {et~al.}(2025)\citenamefont {Kattel},
  \citenamefont {Zhakenov},\ and\ \citenamefont {Andrei}}]{rg1x-bztv}%
  \BibitemOpen
  \bibfield  {author} {\bibinfo {author} {\bibfnamefont {P.}~\bibnamefont
  {Kattel}}, \bibinfo {author} {\bibfnamefont {A.}~\bibnamefont {Zhakenov}},\
  and\ \bibinfo {author} {\bibfnamefont {N.}~\bibnamefont {Andrei}},\
  }\bibfield  {title} {\bibinfo {title} {Kondo overscreening in the presence of
  superconductivity},\ }\href@noop {} {\bibfield  {journal} {\bibinfo
  {journal} {Phys. Rev. B}\ }\textbf {\bibinfo {volume} {112}},\ \bibinfo
  {pages} {085103} (\bibinfo {year} {2025})}\BibitemShut {NoStop}%
\bibitem [{\citenamefont {Moca}\ \emph {et~al.}(2021)\citenamefont {Moca},
  \citenamefont {Weymann}, \citenamefont {Werner},\ and\ \citenamefont
  {Zar\'and}}]{PhysRevLett.127.186804}%
  \BibitemOpen
  \bibfield  {author} {\bibinfo {author} {\bibfnamefont {C.~P.}\ \bibnamefont
  {Moca}}, \bibinfo {author} {\bibfnamefont {I.}~\bibnamefont {Weymann}},
  \bibinfo {author} {\bibfnamefont {M.~A.}\ \bibnamefont {Werner}},\ and\
  \bibinfo {author} {\bibfnamefont {G.}~\bibnamefont {Zar\'and}},\ }\bibfield
  {title} {\bibinfo {title} {Kondo cloud in a superconductor},\ }\href@noop {}
  {\bibfield  {journal} {\bibinfo  {journal} {Phys. Rev. Lett.}\ }\textbf
  {\bibinfo {volume} {127}},\ \bibinfo {pages} {186804} (\bibinfo {year}
  {2021})}\BibitemShut {NoStop}%
\bibitem [{\citenamefont {Pasnoori}\ \emph {et~al.}(2022)\citenamefont
  {Pasnoori}, \citenamefont {Andrei}, \citenamefont {Rylands},\ and\
  \citenamefont {Azaria}}]{PhysRevB.105.174517}%
  \BibitemOpen
  \bibfield  {author} {\bibinfo {author} {\bibfnamefont {P.~R.}\ \bibnamefont
  {Pasnoori}}, \bibinfo {author} {\bibfnamefont {N.}~\bibnamefont {Andrei}},
  \bibinfo {author} {\bibfnamefont {C.}~\bibnamefont {Rylands}},\ and\ \bibinfo
  {author} {\bibfnamefont {P.}~\bibnamefont {Azaria}},\ }\bibfield  {title}
  {\bibinfo {title} {Rise and fall of yu-shiba-rusinov bound states in
  charge-conserving $s$-wave one-dimensional superconductors},\ }\href@noop {}
  {\bibfield  {journal} {\bibinfo  {journal} {Phys. Rev. B}\ }\textbf {\bibinfo
  {volume} {105}},\ \bibinfo {pages} {174517} (\bibinfo {year}
  {2022})}\BibitemShut {NoStop}%
\bibitem [{\citenamefont {Flatt\'e}\ and\ \citenamefont
  {Byers}(1997{\natexlab{b}})}]{PhysRevB.56.11213}%
  \BibitemOpen
  \bibfield  {author} {\bibinfo {author} {\bibfnamefont {M.~E.}\ \bibnamefont
  {Flatt\'e}}\ and\ \bibinfo {author} {\bibfnamefont {J.~M.}\ \bibnamefont
  {Byers}},\ }\bibfield  {title} {\bibinfo {title} {Local electronic structure
  of defects in superconductors},\ }\href@noop {} {\bibfield  {journal}
  {\bibinfo  {journal} {Phys. Rev. B}\ }\textbf {\bibinfo {volume} {56}},\
  \bibinfo {pages} {11213} (\bibinfo {year} {1997}{\natexlab{b}})}\BibitemShut
  {NoStop}%
\bibitem [{\citenamefont {Liu}\ \emph {et~al.}(2019)\citenamefont {Liu},
  \citenamefont {Huang}, \citenamefont {Chen},\ and\ \citenamefont
  {Ting}}]{PhysRevB.99.174502}%
  \BibitemOpen
  \bibfield  {author} {\bibinfo {author} {\bibfnamefont {C.}~\bibnamefont
  {Liu}}, \bibinfo {author} {\bibfnamefont {Y.}~\bibnamefont {Huang}}, \bibinfo
  {author} {\bibfnamefont {Y.}~\bibnamefont {Chen}},\ and\ \bibinfo {author}
  {\bibfnamefont {C.~S.}\ \bibnamefont {Ting}},\ }\bibfield  {title} {\bibinfo
  {title} {Temperature-dependent spectral function of a kondo impurity in an
  $s$-wave superconductor},\ }\href@noop {} {\bibfield  {journal} {\bibinfo
  {journal} {Phys. Rev. B}\ }\textbf {\bibinfo {volume} {99}},\ \bibinfo
  {pages} {174502} (\bibinfo {year} {2019})}\BibitemShut {NoStop}%
\bibitem [{\citenamefont {M\"uller-Hartmann}\ and\ \citenamefont
  {Zittartz}(1971)}]{PhysRevLett.26.428}%
  \BibitemOpen
  \bibfield  {author} {\bibinfo {author} {\bibfnamefont {E.}~\bibnamefont
  {M\"uller-Hartmann}}\ and\ \bibinfo {author} {\bibfnamefont {J.}~\bibnamefont
  {Zittartz}},\ }\bibfield  {title} {\bibinfo {title} {Kondo effect in
  superconductors},\ }\href@noop {} {\bibfield  {journal} {\bibinfo  {journal}
  {Phys. Rev. Lett.}\ }\textbf {\bibinfo {volume} {26}},\ \bibinfo {pages}
  {428} (\bibinfo {year} {1971})}\BibitemShut {NoStop}%
\bibitem [{\citenamefont {Schuh}\ and\ \citenamefont
  {M{\"u}ller-Hartmann}(1978)}]{Schuh1978}%
  \BibitemOpen
  \bibfield  {author} {\bibinfo {author} {\bibfnamefont {B.}~\bibnamefont
  {Schuh}}\ and\ \bibinfo {author} {\bibfnamefont {E.}~\bibnamefont
  {M{\"u}ller-Hartmann}},\ }\bibfield  {title} {\bibinfo {title}
  {Self-consistent theory of pairbreaking in kondo superconductors},\
  }\href@noop {} {\bibfield  {journal} {\bibinfo  {journal} {Zeitschrift
  f{\"u}r Physik B Condensed Matter}\ }\textbf {\bibinfo {volume} {29}},\
  \bibinfo {pages} {39} (\bibinfo {year} {1978})}\BibitemShut {NoStop}%
\bibitem [{\citenamefont {Jarrell}(1990)}]{PhysRevB.41.4815}%
  \BibitemOpen
  \bibfield  {author} {\bibinfo {author} {\bibfnamefont {M.}~\bibnamefont
  {Jarrell}},\ }\bibfield  {title} {\bibinfo {title} {Universal reduction of
  $t_c$ in strong-coupling superconductors by a small concentration of magnetic
  impurities},\ }\href@noop {} {\bibfield  {journal} {\bibinfo  {journal}
  {Phys. Rev. B}\ }\textbf {\bibinfo {volume} {41}},\ \bibinfo {pages} {4815}
  (\bibinfo {year} {1990})}\BibitemShut {NoStop}%
\bibitem [{\citenamefont {Müller-Hartmann}\ and\ \citenamefont
  {Zittartz}(1972)}]{MULLERHARTMANN1972401}%
  \BibitemOpen
  \bibfield  {author} {\bibinfo {author} {\bibfnamefont {E.}~\bibnamefont
  {Müller-Hartmann}}\ and\ \bibinfo {author} {\bibfnamefont {J.}~\bibnamefont
  {Zittartz}},\ }\bibfield  {title} {\bibinfo {title} {Specific heat of
  superconductors with magnetic impurities},\ }\href@noop {} {\bibfield
  {journal} {\bibinfo  {journal} {Solid State Communications}\ }\textbf
  {\bibinfo {volume} {11}},\ \bibinfo {pages} {401} (\bibinfo {year}
  {1972})}\BibitemShut {NoStop}%
\bibitem [{\citenamefont {Müller-Hartmann}\ \emph {et~al.}(1976)\citenamefont
  {Müller-Hartmann}, \citenamefont {Schuh},\ and\ \citenamefont
  {Zittartz}}]{MULLERHARTMANN1976439}%
  \BibitemOpen
  \bibfield  {author} {\bibinfo {author} {\bibfnamefont {E.}~\bibnamefont
  {Müller-Hartmann}}, \bibinfo {author} {\bibfnamefont {B.}~\bibnamefont
  {Schuh}},\ and\ \bibinfo {author} {\bibfnamefont {J.}~\bibnamefont
  {Zittartz}},\ }\bibfield  {title} {\bibinfo {title} {Pair-breaking in kondo
  superconductors},\ }\href@noop {} {\bibfield  {journal} {\bibinfo  {journal}
  {Solid State Communications}\ }\textbf {\bibinfo {volume} {19}},\ \bibinfo
  {pages} {439} (\bibinfo {year} {1976})}\BibitemShut {NoStop}%
\bibitem [{\citenamefont {Matsuura}\ and\ \citenamefont
  {Nagaoka}(1976)}]{MATSUURA19761583}%
  \BibitemOpen
  \bibfield  {author} {\bibinfo {author} {\bibfnamefont {T.}~\bibnamefont
  {Matsuura}}\ and\ \bibinfo {author} {\bibfnamefont {Y.}~\bibnamefont
  {Nagaoka}},\ }\bibfield  {title} {\bibinfo {title} {Depression of
  superconducting critical temperature by impurities with kondo effect},\
  }\href@noop {} {\bibfield  {journal} {\bibinfo  {journal} {Solid State
  Communications}\ }\textbf {\bibinfo {volume} {18}},\ \bibinfo {pages} {1583}
  (\bibinfo {year} {1976})}\BibitemShut {NoStop}%
\bibitem [{\citenamefont {Matsuura}(1977)}]{10.1143/PTP.57.1823}%
  \BibitemOpen
  \bibfield  {author} {\bibinfo {author} {\bibfnamefont {T.}~\bibnamefont
  {Matsuura}},\ }\bibfield  {title} {\bibinfo {title} {The effects of
  impurities on superconductors with kondo effect},\ }\href@noop {} {\bibfield
  {journal} {\bibinfo  {journal} {Progress of Theoretical Physics}\ }\textbf
  {\bibinfo {volume} {57}},\ \bibinfo {pages} {1823} (\bibinfo {year}
  {1977})}\BibitemShut {NoStop}%
\bibitem [{\citenamefont {Yang}\ and\ \citenamefont
  {Wu}(2024{\natexlab{a}})}]{PhysRevB.109.064508}%
  \BibitemOpen
  \bibfield  {author} {\bibinfo {author} {\bibfnamefont {F.}~\bibnamefont
  {Yang}}\ and\ \bibinfo {author} {\bibfnamefont {M.~W.}\ \bibnamefont {Wu}},\
  }\bibfield  {title} {\bibinfo {title} {Diamagnetic property and optical
  absorption of conventional superconductors with magnetic impurities in linear
  response},\ }\href@noop {} {\bibfield  {journal} {\bibinfo  {journal} {Phys.
  Rev. B}\ }\textbf {\bibinfo {volume} {109}},\ \bibinfo {pages} {064508}
  (\bibinfo {year} {2024}{\natexlab{a}})}\BibitemShut {NoStop}%
\bibitem [{\citenamefont {Abrikosov}\ \emph {et~al.}(1963)\citenamefont
  {Abrikosov}, \citenamefont {Gorkov},\ and\ \citenamefont
  {Dzyaloshinski}}]{abrikosov2012methods}%
  \BibitemOpen
  \bibfield  {author} {\bibinfo {author} {\bibfnamefont {A.~A.}\ \bibnamefont
  {Abrikosov}}, \bibinfo {author} {\bibfnamefont {L.~P.}\ \bibnamefont
  {Gorkov}},\ and\ \bibinfo {author} {\bibfnamefont {I.~E.}\ \bibnamefont
  {Dzyaloshinski}},\ }\href@noop {} {\emph {\bibinfo {title} {Methods of
  quantum field theory in statistical physics}}}\ (\bibinfo  {publisher}
  {Prentice Hall, Englewood Cliffs},\ \bibinfo {year} {1963})\BibitemShut
  {NoStop}%
\bibitem [{\citenamefont {Silaev}(2019)}]{PhysRevB.99.224511}%
  \BibitemOpen
  \bibfield  {author} {\bibinfo {author} {\bibfnamefont {M.}~\bibnamefont
  {Silaev}},\ }\bibfield  {title} {\bibinfo {title} {Nonlinear electromagnetic
  response and higgs-mode excitation in bcs superconductors with impurities},\
  }\href@noop {} {\bibfield  {journal} {\bibinfo  {journal} {Phys. Rev. B}\
  }\textbf {\bibinfo {volume} {99}},\ \bibinfo {pages} {224511} (\bibinfo
  {year} {2019})}\BibitemShut {NoStop}%
\bibitem [{\citenamefont {Yang}\ and\ \citenamefont
  {Wu}(2024{\natexlab{b}})}]{yang2024optical}%
  \BibitemOpen
  \bibfield  {author} {\bibinfo {author} {\bibfnamefont {F.}~\bibnamefont
  {Yang}}\ and\ \bibinfo {author} {\bibfnamefont {M.}~\bibnamefont {Wu}},\
  }\bibfield  {title} {\bibinfo {title} {Optical response of higgs mode in
  superconductors at clean limit: formulation through eilenberger equation and
  ginzburg--landau lagrangian},\ }\href@noop {} {\bibfield  {journal} {\bibinfo
   {journal} {J. Phys.: Condens. Matter.}\ }\textbf {\bibinfo {volume} {36}},\
  \bibinfo {pages} {425701} (\bibinfo {year} {2024}{\natexlab{b}})}\BibitemShut
  {NoStop}%
\bibitem [{\citenamefont {Hugdal}\ \emph {et~al.}(2017)\citenamefont {Hugdal},
  \citenamefont {Linder},\ and\ \citenamefont {Jacobsen}}]{PhysRevB.95.235403}%
  \BibitemOpen
  \bibfield  {author} {\bibinfo {author} {\bibfnamefont {H.~G.}\ \bibnamefont
  {Hugdal}}, \bibinfo {author} {\bibfnamefont {J.}~\bibnamefont {Linder}},\
  and\ \bibinfo {author} {\bibfnamefont {S.~H.}\ \bibnamefont {Jacobsen}},\
  }\bibfield  {title} {\bibinfo {title} {Quasiclassical theory for the
  superconducting proximity effect in dirac materials},\ }\href@noop {}
  {\bibfield  {journal} {\bibinfo  {journal} {Phys. Rev. B}\ }\textbf {\bibinfo
  {volume} {95}},\ \bibinfo {pages} {235403} (\bibinfo {year}
  {2017})}\BibitemShut {NoStop}%
\bibitem [{\citenamefont {Buzdin}(2005)}]{RevModPhys.77.935}%
  \BibitemOpen
  \bibfield  {author} {\bibinfo {author} {\bibfnamefont {A.~I.}\ \bibnamefont
  {Buzdin}},\ }\bibfield  {title} {\bibinfo {title} {Proximity effects in
  superconductor-ferromagnet heterostructures},\ }\href@noop {} {\bibfield
  {journal} {\bibinfo  {journal} {Rev. Mod. Phys.}\ }\textbf {\bibinfo {volume}
  {77}},\ \bibinfo {pages} {935} (\bibinfo {year} {2005})}\BibitemShut
  {NoStop}%
\bibitem [{\citenamefont {Usadel}(1970)}]{PhysRevLett.25.507}%
  \BibitemOpen
  \bibfield  {author} {\bibinfo {author} {\bibfnamefont {K.~D.}\ \bibnamefont
  {Usadel}},\ }\bibfield  {title} {\bibinfo {title} {Generalized diffusion
  equation for superconducting alloys},\ }\href@noop {} {\bibfield  {journal}
  {\bibinfo  {journal} {Phys. Rev. Lett.}\ }\textbf {\bibinfo {volume} {25}},\
  \bibinfo {pages} {507} (\bibinfo {year} {1970})}\BibitemShut {NoStop}%
\bibitem [{\citenamefont {Espedal}\ \emph {et~al.}(2016)\citenamefont
  {Espedal}, \citenamefont {Yokoyama},\ and\ \citenamefont
  {Linder}}]{PhysRevLett.116.127002}%
  \BibitemOpen
  \bibfield  {author} {\bibinfo {author} {\bibfnamefont {C.}~\bibnamefont
  {Espedal}}, \bibinfo {author} {\bibfnamefont {T.}~\bibnamefont {Yokoyama}},\
  and\ \bibinfo {author} {\bibfnamefont {J.}~\bibnamefont {Linder}},\
  }\bibfield  {title} {\bibinfo {title} {Anisotropic paramagnetic meissner
  effect by spin-orbit coupling},\ }\href@noop {} {\bibfield  {journal}
  {\bibinfo  {journal} {Phys. Rev. Lett.}\ }\textbf {\bibinfo {volume} {116}},\
  \bibinfo {pages} {127002} (\bibinfo {year} {2016})}\BibitemShut {NoStop}%
\bibitem [{\citenamefont {Suhl}\ and\ \citenamefont
  {Matthias}(1959)}]{suhl1959impurity}%
  \BibitemOpen
  \bibfield  {author} {\bibinfo {author} {\bibfnamefont {H.}~\bibnamefont
  {Suhl}}\ and\ \bibinfo {author} {\bibfnamefont {B.}~\bibnamefont
  {Matthias}},\ }\bibfield  {title} {\bibinfo {title} {Impurity scattering in
  superconductors},\ }\href@noop {} {\bibfield  {journal} {\bibinfo  {journal}
  {Phys. Rev.}\ }\textbf {\bibinfo {volume} {114}},\ \bibinfo {pages} {977}
  (\bibinfo {year} {1959})}\BibitemShut {NoStop}%
\bibitem [{\citenamefont {Skalski}\ \emph {et~al.}(1964)\citenamefont
  {Skalski}, \citenamefont {Betbeder-Matibet},\ and\ \citenamefont
  {Weiss}}]{skalski1964properties}%
  \BibitemOpen
  \bibfield  {author} {\bibinfo {author} {\bibfnamefont {S.}~\bibnamefont
  {Skalski}}, \bibinfo {author} {\bibfnamefont {O.}~\bibnamefont
  {Betbeder-Matibet}},\ and\ \bibinfo {author} {\bibfnamefont {P.}~\bibnamefont
  {Weiss}},\ }\bibfield  {title} {\bibinfo {title} {Properties of
  superconducting alloys containing paramagnetic impurities},\ }\href@noop {}
  {\bibfield  {journal} {\bibinfo  {journal} {Phys. Rev.}\ }\textbf {\bibinfo
  {volume} {136}},\ \bibinfo {pages} {A1500} (\bibinfo {year}
  {1964})}\BibitemShut {NoStop}%
\bibitem [{\citenamefont {Andersen}\ \emph {et~al.}(2020)\citenamefont
  {Andersen}, \citenamefont {Ramires}, \citenamefont {Wang}, \citenamefont
  {Lorenz},\ and\ \citenamefont {Ando}}]{andersen2020generalized}%
  \BibitemOpen
  \bibfield  {author} {\bibinfo {author} {\bibfnamefont {L.}~\bibnamefont
  {Andersen}}, \bibinfo {author} {\bibfnamefont {A.}~\bibnamefont {Ramires}},
  \bibinfo {author} {\bibfnamefont {Z.}~\bibnamefont {Wang}}, \bibinfo {author}
  {\bibfnamefont {T.}~\bibnamefont {Lorenz}},\ and\ \bibinfo {author}
  {\bibfnamefont {Y.}~\bibnamefont {Ando}},\ }\bibfield  {title} {\bibinfo
  {title} {Generalized {Anderson's} theorem for superconductors derived from
  topological insulators},\ }\href@noop {} {\bibfield  {journal} {\bibinfo
  {journal} {Sci. Adv.}\ }\textbf {\bibinfo {volume} {6}},\ \bibinfo {pages}
  {eaay6502} (\bibinfo {year} {2020})}\BibitemShut {NoStop}%
\bibitem [{\citenamefont {Anderson}(1959)}]{anderson1959theory}%
  \BibitemOpen
  \bibfield  {author} {\bibinfo {author} {\bibfnamefont {P.~W.}\ \bibnamefont
  {Anderson}},\ }\bibfield  {title} {\bibinfo {title} {Theory of dirty
  superconductors},\ }\href@noop {} {\bibfield  {journal} {\bibinfo  {journal}
  {J. Phys. Chem. Solids}\ }\textbf {\bibinfo {volume} {11}},\ \bibinfo {pages}
  {26} (\bibinfo {year} {1959})}\BibitemShut {NoStop}%
\bibitem [{\citenamefont {Skowron}\ and\ \citenamefont
  {Gould}(2012)}]{skowron2012generalcomplexpolynomialroot}%
  \BibitemOpen
  \bibfield  {author} {\bibinfo {author} {\bibfnamefont {J.}~\bibnamefont
  {Skowron}}\ and\ \bibinfo {author} {\bibfnamefont {A.}~\bibnamefont
  {Gould}},\ }\href@noop {} {\bibinfo {title} {General complex polynomial root
  solver and its further optimization for binary microlenses}} (\bibinfo {year}
  {2012}),\ \Eprint {https://arxiv.org/abs/1203.1034} {arXiv:1203.1034}
  \BibitemShut {NoStop}%
\bibitem [{\citenamefont {Rammer}\ and\ \citenamefont
  {Smith}(1986)}]{RevModPhys.58.323}%
  \BibitemOpen
  \bibfield  {author} {\bibinfo {author} {\bibfnamefont {J.}~\bibnamefont
  {Rammer}}\ and\ \bibinfo {author} {\bibfnamefont {H.}~\bibnamefont {Smith}},\
  }\bibfield  {title} {\bibinfo {title} {Quantum field-theoretical methods in
  transport theory of metals},\ }\href@noop {} {\bibfield  {journal} {\bibinfo
  {journal} {Rev. Mod. Phys.}\ }\textbf {\bibinfo {volume} {58}},\ \bibinfo
  {pages} {323} (\bibinfo {year} {1986})}\BibitemShut {NoStop}%
\bibitem [{dis()}]{discussion}%
  \BibitemOpen
  \href@noop {} {}\bibinfo {note} {{We note that the $T_c(n_i)$ predicted by
  M\"uller–Hartmann and Zittartz~\cite{PhysRevLett.26.428} is, both
  qualitatively and quantitatively, inconsistent with the classical
  Abrikosov–Gor'kov critical theory~\cite{osti_4097498} as well as with
  Shiba's self-consistent framework~\cite{shiba1968classical}. Moreover, their
  result was subsequently challenged on theoretical grounds by M.
  Jarrell~\cite{PhysRevB.41.4815} and by T. Matsuura and Y.
  Nagaoka~\cite{MATSUURA19761583,10.1143/PTP.57.1823}. To date, no reliable
  experimental evidence has been reported for the low-temperature tail with
  positive curvature or the associated third transition temperature predicted
  by M\"uller–Hartmann and Zittartz. In fact, Jarrell's analysis indicated
  that these features do not persist in more accurate microscopic models. This
  further justifies our emphasis on a rigorous self-consistent treatment rather
  than relying on such speculative theoretical constructs}}\BibitemShut
  {NoStop}%
\bibitem [{\citenamefont {Mattis}\ and\ \citenamefont
  {Bardeen}(1958)}]{PhysRev.111.412}%
  \BibitemOpen
  \bibfield  {author} {\bibinfo {author} {\bibfnamefont {D.~C.}\ \bibnamefont
  {Mattis}}\ and\ \bibinfo {author} {\bibfnamefont {J.}~\bibnamefont
  {Bardeen}},\ }\bibfield  {title} {\bibinfo {title} {Theory of the anomalous
  skin effect in normal and superconducting metals},\ }\href@noop {} {\bibfield
   {journal} {\bibinfo  {journal} {Phys. Rev.}\ }\textbf {\bibinfo {volume}
  {111}},\ \bibinfo {pages} {412} (\bibinfo {year} {1958})}\BibitemShut
  {NoStop}%
\bibitem [{\citenamefont {Nam}(1967)}]{PhysRev.156.470}%
  \BibitemOpen
  \bibfield  {author} {\bibinfo {author} {\bibfnamefont {S.~B.}\ \bibnamefont
  {Nam}},\ }\bibfield  {title} {\bibinfo {title} {Theory of electromagnetic
  properties of superconducting and normal systems. i},\ }\href@noop {}
  {\bibfield  {journal} {\bibinfo  {journal} {Phys. Rev.}\ }\textbf {\bibinfo
  {volume} {156}},\ \bibinfo {pages} {470} (\bibinfo {year}
  {1967})}\BibitemShut {NoStop}%
\bibitem [{\citenamefont {Yang}\ and\ \citenamefont
  {Wu}(2020{\natexlab{a}})}]{PhysRevB.102.144508}%
  \BibitemOpen
  \bibfield  {author} {\bibinfo {author} {\bibfnamefont {F.}~\bibnamefont
  {Yang}}\ and\ \bibinfo {author} {\bibfnamefont {M.~W.}\ \bibnamefont {Wu}},\
  }\bibfield  {title} {\bibinfo {title} {Influence of scattering on the optical
  response of superconductors},\ }\href@noop {} {\bibfield  {journal} {\bibinfo
   {journal} {Phys. Rev. B}\ }\textbf {\bibinfo {volume} {102}},\ \bibinfo
  {pages} {144508} (\bibinfo {year} {2020}{\natexlab{a}})}\BibitemShut
  {NoStop}%
\bibitem [{\citenamefont {Fominov}\ \emph {et~al.}(2011)\citenamefont
  {Fominov}, \citenamefont {Houzet},\ and\ \citenamefont
  {Glazman}}]{PhysRevB.84.224517}%
  \BibitemOpen
  \bibfield  {author} {\bibinfo {author} {\bibfnamefont {Y.~V.}\ \bibnamefont
  {Fominov}}, \bibinfo {author} {\bibfnamefont {M.}~\bibnamefont {Houzet}},\
  and\ \bibinfo {author} {\bibfnamefont {L.~I.}\ \bibnamefont {Glazman}},\
  }\bibfield  {title} {\bibinfo {title} {Surface impedance of superconductors
  with weak magnetic impurities},\ }\href@noop {} {\bibfield  {journal}
  {\bibinfo  {journal} {Phys. Rev. B}\ }\textbf {\bibinfo {volume} {84}},\
  \bibinfo {pages} {224517} (\bibinfo {year} {2011})}\BibitemShut {NoStop}%
\bibitem [{\citenamefont {Wang}\ \emph {et~al.}(2025)\citenamefont {Wang},
  \citenamefont {Boyack},\ and\ \citenamefont {Levin}}]{PhysRevB.111.144512}%
  \BibitemOpen
  \bibfield  {author} {\bibinfo {author} {\bibfnamefont {K.}~\bibnamefont
  {Wang}}, \bibinfo {author} {\bibfnamefont {R.}~\bibnamefont {Boyack}},\ and\
  \bibinfo {author} {\bibfnamefont {K.}~\bibnamefont {Levin}},\ }\bibfield
  {title} {\bibinfo {title} {Higgs amplitude mode in optical conductivity in
  the presence of a supercurrent: Gauge-invariant formulation with disorder},\
  }\href@noop {} {\bibfield  {journal} {\bibinfo  {journal} {Phys. Rev. B}\
  }\textbf {\bibinfo {volume} {111}},\ \bibinfo {pages} {144512} (\bibinfo
  {year} {2025})}\BibitemShut {NoStop}%
\bibitem [{\citenamefont {Li}\ and\ \citenamefont
  {Dzero}(2024)}]{PhysRevB.109.054520}%
  \BibitemOpen
  \bibfield  {author} {\bibinfo {author} {\bibfnamefont {Y.}~\bibnamefont
  {Li}}\ and\ \bibinfo {author} {\bibfnamefont {M.}~\bibnamefont {Dzero}},\
  }\bibfield  {title} {\bibinfo {title} {Amplitude higgs mode in
  superconductors with magnetic impurities},\ }\href@noop {} {\bibfield
  {journal} {\bibinfo  {journal} {Phys. Rev. B}\ }\textbf {\bibinfo {volume}
  {109}},\ \bibinfo {pages} {054520} (\bibinfo {year} {2024})}\BibitemShut
  {NoStop}%
\bibitem [{\citenamefont {Yang}\ and\ \citenamefont
  {Wu}(2022)}]{PhysRevB.106.144509}%
  \BibitemOpen
  \bibfield  {author} {\bibinfo {author} {\bibfnamefont {F.}~\bibnamefont
  {Yang}}\ and\ \bibinfo {author} {\bibfnamefont {M.~W.}\ \bibnamefont {Wu}},\
  }\bibfield  {title} {\bibinfo {title} {Impurity scattering in superconductors
  revisited: Diagrammatic formulation of the supercurrent-supercurrent
  correlation and higgs-mode damping},\ }\href@noop {} {\bibfield  {journal}
  {\bibinfo  {journal} {Phys. Rev. B}\ }\textbf {\bibinfo {volume} {106}},\
  \bibinfo {pages} {144509} (\bibinfo {year} {2022})}\BibitemShut {NoStop}%
\bibitem [{\citenamefont {Dzero}(2024)}]{PhysRevB.109.L100503}%
  \BibitemOpen
  \bibfield  {author} {\bibinfo {author} {\bibfnamefont {M.}~\bibnamefont
  {Dzero}},\ }\bibfield  {title} {\bibinfo {title} {Collisionless dynamics of
  the pairing amplitude in disordered superconductors},\ }\href@noop {}
  {\bibfield  {journal} {\bibinfo  {journal} {Phys. Rev. B}\ }\textbf {\bibinfo
  {volume} {109}},\ \bibinfo {pages} {L100503} (\bibinfo {year}
  {2024})}\BibitemShut {NoStop}%
\bibitem [{\citenamefont {Li}\ and\ \citenamefont {Dzero}(2025)}]{Li_2025}%
  \BibitemOpen
  \bibfield  {author} {\bibinfo {author} {\bibfnamefont {Y.}~\bibnamefont
  {Li}}\ and\ \bibinfo {author} {\bibfnamefont {M.}~\bibnamefont {Dzero}},\
  }\bibfield  {title} {\bibinfo {title} {Collective modes in terahertz field
  response of disordered superconductors},\ }\href@noop {} {\bibfield
  {journal} {\bibinfo  {journal} {J. Phys.: Conden. Matter}\ }\textbf {\bibinfo
  {volume} {37}},\ \bibinfo {pages} {115602} (\bibinfo {year}
  {2025})}\BibitemShut {NoStop}%
\bibitem [{\citenamefont {Yang}\ and\ \citenamefont
  {Wu}(2018)}]{yang2018gauge}%
  \BibitemOpen
  \bibfield  {author} {\bibinfo {author} {\bibfnamefont {F.}~\bibnamefont
  {Yang}}\ and\ \bibinfo {author} {\bibfnamefont {M.}~\bibnamefont {Wu}},\
  }\bibfield  {title} {\bibinfo {title} {Gauge-invariant microscopic kinetic
  theory of superconductivity in response to electromagnetic fields},\
  }\href@noop {} {\bibfield  {journal} {\bibinfo  {journal} {Phys. Rev. B}\
  }\textbf {\bibinfo {volume} {98}},\ \bibinfo {pages} {094507} (\bibinfo
  {year} {2018})}\BibitemShut {NoStop}%
\bibitem [{\citenamefont {Yang}\ \emph {et~al.}(2025)\citenamefont {Yang},
  \citenamefont {Li}, \citenamefont {Talbayev},\ and\ \citenamefont
  {Chen}}]{74d5-4hsw}%
  \BibitemOpen
  \bibfield  {author} {\bibinfo {author} {\bibfnamefont {F.}~\bibnamefont
  {Yang}}, \bibinfo {author} {\bibfnamefont {X.~J.}\ \bibnamefont {Li}},
  \bibinfo {author} {\bibfnamefont {D.}~\bibnamefont {Talbayev}},\ and\
  \bibinfo {author} {\bibfnamefont {L.~Q.}\ \bibnamefont {Chen}},\ }\bibfield
  {title} {\bibinfo {title} {Terahertz-induced second-harmonic generation in
  quantum paraelectrics: Hot-phonon effect},\ }\href@noop {} {\bibfield
  {journal} {\bibinfo  {journal} {Phys. Rev. Lett.}\ }\textbf {\bibinfo
  {volume} {135}},\ \bibinfo {pages} {056901} (\bibinfo {year}
  {2025})}\BibitemShut {NoStop}%
\bibitem [{\citenamefont {Yanagisawa}(2018)}]{yanagisawa2018theory}%
  \BibitemOpen
  \bibfield  {author} {\bibinfo {author} {\bibfnamefont {T.}~\bibnamefont
  {Yanagisawa}},\ }\bibfield  {title} {\bibinfo {title} {Theory of spontaneous
  symmetry breaking and an application to superconductivity: Nambu-goldstone
  and higgs excitation modes},\ }\href@noop {} {\bibfield  {journal} {\bibinfo
  {journal} {Commun. Comput. Phys}\ }\textbf {\bibinfo {volume} {23}},\
  \bibinfo {pages} {459} (\bibinfo {year} {2018})}\BibitemShut {NoStop}%
\bibitem [{\citenamefont {Lee}\ \emph {et~al.}(1974)\citenamefont {Lee},
  \citenamefont {Rice},\ and\ \citenamefont {Anderson}}]{lee1974conductivity}%
  \BibitemOpen
  \bibfield  {author} {\bibinfo {author} {\bibfnamefont {P.}~\bibnamefont
  {Lee}}, \bibinfo {author} {\bibfnamefont {T.}~\bibnamefont {Rice}},\ and\
  \bibinfo {author} {\bibfnamefont {P.}~\bibnamefont {Anderson}},\ }\bibfield
  {title} {\bibinfo {title} {Conductivity from charge or spin density waves},\
  }\href@noop {} {\bibfield  {journal} {\bibinfo  {journal} {Solid State
  Commun.}\ }\textbf {\bibinfo {volume} {14}},\ \bibinfo {pages} {703}
  (\bibinfo {year} {1974})}\BibitemShut {NoStop}%
\bibitem [{\citenamefont {Sun}\ \emph {et~al.}(2020)\citenamefont {Sun},
  \citenamefont {Fogler}, \citenamefont {Basov},\ and\ \citenamefont
  {Millis}}]{PhysRevResearch.2.023413}%
  \BibitemOpen
  \bibfield  {author} {\bibinfo {author} {\bibfnamefont {Z.}~\bibnamefont
  {Sun}}, \bibinfo {author} {\bibfnamefont {M.~M.}\ \bibnamefont {Fogler}},
  \bibinfo {author} {\bibfnamefont {D.~N.}\ \bibnamefont {Basov}},\ and\
  \bibinfo {author} {\bibfnamefont {A.~J.}\ \bibnamefont {Millis}},\ }\bibfield
   {title} {\bibinfo {title} {Collective modes and terahertz near-field
  response of superconductors},\ }\href@noop {} {\bibfield  {journal} {\bibinfo
   {journal} {Phys. Rev. Res.}\ }\textbf {\bibinfo {volume} {2}},\ \bibinfo
  {pages} {023413} (\bibinfo {year} {2020})}\BibitemShut {NoStop}%
\bibitem [{\citenamefont {Dzero}\ and\ \citenamefont
  {Kamenev}(2025)}]{PhysRevB.111.174502}%
  \BibitemOpen
  \bibfield  {author} {\bibinfo {author} {\bibfnamefont {M.}~\bibnamefont
  {Dzero}}\ and\ \bibinfo {author} {\bibfnamefont {A.}~\bibnamefont
  {Kamenev}},\ }\bibfield  {title} {\bibinfo {title} {Schmid-higgs mode in the
  presence of pair-breaking interactions},\ }\href@noop {} {\bibfield
  {journal} {\bibinfo  {journal} {Phys. Rev. B}\ }\textbf {\bibinfo {volume}
  {111}},\ \bibinfo {pages} {174502} (\bibinfo {year} {2025})}\BibitemShut
  {NoStop}%
\bibitem [{\citenamefont {Yang}\ and\ \citenamefont
  {Wu}(2020{\natexlab{b}})}]{PhysRevB.102.014511}%
  \BibitemOpen
  \bibfield  {author} {\bibinfo {author} {\bibfnamefont {F.}~\bibnamefont
  {Yang}}\ and\ \bibinfo {author} {\bibfnamefont {M.~W.}\ \bibnamefont {Wu}},\
  }\bibfield  {title} {\bibinfo {title} {Theory of higgs modes in $d$-wave
  superconductors},\ }\href@noop {} {\bibfield  {journal} {\bibinfo  {journal}
  {Phys. Rev. B}\ }\textbf {\bibinfo {volume} {102}},\ \bibinfo {pages}
  {014511} (\bibinfo {year} {2020}{\natexlab{b}})}\BibitemShut {NoStop}%
\bibitem [{\citenamefont {Schwarz}\ \emph {et~al.}(2020)\citenamefont
  {Schwarz}, \citenamefont {Fauseweh}, \citenamefont {Tsuji}, \citenamefont
  {Cheng}, \citenamefont {Bittner}, \citenamefont {Krull}, \citenamefont
  {Berciu}, \citenamefont {Uhrig}, \citenamefont {Schnyder}, \citenamefont
  {Kaiser} \emph {et~al.}}]{schwarz2020classification}%
  \BibitemOpen
  \bibfield  {author} {\bibinfo {author} {\bibfnamefont {L.}~\bibnamefont
  {Schwarz}}, \bibinfo {author} {\bibfnamefont {B.}~\bibnamefont {Fauseweh}},
  \bibinfo {author} {\bibfnamefont {N.}~\bibnamefont {Tsuji}}, \bibinfo
  {author} {\bibfnamefont {N.}~\bibnamefont {Cheng}}, \bibinfo {author}
  {\bibfnamefont {N.}~\bibnamefont {Bittner}}, \bibinfo {author} {\bibfnamefont
  {H.}~\bibnamefont {Krull}}, \bibinfo {author} {\bibfnamefont
  {M.}~\bibnamefont {Berciu}}, \bibinfo {author} {\bibfnamefont
  {G.}~\bibnamefont {Uhrig}}, \bibinfo {author} {\bibfnamefont
  {A.}~\bibnamefont {Schnyder}}, \bibinfo {author} {\bibfnamefont
  {S.}~\bibnamefont {Kaiser}}, \emph {et~al.},\ }\bibfield  {title} {\bibinfo
  {title} {Classification and characterization of nonequilibrium higgs modes in
  unconventional superconductors},\ }\href@noop {} {\bibfield  {journal}
  {\bibinfo  {journal} {Nat. Commun.}\ }\textbf {\bibinfo {volume} {11}},\
  \bibinfo {pages} {287} (\bibinfo {year} {2020})}\BibitemShut {NoStop}%
\bibitem [{\citenamefont {Yang}\ and\ \citenamefont
  {Chen}(2024)}]{yang24thermodynamic}%
  \BibitemOpen
  \bibfield  {author} {\bibinfo {author} {\bibfnamefont {F.}~\bibnamefont
  {Yang}}\ and\ \bibinfo {author} {\bibfnamefont {L.~Q.}\ \bibnamefont
  {Chen}},\ }\href@noop {} {\bibinfo {title} {Thermodynamic theory of
  disordered 2d superconductors}} (\bibinfo {year} {2024}),\ \Eprint
  {https://arxiv.org/abs/2410.05216} {arXiv:2410.05216} \BibitemShut {NoStop}%
\bibitem [{\citenamefont {Kowal}\ and\ \citenamefont
  {Ovadyahu}(1994)}]{kowal1994disorder}%
  \BibitemOpen
  \bibfield  {author} {\bibinfo {author} {\bibfnamefont {D.}~\bibnamefont
  {Kowal}}\ and\ \bibinfo {author} {\bibfnamefont {Z.}~\bibnamefont
  {Ovadyahu}},\ }\bibfield  {title} {\bibinfo {title} {Disorder induced
  granularity in an amorphous superconductor},\ }\href@noop {} {\bibfield
  {journal} {\bibinfo  {journal} {Solid State Commun.}\ }\textbf {\bibinfo
  {volume} {90}},\ \bibinfo {pages} {783} (\bibinfo {year} {1994})}\BibitemShut
  {NoStop}%
\bibitem [{\citenamefont {Ghosal}\ \emph {et~al.}(1998)\citenamefont {Ghosal},
  \citenamefont {Randeria},\ and\ \citenamefont {Trivedi}}]{ghosal1998role}%
  \BibitemOpen
  \bibfield  {author} {\bibinfo {author} {\bibfnamefont {A.}~\bibnamefont
  {Ghosal}}, \bibinfo {author} {\bibfnamefont {M.}~\bibnamefont {Randeria}},\
  and\ \bibinfo {author} {\bibfnamefont {N.}~\bibnamefont {Trivedi}},\
  }\bibfield  {title} {\bibinfo {title} {Role of spatial amplitude fluctuations
  in highly disordered s-wave superconductors},\ }\href@noop {} {\bibfield
  {journal} {\bibinfo  {journal} {Phys. Rev. Lett.}\ }\textbf {\bibinfo
  {volume} {81}},\ \bibinfo {pages} {3940} (\bibinfo {year}
  {1998})}\BibitemShut {NoStop}%
\bibitem [{\citenamefont {Dubi}\ \emph {et~al.}(2007)\citenamefont {Dubi},
  \citenamefont {Meir},\ and\ \citenamefont {Avishai}}]{dubi2007nature}%
  \BibitemOpen
  \bibfield  {author} {\bibinfo {author} {\bibfnamefont {Y.}~\bibnamefont
  {Dubi}}, \bibinfo {author} {\bibfnamefont {Y.}~\bibnamefont {Meir}},\ and\
  \bibinfo {author} {\bibfnamefont {Y.}~\bibnamefont {Avishai}},\ }\bibfield
  {title} {\bibinfo {title} {Nature of the superconductor--insulator transition
  in disordered superconductors},\ }\href@noop {} {\bibfield  {journal}
  {\bibinfo  {journal} {Nature}\ }\textbf {\bibinfo {volume} {449}},\ \bibinfo
  {pages} {876} (\bibinfo {year} {2007})}\BibitemShut {NoStop}%
\bibitem [{\citenamefont {Sac{\'e}p{\'e}}\ \emph {et~al.}(2020)\citenamefont
  {Sac{\'e}p{\'e}}, \citenamefont {Feigel’man},\ and\ \citenamefont
  {Klapwijk}}]{sacepe2020quantum}%
  \BibitemOpen
  \bibfield  {author} {\bibinfo {author} {\bibfnamefont {B.}~\bibnamefont
  {Sac{\'e}p{\'e}}}, \bibinfo {author} {\bibfnamefont {M.}~\bibnamefont
  {Feigel’man}},\ and\ \bibinfo {author} {\bibfnamefont {T.~M.}\ \bibnamefont
  {Klapwijk}},\ }\bibfield  {title} {\bibinfo {title} {Quantum breakdown of
  superconductivity in low-dimensional materials},\ }\href@noop {} {\bibfield
  {journal} {\bibinfo  {journal} {Nat. Phys.}\ }\textbf {\bibinfo {volume}
  {16}},\ \bibinfo {pages} {734} (\bibinfo {year} {2020})}\BibitemShut
  {NoStop}%
\bibitem [{\citenamefont {Sac{\'e}p{\'e}}\ \emph {et~al.}(2011)\citenamefont
  {Sac{\'e}p{\'e}}, \citenamefont {Dubouchet}, \citenamefont {Chapelier},
  \citenamefont {Sanquer}, \citenamefont {Ovadia}, \citenamefont {Shahar},
  \citenamefont {Feigel'Man},\ and\ \citenamefont
  {Ioffe}}]{sacepe2011localization}%
  \BibitemOpen
  \bibfield  {author} {\bibinfo {author} {\bibfnamefont {B.}~\bibnamefont
  {Sac{\'e}p{\'e}}}, \bibinfo {author} {\bibfnamefont {T.}~\bibnamefont
  {Dubouchet}}, \bibinfo {author} {\bibfnamefont {C.}~\bibnamefont
  {Chapelier}}, \bibinfo {author} {\bibfnamefont {M.}~\bibnamefont {Sanquer}},
  \bibinfo {author} {\bibfnamefont {M.}~\bibnamefont {Ovadia}}, \bibinfo
  {author} {\bibfnamefont {D.}~\bibnamefont {Shahar}}, \bibinfo {author}
  {\bibfnamefont {M.}~\bibnamefont {Feigel'Man}},\ and\ \bibinfo {author}
  {\bibfnamefont {L.}~\bibnamefont {Ioffe}},\ }\bibfield  {title} {\bibinfo
  {title} {Localization of preformed {Cooper} pairs in disordered
  superconductors},\ }\href@noop {} {\bibfield  {journal} {\bibinfo  {journal}
  {Nat. Phys.}\ }\textbf {\bibinfo {volume} {7}},\ \bibinfo {pages} {239}
  (\bibinfo {year} {2011})}\BibitemShut {NoStop}%
\bibitem [{\citenamefont {Sac{\'e}p{\'e}}\ \emph {et~al.}(2008)\citenamefont
  {Sac{\'e}p{\'e}}, \citenamefont {Chapelier}, \citenamefont {Baturina},
  \citenamefont {Vinokur}, \citenamefont {Baklanov},\ and\ \citenamefont
  {Sanquer}}]{sacepe2008disorder}%
  \BibitemOpen
  \bibfield  {author} {\bibinfo {author} {\bibfnamefont {B.}~\bibnamefont
  {Sac{\'e}p{\'e}}}, \bibinfo {author} {\bibfnamefont {C.}~\bibnamefont
  {Chapelier}}, \bibinfo {author} {\bibfnamefont {T.}~\bibnamefont {Baturina}},
  \bibinfo {author} {\bibfnamefont {V.}~\bibnamefont {Vinokur}}, \bibinfo
  {author} {\bibfnamefont {M.}~\bibnamefont {Baklanov}},\ and\ \bibinfo
  {author} {\bibfnamefont {M.}~\bibnamefont {Sanquer}},\ }\bibfield  {title}
  {\bibinfo {title} {Disorder-induced inhomogeneities of the superconducting
  state close to the superconductor-insulator transition},\ }\href@noop {}
  {\bibfield  {journal} {\bibinfo  {journal} {Phys. Rev. Lett.}\ }\textbf
  {\bibinfo {volume} {101}},\ \bibinfo {pages} {157006} (\bibinfo {year}
  {2008})}\BibitemShut {NoStop}%
\bibitem [{\citenamefont {Pracht}\ \emph {et~al.}(2016)\citenamefont {Pracht},
  \citenamefont {Bachar}, \citenamefont {Benfatto}, \citenamefont {Deutscher},
  \citenamefont {Farber}, \citenamefont {Dressel},\ and\ \citenamefont
  {Scheffler}}]{pracht2016enhanced}%
  \BibitemOpen
  \bibfield  {author} {\bibinfo {author} {\bibfnamefont {U.~S.}\ \bibnamefont
  {Pracht}}, \bibinfo {author} {\bibfnamefont {N.}~\bibnamefont {Bachar}},
  \bibinfo {author} {\bibfnamefont {L.}~\bibnamefont {Benfatto}}, \bibinfo
  {author} {\bibfnamefont {G.}~\bibnamefont {Deutscher}}, \bibinfo {author}
  {\bibfnamefont {E.}~\bibnamefont {Farber}}, \bibinfo {author} {\bibfnamefont
  {M.}~\bibnamefont {Dressel}},\ and\ \bibinfo {author} {\bibfnamefont
  {M.}~\bibnamefont {Scheffler}},\ }\bibfield  {title} {\bibinfo {title}
  {Enhanced {Cooper} pairing versus suppressed phase coherence shaping the
  superconducting dome in coupled aluminum nanograins},\ }\href@noop {}
  {\bibfield  {journal} {\bibinfo  {journal} {Phys. Rev. B}\ }\textbf {\bibinfo
  {volume} {93}},\ \bibinfo {pages} {100503} (\bibinfo {year}
  {2016})}\BibitemShut {NoStop}%
\bibitem [{\citenamefont {Dubouchet}\ \emph {et~al.}(2019)\citenamefont
  {Dubouchet}, \citenamefont {Sac{\'e}p{\'e}}, \citenamefont {Seidemann},
  \citenamefont {Shahar}, \citenamefont {Sanquer},\ and\ \citenamefont
  {Chapelier}}]{dubouchet2019collective}%
  \BibitemOpen
  \bibfield  {author} {\bibinfo {author} {\bibfnamefont {T.}~\bibnamefont
  {Dubouchet}}, \bibinfo {author} {\bibfnamefont {B.}~\bibnamefont
  {Sac{\'e}p{\'e}}}, \bibinfo {author} {\bibfnamefont {J.}~\bibnamefont
  {Seidemann}}, \bibinfo {author} {\bibfnamefont {D.}~\bibnamefont {Shahar}},
  \bibinfo {author} {\bibfnamefont {M.}~\bibnamefont {Sanquer}},\ and\ \bibinfo
  {author} {\bibfnamefont {C.}~\bibnamefont {Chapelier}},\ }\bibfield  {title}
  {\bibinfo {title} {Collective energy gap of preformed {Cooper} pairs in
  disordered superconductors},\ }\href@noop {} {\bibfield  {journal} {\bibinfo
  {journal} {Nat. Phys.}\ }\textbf {\bibinfo {volume} {15}},\ \bibinfo {pages}
  {233} (\bibinfo {year} {2019})}\BibitemShut {NoStop}%
\bibitem [{\citenamefont {Crane}\ \emph {et~al.}(2007)\citenamefont {Crane},
  \citenamefont {Armitage}, \citenamefont {Johansson}, \citenamefont
  {Sambandamurthy}, \citenamefont {Shahar},\ and\ \citenamefont
  {Gr{\"u}ner}}]{crane2007survival}%
  \BibitemOpen
  \bibfield  {author} {\bibinfo {author} {\bibfnamefont {R.}~\bibnamefont
  {Crane}}, \bibinfo {author} {\bibfnamefont {N.}~\bibnamefont {Armitage}},
  \bibinfo {author} {\bibfnamefont {A.}~\bibnamefont {Johansson}}, \bibinfo
  {author} {\bibfnamefont {G.}~\bibnamefont {Sambandamurthy}}, \bibinfo
  {author} {\bibfnamefont {D.}~\bibnamefont {Shahar}},\ and\ \bibinfo {author}
  {\bibfnamefont {G.}~\bibnamefont {Gr{\"u}ner}},\ }\bibfield  {title}
  {\bibinfo {title} {Survival of superconducting correlations across the
  two-dimensional superconductor-insulator transition: {A} finite-frequency
  study},\ }\href@noop {} {\bibfield  {journal} {\bibinfo  {journal} {Phys.
  Rev. B}\ }\textbf {\bibinfo {volume} {75}},\ \bibinfo {pages} {184530}
  (\bibinfo {year} {2007})}\BibitemShut {NoStop}%
\bibitem [{\citenamefont {Noat}\ \emph {et~al.}(2013)\citenamefont {Noat},
  \citenamefont {Cherkez}, \citenamefont {Brun}, \citenamefont {Cren},
  \citenamefont {Carbillet}, \citenamefont {Debontridder}, \citenamefont
  {Ilin}, \citenamefont {Siegel}, \citenamefont {Semenov}, \citenamefont
  {H{\"u}bers} \emph {et~al.}}]{noat2013unconventional}%
  \BibitemOpen
  \bibfield  {author} {\bibinfo {author} {\bibfnamefont {Y.}~\bibnamefont
  {Noat}}, \bibinfo {author} {\bibfnamefont {V.}~\bibnamefont {Cherkez}},
  \bibinfo {author} {\bibfnamefont {C.}~\bibnamefont {Brun}}, \bibinfo {author}
  {\bibfnamefont {T.}~\bibnamefont {Cren}}, \bibinfo {author} {\bibfnamefont
  {C.}~\bibnamefont {Carbillet}}, \bibinfo {author} {\bibfnamefont
  {F.}~\bibnamefont {Debontridder}}, \bibinfo {author} {\bibfnamefont
  {K.}~\bibnamefont {Ilin}}, \bibinfo {author} {\bibfnamefont {M.}~\bibnamefont
  {Siegel}}, \bibinfo {author} {\bibfnamefont {A.}~\bibnamefont {Semenov}},
  \bibinfo {author} {\bibfnamefont {H.-W.}\ \bibnamefont {H{\"u}bers}}, \emph
  {et~al.},\ }\bibfield  {title} {\bibinfo {title} {Unconventional
  superconductivity in ultrathin superconducting {NbN} films studied by
  scanning tunneling spectroscopy},\ }\href@noop {} {\bibfield  {journal}
  {\bibinfo  {journal} {Phys. Rev. B}\ }\textbf {\bibinfo {volume} {88}},\
  \bibinfo {pages} {014503} (\bibinfo {year} {2013})}\BibitemShut {NoStop}%
\bibitem [{\citenamefont {Eilenberger}(1968)}]{eilenberger1968transformation}%
  \BibitemOpen
  \bibfield  {author} {\bibinfo {author} {\bibfnamefont {G.}~\bibnamefont
  {Eilenberger}},\ }\bibfield  {title} {\bibinfo {title} {Transformation of
  gorkov's equation for type ii superconductors into transport-like
  equations},\ }\href@noop {} {\bibfield  {journal} {\bibinfo  {journal} {Z.
  Phys.}\ }\textbf {\bibinfo {volume} {214}},\ \bibinfo {pages} {195} (\bibinfo
  {year} {1968})}\BibitemShut {NoStop}%
\end{thebibliography}
%

\end{document}